\documentclass[a4paper,11pt]{article}
\pdfoutput=1 

\usepackage{jheppub} 

\usepackage{fancyhdr}
\usepackage{graphicx}

\usepackage{braket}
\usepackage{slashed}

\newcommand{\be}{\begin{equation}}
\newcommand{\ee}{\end{equation}}
\newcommand{\bea}{\begin{eqnarray}}

\newcommand{\eea}{\end{eqnarray}}

\newcommand{\refe}[1]{Eqn.~(\ref{#1})}
\newcommand{\adj}{\dagger}

\newcommand{\Rmnum}[1]{\expandafter\@slowromancap\romannumeral #1@}

\begin{document}

 \title{\boldmath En-gauging Naturalness}

\begin{flushright}DESY 13-171$\quad$$\quad$\\ LAPTH-059/13$\quad$\\ TUM-HEP-909-13\end{flushright} 




\author[a,b]{Aoife Bharucha,}
\author[c]{Andreas Goudelis}
\author[d]{and Moritz McGarrie}


\affiliation[a]{II. Institut f\"ur Theoretische Physik, Universit\"at Hamburg,
Luruper Chaussee
149, D-22761 \\Hamburg, Germany}
\affiliation[b]{Physik Department T31, 
Technische Universit\"at M\"unchen,
James-Franck-Stra\ss e~1, D-85748 \\Garching, Germany}
\affiliation[c]{LAPTh, Universit\'e de Savoie, CNRS, 9 Chemin de Bellevue, B.P.\
110,
F-74941 Annecy-le-Vieux, France
}
\affiliation[d]{Deutsches Elektronen-Synchrotron, DESY,\ Notkestrasse 85, D-22607
Hamburg, Germany}
\emailAdd{aoife.bharucha@tum.de}
\emailAdd{andreas.goudelis@lapth.cnrs.fr}
\emailAdd{moritz.mcgarrie@desy.de}

\abstract{ The discovery of a 125.5 GeV Higgs with standard model-like couplings
and naturalness considerations motivate gauge extensions of the MSSM. We analyse 
two variants of such an extension and carry out a
phenomenological study of regions of the parameter space 
satisfying current direct and indirect constraints, employing state-of-the-art two-loop 
RGE evolution
and GMSB boundary conditions. We find that due to the appearance of non-decoupled 
D-terms it is possible to obtain a $125.5$ GeV
Higgs with stops below 2 TeV, while the uncolored sparticles could still lie 
within reach of the LHC. We compare the contributions of the stop sector and the 
non-decoupled D-terms to the Higgs mass, and study 
their effect on the Higgs couplings. We further investigate the 
nature of the next-to lightest supersymmetric particle, in light of the GMSB 
motivated searches currently being pursued by ATLAS and CMS.
}

\maketitle
\flushbottom

\section{Introduction} \label{sec:intro}
The recent discovery of  a $\sim 125.5$ GeV particle consistent with the
properties of the standard model Higgs boson \cite{Aad:2012tfa,Chatrchyan:2012ufa} and no direct
evidence of supersymmetry (SUSY) in the current LHC data are pushing 
traditional setups of gauge-mediated supersymmetry breaking (GMSB) 
for the minimal supersymmetric standard model (MSSM) into fine-tuned territory. 
In the MSSM, to obtain a lightest CP-even scalar of the
observed mass requires either heavy stops, thereby introducing a \emph{naturalness} or fine-tuning
problem, or substantial left-right stop mixing, which, being strongly
dependent on the trilinear soft term $A_t$, is heavily influenced by the
mechanism of supersymmetry-breaking mediation that is invoked. In particular, in
GMSB trilinear terms such as $A_t$ are
vanishing at the supersymmetry breaking scale $M$, and a large $A_t$ can only be 
generated via renormalisation group evolution. This requires the scale $M$ to be very 
high, which is also detrimental to the naturalness of the theory.
Moreover, in minimal SUSY models, choosing heavy stops results in the sparticle
spectrum becoming heavier and beyond the reach of the LHC, and consequently 
phenomenologically  less interesting.

The heart of the problem appears to be that in the MSSM, the tree-level Higgs
mass is simply too small, the upper bound being the mass of the $Z$ boson. 
If however we drop minimality from our criteria, natural scenarios of
supersymmetry breaking \emph{with} discovery potential still persist. These 
usually involve lifting the Higgs potential at tree-level through
non-decoupled F or D type terms \cite{Batra:2003nj,Dine:2007xi,Blum:2012ii}, the later of which arise 
when the MSSM is extended by additional gauge groups. The less-studied 
gauge extensions involving non-decoupled D-terms could enhance the tree-level Higgs mass, 
resulting in detectable deviations in the Higgs couplings, 
and further induce suppressions to the scalar masses compared to minimal GMSB.
Such models may therefore have a direct impact on phenomenology and, unsurprisingly, 
have of late found increasing interest~\cite{Csaki:2001em,Cheng:2001an,Batra:2004vc,Delgado:2004pr,DeSimone:2008gm, Medina:2009ey,McGarrie:2010qr,
Craig:2011yk,Auzzi:2012dv,Huo:2012tw,d2013fitting}.

In this paper we study a quiver model or gauge extension of the MSSM within
the framework of GMSB, to determine whether it is indeed possible to reproduce
the observed Higgs boson mass while keeping the stop masses below roughly $2$
TeV. In particular we build a tailor-made spectrum generator for our model using
the publicly available tool \texttt{SARAH} \cite{Staub:2008uz,Staub:2010jh,Staub:2012pb}. This  allows us to perform the renormalization group
evolution at two loops and analyze several aspects of the model's phenomenology.
Although we do not carry out a thorough ``naturalness" or fine-tuning study, it
is at least clear that qualitively, having stops lighter than benchmark minimal
GMSB certainly improves the relative naturalness of the model. We therefore
study the resulting spectra for the model, consistent with experimental results,
in particular the Higgs sector. We further
investigate possible signatures of this model at the LHC, taking into 
account the latest results of GMSB motivated searches. 

It is useful to summarise the current status of the literature that explores two site or minimal quiver models.  Initially a simple two site deconstruction, similar to  our model MI (described later in section \ref{sec:superpotential}), with gauge mediated boundary conditions was proposed in \cite{Cheng:2001an}, in particular in which both quiver gauge groups $G_A$ and $G_B$ are $U(1)\times SU(2)\times SU(3)$  and in which both site A and B  gauge couplings unify separately. As in such models sfermion soft masses are typically smaller than gaugino soft masses it unwittingly allows for the foundations of a natural spectrum.  However the resulting tree-level Higgs mass is described by the MSSM, such that the observed Higgs mass \cite{Aad:2012tfa,Chatrchyan:2012ufa} must therefore be generated by large $A_t$ or heavy stops.  Later in \cite{DeSimone:2008gm} the non decoupled D-terms of \cite{Batra:2003nj} in the Higgs sector were included by hand, although they did not contribute significantly and the resulting Higgs mass of their benchmarks, of around $116$ GeV, are also now firmly excluded.   In \cite{Batra:2004vc,Delgado:2004pr} a model similar to our model MII\footnote{The first references of split families we are aware of are in \cite{Muller:1996dj,Malkawi:1996fs,Malkawi:1999sa}, in a non-SUSY context.}, was sketched in which it is argued that one could obtain a) a linking field vev $v<10$ TeV b) unification of each gauge site separately, as well as c) lighter 3rd generation scalars relative to 1st and 2nd, due to location of generations in different sites. Most of the studies so far mentioned were carried out at tree-level. The formulas for the soft masses in the case of gauge mediation were also given in \cite{McGarrie:2010qr,Auzzi:2010mb,McGarrie:2011dc} where some interest had developed in finding and making precise models where sfermions were lighter than gauginos at the messenger scale.   In light of the Higgs discovery \cite{Aad:2012tfa,Chatrchyan:2012ufa} and naturalness, the split families models of \cite{Batra:2004vc,Delgado:2004pr} have re-emerged \cite{Craig:2011yk,Craig:2012hc,Auzzi:2012dv}. 

In \cite{Auzzi:2012dv} a study using the MSSM spectrum generator, \texttt{SOFTSUSY} \cite{Allanach:2001kg} combined with some private codes found that the model based on \cite{Batra:2003nj,Batra:2004vc,Delgado:2004pr,DeSimone:2008gm,McGarrie:2010qr,Auzzi:2010mb,McGarrie:2011dc,Craig:2011yk,Craig:2012hc} could not simultaneously achieve unification and obtain the necessary enhancement of the Higgs mass. If the necessary enhancement was obtained then the gauge groups hit a Landau pole much before the GUT scale. Some models involving three sites were proposed to alleviate the problem.  The study of  \cite{Auzzi:2012dv} misses some leading order effects to some one loop RGEs and as such a more comprehensive study and implementation of these models is necessary.  Furthermore, simpler (more minimal) variants of this class of model may fair better.  The core issue is that the $SU(3)_A\times SU(3)_B$ matter  component of the quiver introduces additional flavours to each side respectively, charged also under the electroweak groups, and this additional matter content can cause the Landau pole problem, if one wishes for $g_A>g_B$ to increase the effect of the Higgs enhancement.  Given the importance of the Higgs enhancement, perhaps it is worth removing the copy of $SU(3)$ (and possibly complete unification) in favour of naturalness. Such a setup should anyway test what is the maximum allowed enhancement to the Higgs from the electroweak quiver without worrying about unification or Landau poles.

There are clearly a number of important and unresolved issues which our paper addresses.  The most detailed study so far, in \cite{Auzzi:2012dv}, although useful does not include some leading one-loop contributions to the RGEs (see appendix \ref{sec:moreimplementation} and \cite{modelpdf}), it is essentially based on an MSSM spectrum generator \cite{Allanach:2001kg} such that the Higgs enhancement is added afterwards by hand.  Considering the importance not only on Higgs observables but the pressing issue of Naturalness this is not sufficient and a dedicated (and publicly available) spectrum generator and study of this type of model is required. It is of course essential that the HEP community has access to a myriad of spectrum generators, particularly those relevant to the Higgs sector, and more ideally several spectrum generators dedicated to the exact same model.  This has been incredibly fruitful in studies based on the MSSM and NMSSM. This paper is an important step in that direction as it supplies a full two loop spectrum generator, with the non decoupled D-term Higgs enhancements included in the model and in the self energies for the calculation of the masses. What is particularly exciting and novel regarding our work is that our spectrum generator is on a par with all currently publicly available spectrum generators (such as \cite{Baer:1993ae,Allanach:2001kg,Djouadi:2002ze,Porod:2003um}) and is the first dedicated spectrum generator to include the non decoupled D-term enhancement to the Higgs sector.  Secondly a dedicated study can address the important question of, if by removing the quiver of $SU(3)$,  one can then attempt to maximise the Higgs enhancement and achieve a linking field vev, $v$,  below $10$ TeV, whilst keeping all gauge couplings perturbative to the Planck scale.  In fact we found that this cannot be done.  In addition this paper supplies a number of important and new results: it supplies a full derivation of the non decoupled D-term effect to all scalars and not just the Higgs fields, these contributions have so far gone unnoticed in the literature, but may be quite relevent to precisions studies at $e^{+},e^{-}$ colliders.  Further, this implementation includes all RGEs to two loops for all parameters and all anomalous dimensions for all fields, supplying for the first time, the full anatomy of such a setup.

The outline of the paper is as follows: In section \ref{sec:keyresults} we give
a first account of our key findings. In section \ref{sec:TheModel} we present
the specific realization of a quiver gauge theory that we will be studying in
the following sections, discuss its general features and our choice of
parameterisation of the soft term boundary conditions. Amusingly,
these setups in some sense also model the effects of a truncated theory of gauge
fields in an extra dimension and we make
a small digression in section~\ref{digression} to discuss this connection. In
section \ref{sec:ToolsObservables} we discuss the concrete implementation of our
quiver setup, the model's parameter space and the phenomenological constraints
it is subject to. Then, in section~\ref{sec:results} we present our results on
the phenomenology of the model  and discuss the features of its particle
spectrum. Finally, in section \ref{sec:conclusions} we 
conclude. The appendix~\ref{sec:moreimplementation} contains details about the
implementation of our framework along with some important relations. 
This paper is accompanied by a supplementary document containing details with 
regards to the implementation~\cite{modelpdf}.

\begin{figure}[t!]
\begin{center}
\includegraphics[scale=0.5]{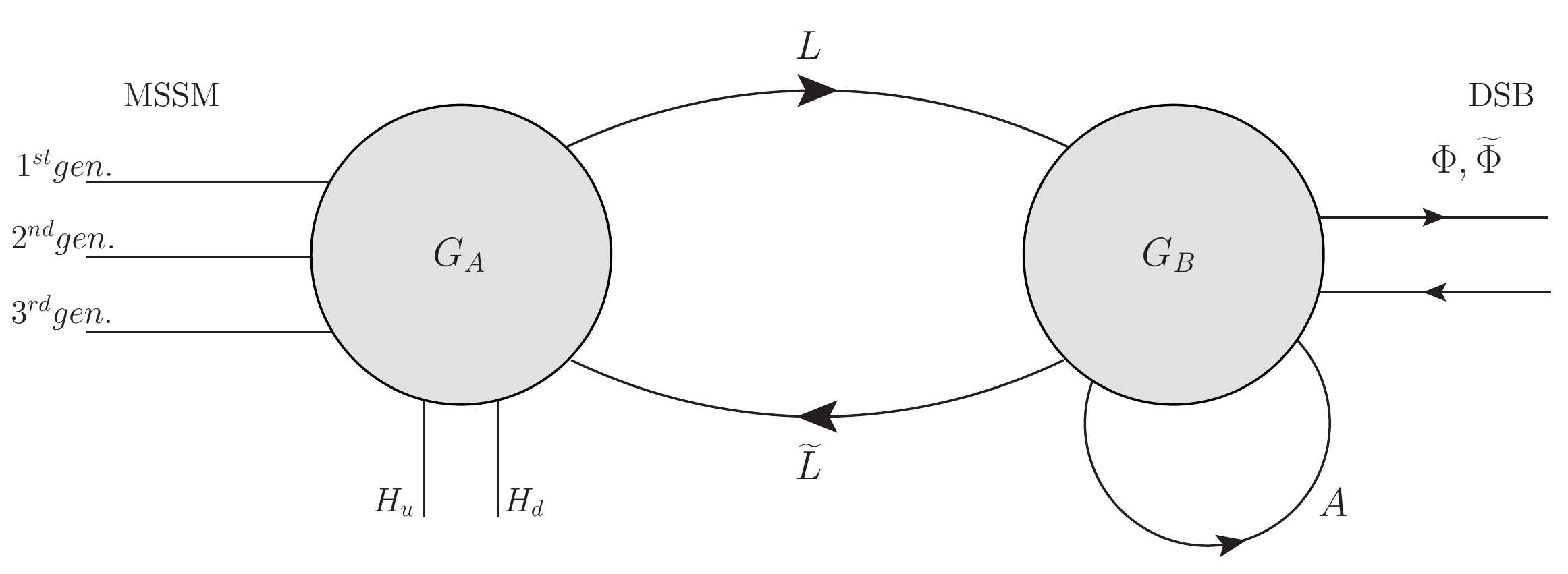}
\caption{A picture to represent the quiver module of the electroweak sector for
MI as in table \ref{table:matterfieldsV1}.   The electroweak part of the
supersymmetric standard model is on site A, with messenger
fields ($\Phi,\tilde{\Phi}$) coupled to another site, site B.  The linking
fields ($L,\tilde{L}$) connect the two sites.  The adjoint field ($A$) is
charged on the second site, site B. The singlet field ($K$) is not shown.} 
\label{fig:MI}
\end{center}
\end{figure}
\section{Key results}\label{sec:keyresults}
It is useful here to give a first summary of our key results. 
In this work we will show that
\begin{itemize}
  \item With the setup we adopt it is perfectly possible to obtain a $125.5$ GeV
Higgs with stops lighter than 2 TeV, gluinos of order 1600 GeV, light
electroweakinos and sleptons potentially within the reach of the LHC, all within
a GMSB framework (see benchmarks in table \ref{tab:ExampleSpectra}).
  \item The correct Higgs mass is obtained through \emph{non decoupled D-term}
contributions in the low-energy lagrangian that lift the Higgs boson mass at
tree level, as shown in figure~\ref{fig:mHvsTb}.
  \item These terms also modify the Higgs branching ratios but well within
current LHC bounds, and could be probed by the ILC as seen in figure~\ref{fig:mHvsRb}.
  \item The light uncoloured sparticle spectrum is achieved primarily due to the
specific supersymmetry breaking mediation mechanism we employ (see figure~\ref{fig:mgVmnlsp}).
\end{itemize}
These results are obtained by implementing our model into the publicly available
package \texttt{SARAH}, which enables us to create a spectrum generator in order to
perform the RGE evolution of all model parameters at two loops, from the
messenger scale $M$ down to the TeV scale, with GMSB boundary conditions.

We have implemented five gauge groups at full one- and two-loop running, plus
one-loop self energies, from the GUT or messenger scale, Higgsing and breaking
to the diagonal subgroup of 4 gauge groups, while finite shifts and threshold
corrections are carefully applied for each degree of freedom.

We finally stress that we have implemented a conservative (precise) formulation
of GMSB with full two loop equations for soft masses in the hidden sector. Still
at this level of specification a reasonably natural spectrum is obtained, which
demonstrates the ease with which much lighter spectra would be obtainable if
these high standards were relaxed or some more phenomenological parameterisation
adopted. The framework that we have developed quite straightforwardly admits
numerous extensions such as inclusion of $U(1)$ kinetic mixing or quivering the
$SU(3)$ sector, tasks  which are left for future work. All of these remarks will
be clarified in the following sections.


\section{An electroweak quiver}\label{sec:TheModel}
In this paper we wish to explore two different quiver models for comparison. 
The first carries the generic features of non-decoupled D-terms and, in the case
of GMSB, suppressed scalar soft masses versus gauginos. The second is a
flavourful extension of the first model to achieve lighter stops than the first
two generations, which still obeys all anomaly cancellations. A common feature
is that we will apply a gauge mediated supersymmetry breaking scenario to both
and both are characterised by the scale of supersymmetry breaking, $M$, and the
vevs of the linking fields\footnote{not to be confused with $v_{ew}$.}, $v$. In
this section we outline these models and their features.
\subsection{The models and features}\label{sec:superpotential}

\begin{figure}[t!]
\begin{center}
\includegraphics[scale=0.5]{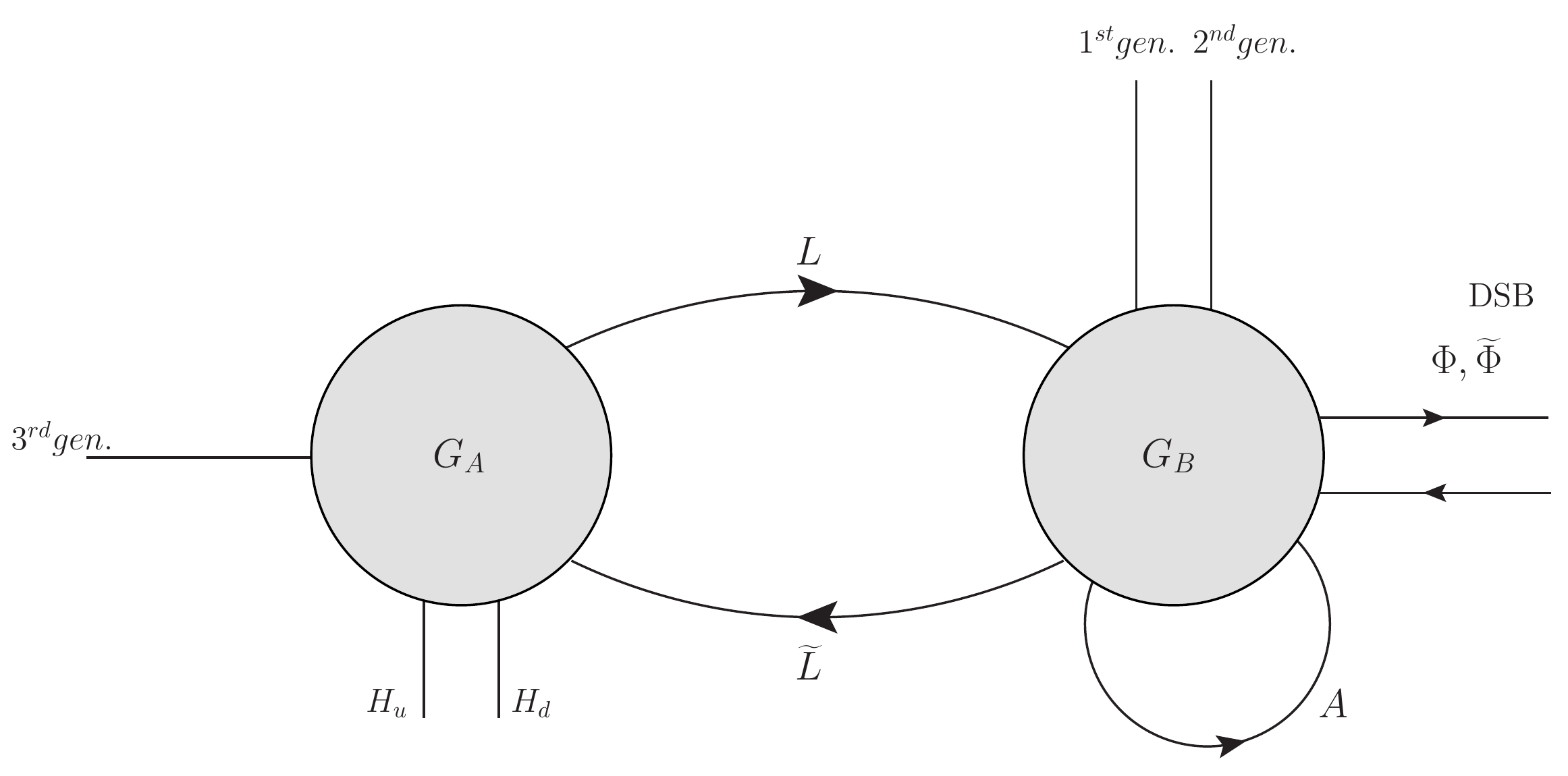}
\caption{The quiver module of the electroweak sector for MII as in table
\ref{table:matterfieldsV2}.   The first and second generation matter is charged
under site B and the third generation and MSSM Higgs fields are charged under
site A. The messenger fields ($\Phi,\tilde{\Phi}$) are charged under site B. 
The linking fields ($L,\tilde{L}$) connect the two sites.  } 
\label{fig:MII}
\end{center}
\end{figure}

Let us consider an electroweak two-site quiver with gauge group $G_A \times G_B
\times SU(3)_c$, where $G_A = SU(2)_A \times U(1)_A$ and  $G_B = SU(2)_B \times
U(1)_B$ as in table \ref{table:vectorfields}.
\renewcommand{\arraystretch}{1.1}
\begin{table}
\begin{center} 
\begin{tabular}{|c|c|c|c|c|} 
\hline \hline 
SF & Spin \(\frac{1}{2}\) & Spin 1 & \(SU(N)\) & Coupling \\ 
 \hline 
\(\hat{B}_{A}\) & \(\tilde{B}_{A}\) & \(B_{A}\) & \(U(1)_{A}\) & \(g_{A1}\) \\ 
\(\hat{W}_B\) & \(\tilde{W}_{B}\) & \(W_{B}\) & \(\text{SU}(2)_{B}\) &
\(g_{B2}\) \\ 
\(\hat{g}\) & \(\tilde{g}\) & \(g\) & \(\text{SU}(3)_c\) & \(g_{3}\) \\ 
\(\hat{B}_{B}\) & \(\tilde{B}_{B}\) & \(B_{B}\) & \(U(1)_{B}\) & \(g_{B1}\) \\ 
\(\hat{W}_{A}\) & \(\tilde{W}_{A}\) & \(W_{A}\) & \(\text{SU}(2)_{A}\) &
\(g_{A2}\) \\ 
\hline \hline
\end{tabular} \caption{Gauge superfields of the model.
\label{table:vectorfields}}
\end{center} 
\end{table}
\renewcommand{\arraystretch}{}
The two sites are connected by means of a pair of linking chiral superfields
$\hat{L}\ ,\hat{\tilde{L}}$. These superfields will play a crucial role both in
the breaking of the enlarged gauge group to the MSSM gauge groups, by obtaining
vevs, and in the mediation of supersymmetry breaking effects. 
Moreover, the setup includes a singlet chiral superfield $K$, whose role will be
clarified shortly, as well as an additional superfield $A$ transforming as the
adjoint of $SU(2)_B$ that serves the role of giving masses to certain
fermionic components of the linking fields upon $G_A \times G_B$ breaking. 
Much below the higgsing scale, $v$, the quiver fields usually decouple and so for
phenomenological purposes at low energies the model is essentially MSSM-like
with the addition of an effective action for the Higgs potential. It will be
useful then to refer to the enlarged gauge groups as regime 1 and the MSSM as
regime 2. This paper is based on two models which are as follows:
\renewcommand{\arraystretch}{1.2}
\begin{table}
\begin{center} 
\begin{tabular}{|c|c|c|c|c|} 
\hline \hline 
SF & Spin 0 & Spin \(\frac{1}{2}\) & $G_A \times G_B \times SU(3)_c$ \ \ (MI)\\
\hline 
\(\hat{q}^{f}\) & \(\tilde{q}^{f}\) & \(q^{f}\) & \(({\bf 2}, \frac{1}{6}, {\bf
1}, 0, {\bf 3}) \)  \\ 
\(\hat{l}^{f}\) & \(\tilde{l}^{f}\) & \(l^{f}\) & \(({\bf 2},-\frac{1}{2}, {\bf
1} ,0, {\bf 1}) \) \\ 
\(\hat{H}_d\)   & \(H_d\) & \(\tilde{H}_d\)     & \(({\bf 2},-\frac{1}{2}, {\bf
1}, 0, {\bf 1}) \)  \\ 
\(\hat{H}_u\) & \(H_u\) & \(\tilde{H}_u\)       & \(({\bf 2}, \frac{1}{2}, {\bf
1}, 0, {\bf 1}) \)  \\ 
\(\hat{d}^{f}\) & \(\tilde{d}_R^{f*}\) & \(d_R^{f*}\) & \(({\bf 1}, \frac{1}{3},
{\bf 1}, 0, {\bf \overline{3}}) \)  \\ 
\(\hat{u}^{f}\) & \(\tilde{u}_R^{f*}\) & \(u_R^{f*}\) & \(({\bf 1},-\frac{2}{3},
{\bf 1}, 0, {\bf \overline{3}}) \)  \\
\(\hat{e}^{f}\) & \(\tilde{e}_R^{f*}\) & \(e_R^{f*}\) & \(({\bf 1}, 1, {\bf 1},
0, {\bf 1}) \)  \\ \hline\hline
\(\hat{L}\) & \(L\) & \(\psi_L\) & \(({\bf 2}, -\frac{1}{2}, {\bf \overline{2}},
\frac{1}{2}, {\bf 1}) \)   \\ 
\(\hat{\tilde{L}}\) & \(\tilde{L}\) & \(\psi_{\tilde{L}}\) & \(({\bf
\overline{2}}, \frac{1}{2}, {\bf 2}, -\frac{1}{2}, {\bf 1}) \)  \\
\(\hat{K}\) & \(K\) & \(\psi_{K}\) & \(({\bf 1}, 0, {\bf 1}, 0, {\bf 1}) \)  \\
\(\hat{A}\) & \(\text{A}\) & \(\psi_A\) & \(({\bf 1}, 0, {\bf 3}, 0, {\bf 1}) \)
 \\
\hline \hline
\end{tabular} \caption{Chiral superfields of the model MI. The index $f$ runs
over all three generations. The representation ordering corresponds to
$(SU(2)_A, U(1)_A, SU(2)_B, U(1)_B, SU(3)_c)$. The superpotential is
\refe{eq:MSSM}  and \refe{eq:Quiver} and the supersymmetry breaking messenger
fields are charged under site B. \label{table:matterfieldsV1}}
\end{center} 
\end{table}
\renewcommand{\arraystretch}{}
\newline
\\  {\bf[Model I]}: The first model (MI) is a basic example of a quiver model
where the MSSM chiral superfields are taken to be charged under site A
identically to their charges under the MSSM gauge group and are neutral under
site B (see figure \ref{fig:MI} and table \ref{table:matterfieldsV1}). The
superpotential of the MSSM-like matter is given by
\begin{align} 
W_{\text{SSM}} = & \,  Y_u\,\hat{u}\,\epsilon_{ij} \hat{q}^i\,\hat{H}^j_u\,- Y_d
\,\hat{d}\,\epsilon_{ij} \hat{q}^i\,\hat{H}^j_d\,- Y_e \,\hat{e}\,\epsilon_{ij}
\hat{l}^i\,\hat{H}^j_d\,+\mu\epsilon_{ij}\,\hat{H}^i_u\,\hat{H}^j_d\,\,
\end{align} \label{eq:MSSM}
with $i,j,k$ labelling $SU(2)$ indices, and as this group is pseudo real, the
$\bar{2}$ $\phi_i$ is simply $\epsilon_{ij}\phi^j$ of the $2$ representation
$\phi^i$ of $SU(2)$. The superpotential of the quiver module is given by
\begin{align}
W_{\text{Quiver}}=\frac{Y_K}{2} \hat{K} (\,\hat{L}_i^j\,\hat{\tilde{L}}^i_j\,-
V^2 )\,+Y_A\,\hat{L}_i^j\,\hat{A}_j^k\,\hat{\tilde{L}}_k^i\, . 
\,\label{eq:Quiver}
\end{align}
A model with a similar structure albeit based on a more enlarged gauge group 
was first introduced in \citep{Cheng:2001an}. 
The general features of this model will be outlined below and unless stated in
the text all RGEs and equations of this paper refer to MI.
\newline
\\
{\bf[Model II]}: The second model (MII) is a flavourful deformation of model I
in that by construction the first and second generation MSSM chiral superfields
are taken to be charged under site B and neutral under site A while the 3rd
generation and the Higgs fields are kept on site A (see figure \ref{fig:MII} and
table \ref{table:matterfieldsV2}). Similar representation assignments have been 
considered, for example, in \cite{Batra:2004vc,Delgado:2004pr} and then later in \cite{Craig:2011yk, Craig:2012hc, Auzzi:2012dv} in the framework of 
models of natural supersymmetry that could potentially further address the flavour problem.
The superpotential we use in regime 1 is
\begin{align} 
W_{\text{MII}} = & \,  Y^3_u\,\hat{u}^3\,\epsilon_{ij}
\hat{q}^{3i}\,\hat{H}^j_u\,- Y^3_d \,\hat{d}^3\,\epsilon_{ij}
\hat{q}^{3i}\,\hat{H}^j_d\,- Y^3_e \,\hat{e}^3\,\epsilon_{ij}
\hat{l}^{3i}\,\hat{H}^j_d\,+\mu\,\epsilon_{ij} \hat{H}^i_u\,\hat{H}^j_d\,\, 
\label{eq:MSS2} \end{align} 
plus the quiver superpotential \refe{eq:Quiver}.  In regime 2 after returning to
the MSSM gauge groups, we adopt the MSSM superpotential by \refe{eq:MSSM}.

\renewcommand{\arraystretch}{1.2}
\begin{table}
\begin{center} 
\begin{tabular}{|c|c|c|c|c|c|} 
\hline \hline 
SF & Spin 0 & Spin \(\frac{1}{2}\) & $G_A \times G_B \times SU(3)_c$ \ \
(MII)\\ 
\hline 
\(\hat{q}^{1,2}\) & \(\tilde{q}^{1,2}\) & \(q^{1,2}\) & \(({\bf 1}, 0, {\bf 2},
\frac{1}{6}, {\bf 3}) \) \\ 
\(\hat{l}^{1,2}\) & \(\tilde{l}^{1,2}\) & \(l^{1,2}\) & \(({\bf 1} ,0, {\bf
2},-\frac{1}{2}, {\bf 1}) \)\\ 
\(\hat{d}^{1,2}\) & \(\tilde{d}_R^{1,2*}\) & \(d_R^{1,2*}\) & \(( {\bf 1}, 0,
{\bf 1}, \frac{1}{3},{\bf \overline{3}}) \) \\ 
\(\hat{u}^{1,2}\) & \(\tilde{u}_R^{1,2*}\) & \(u_R^{1,2*}\) & \(({\bf 1}, 0,
{\bf 1},-\frac{2}{3}, {\bf \overline{3}}) \) \\
\(\hat{e}^{1,2}\) & \(\tilde{e}_R^{1,2*}\) & \(e_R^{1,2*}\) & \(({\bf 1}, 0,
{\bf 1}, 1, {\bf 1}) \) \\
\hline
\(\hat{H}_d\)   & \(H_d\) & \(\tilde{H}_d\)     & \(({\bf 2},-\frac{1}{2}, {\bf
1}, 0, {\bf 1}) \)  \\ 
\(\hat{H}_u\) & \(H_u\) & \(\tilde{H}_u\)       & \(({\bf 2}, \frac{1}{2}, {\bf
1}, 0, {\bf 1}) \)  \\ 
\(\hat{q}^{3}\) & \(\tilde{q}^{3}\) & \(q^{3}\) & \(({\bf 2}, \frac{1}{6}, {\bf
1}, 0, {\bf 3}) \)  \\ 
\(\hat{l}^{3}\) & \(\tilde{l}^{3}\) & \(l^{3}\) & \(({\bf 2},-\frac{1}{2}, {\bf
1} ,0, {\bf 1}) \) \\ 
\(\hat{d}^{3}\) & \(\tilde{d}_R^{3*}\) & \(d_R^{3*}\) & \(({\bf 1}, \frac{1}{3},
{\bf 1}, 0, {\bf \overline{3}}) \)  \\ 
\(\hat{u}^{3}\) & \(\tilde{u}_R^{3*}\) & \(u_R^{3*}\) & \(({\bf 1},-\frac{2}{3},
{\bf 1}, 0, {\bf \overline{3}}) \)  \\
\(\hat{e}^{3}\) & \(\tilde{e}_R^{3*}\) & \(e_R^{3*}\) & \(({\bf 1}, 1, {\bf 1},
0, {\bf 1}) \) \\  \hline\hline
\(\hat{L}\) & \(L\) & \(\psi_L\) & \(({\bf 2}, -\frac{1}{2}, {\bf \overline{2}},
\frac{1}{2}, {\bf 1}) \)   \\ 
\(\hat{\tilde{L}}\) & \(\tilde{L}\) & \(\psi_{\tilde{L}}\) & \(({\bf
\overline{2}}, \frac{1}{2}, {\bf 2}, -\frac{1}{2}, {\bf 1}) \)  \\
\(\hat{K}\) & \(K\) & \(\psi_{K}\) & \(({\bf 1}, 0, {\bf 1}, 0, {\bf 1}) \)  \\
\(\hat{A}\) & \(\text{A}\) & \(\psi_A\) & \(({\bf 1}, 0, {\bf 3}, 0, {\bf 1}) \)
 \\
\hline \hline
\end{tabular} \caption{Chiral superfields of the model MII. The first and second
generation are charged under site B. The third generation are charged under site
A. The representation ordering corresponds to $(SU(2)_A, U(1)_A, SU(2)_B,
U(1)_B, SU(3)_c)$.  The superpotential is given by \refe{eq:MSS2}  and
\refe{eq:Quiver}.  The supersymmetry breaking messenger fields are charged under
site B.  \label{table:matterfieldsV2}}
\end{center} 
\end{table}
\renewcommand{\arraystretch}{}

\subsection{Gauge symmetry breaking}\label{sec:gaugeSB}
We now describe certain features of the general setup. 
The superpotential \refe{eq:Quiver} gives rise to a scalar potential which when
minimized sets a vacuum expectation value for the scalar components of the
linking fields of the model. Denoting 
\be
L=\left( \begin{array}{cc}
\varphi_{L1} &\varphi_{L2} \\
\varphi_{L3} &\varphi_{L4}
\end{array}\right) \ , 
\ee
in the absence of supersymmetry breaking, we write
\be
\braket{L}=\braket{\tilde{L}}=v\mathbb{I}_{2\times2}\label{eq:vevs1}  \  \ \  \
\text{where}  \ \ \ \  \text{tr}(v^2\mathbb{I}_{2\times2})=V^2.
\ee
These break the gauge group $G_A \times G_B$ down to the diagonal subgroup
$G_{\text{Diag}} = SU(2)_L \times U(1)_Y$, which are simply the MSSM gauge
groups. The symmetry breaking pattern takes the form
\be
 SU(2)_{A}\times SU(2)_B \rightarrow SU(2)_L \ \ \ \  \text{and}   \ \ \ \ 
U(1)_{A}\times U(1)_B\rightarrow U(1)_Y.
\ee
Now, including soft breaking masses $m_{L}^2,m_{\tilde{L}}^2$ for the linking fields,
we expand the diagonal complex scalars into real scalar and pseudoscalar
components 
\be
\varphi_{L 1,4} = \, v + i \sigma_{L1,4}  + \phi_{L1,4}.\label{eq:vevs}
\ee
The $\sigma$'s play the role of Goldstone bosons and get eaten by the gauge
fields. Minimizing the scalar potential with respect to these fields leads to
the tadpole equations, which at tree-level read
\be
\frac{\partial V}{\partial \phi_{L1}} = 2 m_{L}^2 v  -\frac{1}{2} v Y_k
\text{Re}[Y_k V^2]  + v^{3} Y_{k}^{2} \label{eq:tadpole}
\ee
and a similar expression for $\phi_{L4}$. The value of the vev $v$ can be
obtained by requiring that the tadpoles should vanish. In practice it turns out
to be much more convenient to take $v$ as an input parameter and compute the
superpotential parameter $V^2$ from \refe{eq:tadpole}. 
\\
\\
As a result of the quiver structure at different
renormalisation scales $Q$ the following occur:
\begin{itemize}
\item Regime 1 is characterised by $M\geq Q> v$ with the full matter content and
gauge groups of the quiver.
\item In Regime 2, characterised by $ v>Q$, the vevs of the linking fields break
the groups to the diagonal and the MSSM superfields transform under
$G_{\text{Diag}} \equiv G_{\rm MSSM}$ in the usual way.
 \item The $U(1)$ gauge bosons $B_A,B_B$ mix to generate a massless and a
massive state $B_0$ and $B_m$, the massless one being then the $U(1)_Y$ boson
and the $B_m$ being a heavy state. Similarly $W_A^i, W_B^i$ mix to form the
massless $SU(2)_L$ $W_0^i$ gauge bosons as well as three heavy states $W_m^i$
with the corresponding mixing angles discussed below. The masses of the heavy
gauge bosons are simply given by
\be
m^2_{v,i}=  2 (g_{A,i}^2+g_{B,i}^2)v^2. \label{eq:vectormass}
\ee
\end{itemize}
In this setup we have not considered the quiver structure for $SU(3)$. Whilst
this is mostly due to practical reasons, given the difficulty of a full and
proper implementation of a Higgsed $SU(3)$ as in~\cite{Auzzi:2012dv,DeSimone:2008gm}, it is also not necessary for our purposes.
Indeed naively we sacrifice a GUT completion, but clearly we are expecting our
model to be valid only up to the messenger scale in the case of GMSB and anyway
it should be quite straightforward to tidy up this setup to restore gauge
unification without sacrificing the key results of this work. In this sense our
setup is both minimal enough, and yet concrete enough to be a ``theoretical
simplified model'' which captures relevant features of a much larger range of
possibilities, for example, it incorporates an extra $SU(2)$ and $U(1)$.
\subsection{Supersymmetry breaking and soft breaking terms for gauge
mediation}\label{sec:SUSYbreak}
In principle the models above may be combined with any supersymmetry breaking
scenario, for example mSUGRA,  or some more phenomenological parameterisation.
The model contains a large number of soft terms. The soft breaking scalar
potential reads
\begin{align} 
\mathcal{V}_{soft}=\, & B_{\mu}\epsilon_{ij}H^i_uH^j_d    + T_d^{ab}H^i_d
\tilde{d}_a\tilde{q}_{bi}+T_e^{ab}H^i_d\tilde{e}_a\tilde{l}_{bi} + T_u^{ab}
H^i_u \tilde{u}_a\tilde{q}_{bi}\nonumber\\
& +\frac{1}{2} L_{V^{2}} K     +T_{A} L^i_j A^j_k \tilde{L}^k_i+T_{K}K
L^i_j\tilde{L}^j_i\\
& +m^2_{IJ}\phi^*_I\phi_J +m^2_A |A|^2+\frac{1}{2}m_L^2 (|L|^2+
|\tilde{L}|^2)+m_{K}^2 |K|^2.\nonumber
\end{align}
$a,b$ are flavour indices and $i,j,k$ $SU(2)$ indices, which are lowered with
$\epsilon_{ij}$. The soft terms for the fermions are
\be
\mathcal{L}_{soft}\supset \frac{1}{2}\left( m_{\tilde{g}}
\tilde{g}\tilde{g}+m_{\tilde{B}_B}\tilde{B}_B\tilde{B}_B+m_{\tilde{W}_B}\tilde{W
}_B\tilde{W}_B
+m_{\tilde{B}_A}\tilde{B}_A\tilde{B}_A+m_{\tilde{W}_A}\tilde{W}_A\tilde{W}
_A\right)+h.c.
\ee
In what follows, the RGE evolution of all parameters in this scalar potential
will be accounted for at two loops.  

In this work we focus on gauge mediation and take the highest scale of the RGE
evolution to be $M$, the characteristic mass scale of supersymmetry breaking
which in perturbative models is the messenger scale. As is typical of these
perturbative gauge mediated supersymmetry breaking scenarios, we model the
supersymmetry breaking sector with a set of  messenger fields coupled to a
spurion.  Such a messenger sector can (and should) be generalised
\cite{McGarrie:2010qr}, although we will conform to this standard paradigm.  The
superpotential we use is of the form
\begin{align}
W_{\text{Messenger}}=X\Phi\tilde{\Phi},
\end{align}
where $X$ is a spurion with $X=M+\theta^2F$ and $\Phi,\tilde{\Phi}$ are
representative of fundamental and antifundamental messenger fields respectively,
charged under $SU(3)_c$ and $SU(2)_B,U(1)_B$, but not under the A-site
electroweak group.  This leads to a scale $\Lambda=F/M$ which may differ for
each gauge group so we can in general write $\Lambda_{1,2,3}$ for the three
gauge groups. The messenger fields and spurion are integrated out at $M$ to
generate the soft terms, the explicit equations for which are supplied
below.\footnote{It would be interesting to extend this work to include explicit
messenger fields and run supersymmetrically from the GUT scale to the messenger
scale, include messenger effects at the scale M and then run down to the electroweak
scale.}
Here we describe the gauge mediation parameterisation of the above soft terms.
\subsubsection{The trilinear, bilinear and linear terms}
The trilinear T-terms (or A-terms) are taken to be zero at the messenger scale,
including those corresponding to $Y_K$ and $Y_A$. The linear soft term 
$L_{V^{2}}$ for the singlet $\phi_K$, is also taken to be vanishing at the
messenger scale.  In GMSB, the bilinear term $B_{\mu}$ is expected to be zero at
the supersymmetry breaking scale  and should be generated by RG running. In what follows 
we will use the standard SUSY Les Houches Accord GMSB conventions
\citep{Skands:2003cj,Allanach:2008qq,Porod:2003um,Porod:2011nf,Allanach:2001kg,Djouadi:2002ze,Baer:1993ae} 
according to which tadpole equations are solved for $\mu$ 
and $B_{\mu}$ and $\tan \beta$ is given as an input. Above the scale of
$G_A \times G_B$ breaking, the $\beta$-function for $B_{\mu}$ reads at one-loop
\begin{align}
\beta_{B\mu}^{(1)}=&   \left(-\frac{3}{5}g_{A1}^2  - 3g_{A2}^2 +       
\text{Tr}(Y_d^{\adj}Y_d) + \text{Tr}(Y_e^{\adj}Y_e) +
\text{Tr}(Y_u^{\adj}Y_u)\right)B_{\mu}  \nonumber \\ & +\frac{2}{5}
\left(3g_{A1}^2 m_{\tilde{B}_{A}} + 15g_{A2}^2 m_{\tilde{W}_{A}} +
5\text{Tr}(Y_d^{\adj}T_d) + 5\text{Tr}(Y_e^{\adj}T_e)  + 5\text{Tr}(Y_u^{\adj}
T_u) \right) \mu 
\end{align}
which  results in a large $B_{\mu}$ if the $g_{Ai}$'s are relatively large. The
equations are similar below the quiver breaking scale. Note that there are also
two-loop contributions in both regimes.

\subsubsection{Gaugino soft masses}
For $SU(3)_c\times SU(2)_B\times U(1)_B  $ the gauginos acquire soft masses
according to the standard GMSB formula
\be
m_{\lambda,r}=N\Lambda  \left(\frac{g^2_r}{16\pi^2} \right) g(x)
\ee
where $x=F/M^2$, $\Lambda=F/M$ and $r$ refers to the corresponding gauge group.
$N=(n_{5plets} + 3n_{10plets})$ is the messenger index  and the function $g(x)$
is the standard function appearing in GMSB gaugino soft masses.

As the messenger sector is not charged under $U(1)_A$ and $SU(2)_A$, the
corresponding gauginos are taken to be massless at the messenger scale:
\be
m_{\tilde{B}_A} = m_{\tilde{W}_A} \equiv0.
\ee
One might imagine that such a feature could be detrimental to the low energy
spectrum. However, the mass matrices of the gauginos are rather complicated including 
supersymmetric Dirac masses as well as the above soft terms, so this turns out not to be 
the case: the mass eigenstates result from a combination of the site $A$ and site $B$ 
gauginos
\footnote{Full details of this matrix and all mass matrices as well as all RGEs and 
tadpole equations may be found in the supplementary material accompanying the { \tt arXiv} version
of our paper or by interfacing through Mathematica with the { \tt SARAH} model file.}. 
The Majorana soft masses of the broken
theory can be found by identifying the 
masses of the relevant components of the mixing matrices, at the threshold scale
$\mathcal{O}(v)$ (see also appendix \ref{app:fermions}).   As the A-site gauginos do not
obtain significant soft masses until the scale $v$, the RGEs of matter charged
under site A will not feel these effects until a scale $Q<v$. This turns to be
advantageous for naturalness as now the threshold scale $T_{scale}=v$ acts as a
cutoff to the leading RGE logarithm.  Such an effect will become especially
important for an $SU(3)_A\times SU(3)_B$ quiver,  as then the A-site gaugino
would only influence the RGEs between $M_{ew}$ and $T_{scale}$ and have
essentially no effect on the RGEs of site A matter above this scale.

\subsubsection{Site A scalar soft masses}
The  scalar soft masses depend on the site under which the corresponding
superfields are actually charged, relative to those of the messenger fields.  We
will always take the messenger fields to be on site B.
Fields charged under $G_A$ get soft masses at two loops from mediation
along the quiver
\be
m_{A}^2 =N \sum_{i=1,2} 2\Lambda^2_i
C_i(r)\left(\frac{g_{i}^2}{16\pi^2}\right)^2 s(x,y_i),\label{eq:SQUIVER}
\ee
where $y=m_v/M$ with $m_v$ being the heavy gauge boson mass, $M$ the messenger
scale and $g_i$ is the corresponding coupling constant. The quadratic Casimir
invariants $C_i(r)$ are $C_1(Y)=3/5\,Y^2$ for fields charged under U(1) with
hypercharge $Y$ and $C_2(2)=3/4$ for doublets under SU(2). In MI,
where all MSSM chiral superfields reside in site A, this formula serves as a
boundary condition for all electroweak contributions to the scalar soft masses. 
In MII, this formula only applies
to the third generation sfermions and the Higgs scalars, whereas the first two
generation sfermions receive their soft masses according to
\refe{GMSBsoftMasses}.

The form of the function $s(x,y)$, associated with gauge mediation along a two
site quiver, can be found in \cite{McGarrie:2010qr,Auzzi:2010mb,McGarrie:2011dc}
and is given in both analytical and graphical form in appendix
\ref{app:SQUIVER}. By inspecting figure \ref{fig:SQUIVER} we can see how the
mediation of supersymmetry breaking along the quiver has the effect of reducing
the site A scalar soft masses with respect to usual gauge-mediated supersymmetry
breaking. In particular, $s(x,y)$ has the limit
\be
s(x,\infty)=f(x),
\ee
where $f(x)$ is the usual GMSB formula. This formula interpolates between
$y\rightarrow \infty$ of GMSB and the suppressed scalar regime as
$y\rightarrow0$, where scalars get their leading soft mass at three loops from
additional contributions,  which arise anyway from RG evolution.

\subsubsection{Site B scalar soft masses}
Scalar fields charged under $G_B$ and $SU(3)_c$ receive standard GMSB soft
masses according to
\be
m_{B}^2=N \sum_{i=1,2,3}2\Lambda^2_i  
C_i(r)\left(\frac{g_{Bi}^2}{16\pi^2}\right)^2f(x).
\label{GMSBsoftMasses}
\ee
The quadratic Casimir invariants are $C_1(Y_B)=3/5 Y_B^2$ for fields with charge 
$Y_B$ under $U(1)_B$ whereas $C_2(2)=3/4$ for
the linking fields $L$ and $C_2(3)=2$ for the $SU(2)_B$ adjoint $A$ field. All coloured 
scalars in our setup are $SU(3)$ triplets, for which $C_3(3)=4/3$.
Note that in the case of site A coloured scalars, the full soft mass is given by the 
sum of \refe{eq:SQUIVER} and the third term of \refe{GMSBsoftMasses}.
\\ 
\subsubsection{The singlet scalar $K$ soft mass}
The superfield $K$ is a gauge singlet and its scalar soft mass is vanishing at
the messenger scale,
\be
 m_K^2=0.
\ee
It evolves a positive value through \refe{Mk} and so does not pose a
phenomenological issue, but if one did wish to assign a tree level soft mass,
two approaches are possible: it may be interesting to consider that it is not a
singlet under some other group or that it couples directly to messenger fields
through a term of the form $K\Phi\tilde{\Phi}$. In this later case it can
develop a one-loop soft mass.
\subsubsection{Linking scalar soft masses}
The linking fields formally get their soft masses from applying
\refe{GMSBsoftMasses} to describe the soft terms for $m_{L}^2$ and
$m_{\tilde{L}}^2$. We will however not be using this formula in order to compute
the linking field soft masses. Instead, we will promote them to free parameters
of the model. The reasons for this choice will be clarified in the following
section. To be noted is that in this setup the two linking field masses are
equal, $m_{L}^2=m^2_{\tilde{L}}$.

\subsection{Linking to the MSSM}\label{sec:linkingtoMSSM}
Below the quiver breaking scale the gauge group and particle content of the
model are those of the minimal supersymmetric standard model with gauge groups
$SU(3)_c\times SU(2)_L \times U(1)_Y$.  The gauge couplings between the unbroken
and the broken theory are matched as 
\be
\frac{1}{g^2_i}=\frac{1}{g^2_{Ai}}+\frac{1}{g^2_{Bi}} \  \Leftrightarrow  \
g^2_i= \frac{g^2_{Ai}g^2_{Bi}}{g^2_{Ai}+g^2_{Bi}} \ ,
\ee
with $i=1,2$ for $U(1)_Y\times SU(2)_L$.   If one of the two gauge couplings is
strong, the other should be weak. Then at low energies the diagonal or MSSM
gauge coupling will be of the order of whichever is weaker.  This is a key
feature which allows for these models to lift the tree-level Higgs mass whilst
being consistent with perturbative unification. The various gauge couplings of
the model are simply related through two rotation angles
\footnote{Here we should stress an important notation subtlety. In all relations applying to the messenger
scale as well as in all RGE expressions, the $U(1)$ coupling constants are taken to be
$SU(5)$ GUT-normalized, so for example $g_1 = g_{1,GUT}=\sqrt{5/3}g'$, with $g'$ being the usual
Standard Model hypercharge coupling constant. In all other relations,  the GUT normalization is 
dropped and $g_1$ identical to $g'$. This is done in order to follow the \texttt{SARAH} package conventions.}
\be
\cos \theta_i=\frac{g_{i}}{g_{Ai}},   \ \ \ \ \sin
\theta_i=\frac{g_{i}}{g_{Bi}}.
\label{eq:couplings}
\ee
In appendix \ref{app:thresholdeffects} we present additional comments on threshold
effects that enter the coupling constants and other parameters calculation below
the quiver breaking scale.  The angles $\theta_1,\theta_2$ are free parameters
of our setup. Varying these amounts to changing the relative strengths between
each site and we typically choose the A-sites to be stronger to enhance
additional contributions to the Higgs mass, as we will explain below.
\\
\\

\noindent One of the most interesting features of this class of models is
that non-decoupled D-terms may arise \cite{Batra:2003nj,Maloney:2004rc} in the
low energy Lagrangian. The real uneaten scalar components of the linking fields
appear in both the A and B site scalar D-term potential and when integrated out
generate an effective action which includes the terms
\be
\delta \mathcal{L}=-\frac{g_1^2\Delta_1}{8}  (H^{\dagger}_u
H_u-{H}^{\dagger}_d H_d)^2-\frac{g_2^2\Delta_2 }{8}\sum_a
(H^{\dagger}_u\sigma^a H_u+{H}^{\dagger}_d\sigma^aH_d)^2,
\ee
where
\be
\Delta_1=\left(\frac{g^2_{A1}}{g^2_{B1}}\right)\frac{m_L^2}{m_{v1}^2+m_{L}^2} \
\ , \ \
\Delta_2=\left(\frac{g^2_{A2}}{g^2_{B2}}\right) \frac{m_L^2}{m_{v2}^2+m_{L}^2}. 
 \label{eq:nondecoupled}
\ee
It is particularly informative to see how in this class of models, these terms
can work to lift the Higgs mass 
without large radiative corrections. In the MSSM, the one-loop Higgs mass in the limit 
$m_{A^0}\gg m_Z$
can be written as~\cite{Drees:2004jm}
\be
m_{h,1}^2\simeq m_{z}^2\cos^2 2\beta +
\frac{3}{4\pi^2}\frac{m_t^4}{v^2_{ew}}\left[\ln
\frac{M^2_{S}}{m_t^2}+\frac{X^2_t}{M_S^2}\left(1-\frac{X^2_t}{12M_S^2}
\right)\right]\label{eq:higgsmass}
\ee
where $M^2_S=m_{\tilde{t}_1}m_{\tilde{t}_2}$ for $m_{\tilde{t}_1}$, $m_{\tilde{t}_2}$ 
as defined in 
appendix~\ref{app:fermions},  and $v_{ew}$ is the electroweak Higgs vev, such that the upper limit on the
tree level Higgs mass ($m_{h,0}$) is set by the Z boson mass $m_z$.
This expression assumes that the left and right-handed soft masses of the stops
are equal.
Note that $X_t=A_t -\mu \cot \beta$, and for convenience the sfermion mixing
matrices are provided explicitly in 
Eqn.~\eqref{eq:sfermion} of appendix~\ref{app:HiggsSfermions}.
In our case however, there may in principle be
a sizeable shift to the Higgs mass at tree level $m_{h,0}$, which takes the
precise form,
\be
m^2_{h,0}= \left[m_z^2
+\left(\frac{g^2_1\Delta_1+g^2_2\Delta_2}{2}\right)v^2_{ew}\right]\cos^2 2\beta.
\label{eq:mh0QEW}
\ee

\noindent Arguably this enhancement is favoured over that of the NMSSM for
a simple reason: in the NMSSM typically
\be
m_{h,0}^2= m_{z}^2\cos 2\beta + \lambda^2 v^2_{ew} \sin 2\beta
\label{eq:mh0NMSSM}
\ee
where $\lambda$ is the coupling between the Higgs singlet and doublet fields
appearing in the superpotential term $\lambda SH_uH_d$. This creates a tension
between wanting a large  $\tan \beta$ to enhance the first term, but a small
$\tan \beta$ for the second, forcing one to accept very large values of
$\lambda$. As a result $\lambda$ ends up non perturbative before the GUT
scale.

It is the above observation that forms the basis for the construction of
natural spectra in the class of models that we examine: the Higgs mass can now
be substantially increased already at tree-level when these new contributions
become large. Of course this enhancement is completely independent of the method
by which supersymmetry  breaking effects are  transmitted to the MSSM.  We have
simply chosen GMSB in this paper on the one hand to demonstrate that it is still
a natural candidate for supersymmetry breaking mediation, and on the other hand
because in our electroweak GMSB quiver the sleptons can be naturally lighter
than their coloured counterparts. This potential enhancement of the tree level
Higgs mass is only significant in certains areas of the model's full parameter
space. Concretely, for this contribution to be sizeable we must have
$g^2_{Ai}/g^2_{Bi} \ge 1$ and $m_{L} \sim {\cal{O}}(m_{v,i})$.

However, this mechanism introduces some additional fine-tuning to the
theory, since the Higgs mass now receives an additional quadratically divergent
contribution at one loop, induced by the linking fields and cut off by $m_L^2$.
This additional fine-tuning should be kept under control in order to not
counterbalance the improved naturalness of the model with respect to traditional
mGMSB. An estimate of the maximal size of $m_L$ can be found following the
arguments of \cite{Blum:2012ii}: demanding less than $10\%$ additional fine
tuning approximately bounds
\be
\frac{g_{SM}^2\Delta}{16\pi^2}\frac{m_{L}^2}{m_h^2}<10.
\ee
Requiring, for example, $\Delta=0.2$ in \refe{eq:nondecoupled} allows 
$m_L^2$ in the $10^6-10^8 \text{GeV}^2$ range and sets an upper value
$v<10^5\phantom{0} \text{GeV}$ so that the additional $D$-terms do not decouple.
So in summary we would ideally want $v<10^5\phantom{0} \text{GeV}$ and
$m_{L}< 10 \phantom{0} \text{TeV}$.  Note that $v$ is also bounded from below
both by electroweak precision tests and direct searches for new gauge bosons.
\\

\noindent The non-decoupled D terms also appear in the tadpole equations 
\begin{align} 
\frac{\partial V}{\partial \phi_{d}} &= \frac{1}{8} \Big(-8 v_u
\text{Re}[B_{\mu}] + (g_{1}^{2} + g_{2}^{2} + g_1^2\Delta_1^{2} +
g_2^2\Delta_2^{2})v_{d}^{3}  \\&\phantom{00000000000}+ v_d [8 m_{H_d}^2  + 8
|\mu|^2  - (g_{1}^{2} + g_{2}^{2} +g_1^2\Delta_1^{2} +g_2^2
\Delta_2^{2})v_{u}^{2} ]\Big)\nonumber \\ 
\frac{\partial V}{\partial \phi_{u}} &= \frac{1}{8} \Big(-8 v_d
\text{Re}[B_{\mu}] + 8 v_u |\mu|^2  \\ &\phantom{00000000000}+ v_u [8 m_{H_u}^2 
- (g_{1}^{2} + g_{2}^{2} + g_1^2\Delta_1^{2} + g_2^2\Delta_2^{2})(- v_{u}^{2}  +
v_{d}^{2})]\Big)\nonumber
\end{align} 
modifying the vacuum structure, as well as the Higgs mixing matrices, which may
be found in appendix \ref{app:HiggsSfermions}.

Additional soft mass terms for all scalars appear in regime 2 of the model at
effective one loop, from integrating out the heavy gauge and linking fields
\cite{DeSimone:2008gm}
\be
\delta m_{\tilde{f}}^2=\sum_i \left(\frac{g_i}{4\pi}\right)^2 C^{\tilde{f}}_k
\left[ m_{v_i}^2\tan^2\theta_i \log \left(1+\frac{2m_L^2}{m_{v_i}^2}\right)
+2\sin^2\theta_i (1-3\sin^2\theta_i)m^2_{i,B} \right]\label{thresholdsoft}
\ee
which are also implemented into the model and importantly the soft mass
parameters are matched across the threshold scale.\footnote{There are further
three loop terms if the $SU(3)$ sector is quivered.}


\subsection{An extra dimensional digression}\label{digression}
Quiver models are naturally related to extra dimensional setups through
deconstruction. For early ideas on the topic we refer the interested reader
to \cite{Bando:1987br}. The contemporary formulation of the topic was initiated
in \cite{ArkaniHamed:2001ca,Hill:2000mu,ArkaniHamed:2001nc, Csaki:2001em}, 
whereas for recent work relating to $\mathcal{N}=1$ see for instance \cite{McGarrie:2010qr,McGarrie:2011dc,McGarrie:2012ks}. 
It should therefore be
expected that these non decoupled D-terms, \refe{eq:nondecoupled}, have a
natural interpretation in terms of extra dimensional models.  We swiftly sketch
and motivate this relationship which certainly warrants further study on its
own.  The quiver construction may be related to $\mathcal{N}=1$ super Yang-Mills
(or $\mathcal{N}=2$ \cite{McGarrie:2013hca,McGarrie:2010yk,McGarrie:2012fi}) in
five dimensions \cite{McGarrie:2010kh}, which contains a vector multiplet and
chiral adjoint $V+\Phi$.  Suppose we compactify on four flat dimensions times a
small interval of length $R$.  The scalar component of $\Phi=(\Sigma+iA_5)$ and
in analogy to the quiver, $A_5$ plays the role of the Goldstone bosons and are
eaten to generate the Kaluza-Klein masses such that we may identify $1/R\sim v$
of \refe{eq:vevs}. To obtain the non decoupled D-terms we write the lagrangian
in 
the off-shell formulation 
\be
\mathcal{L}_5=\frac{1}{2}D^2 + D(\partial_5 \Sigma) -
\frac{1}{2}\partial_{\mu}\Phi\partial^{\mu}\Phi^* + m_{\text{soft}}^2\Sigma
\Sigma+...
\ee
The ellipses denote not just the rest of the bulk action but also any terms
generated on bulks and boundaries that may be of use such as the boundary terms
\be
\int_{\partial \mathcal{M}}(\Sigma D+...).
\ee
To see how this action may generate a non decoupled D-term we define
$\mathcal{H}=(H^{\dagger}_uH_u-H^{\dagger}_dH_d)$. Then, there may be bulk or
boundary terms of the usual form 
\be
\mathcal{L}\supset \frac{1}{2}\mathcal{H}D.
\ee
Integrating out the auxiliary scalar field $D$ gives rise to the D-term scalar
potential.  The field $\Sigma$, the real uneaten scalar degrees of freedom,
corresponds to the real uneaten degrees of freedom in the linking fields
$L,\tilde{L}$ of the quiver.   It is this field $\Sigma$, when integrated out
which generates the non decoupled D-term \refe{eq:nondecoupled}.  In such a
scenario, the $m_v$ of the quiver is related to the Kaluza-Klein mass scale
$m_{kk}$ which is $O(1/R)$, the effective length scale of the extra dimension.
As such, for these terms to be of relevance $\pi/R \leq m_{\text{soft}}$. We
hope to return to this topic in a further publication, but for now we
effectively model this feature with quiver models as they are a more controlled
environment which are more amenable to spectrum generators.  It is certainly
interesting to speculate that as our model has a $v$ of ${\cal{O}}(10^4)$ GeV, that this
corresponds to an ``effective'' extra dimensional length scale of roughly 
${\cal{O}}(10^{-18})$ cm.  

\section{Tools and Observables}\label{sec:ToolsObservables}
In order to study the low-energy phenomenology of our setup in a consistent
manner, it is necessary to perform the RGE evolution of all couplings and mass
parameters from the highest energy scale of the theory down to the TeV scale,
properly imposing all boundary conditions. Here we describe the construction of
a tailor-made spectrum generator for the quiver model. We further
discuss the parameter space we adopt as well as the constraints it is subject
to.

\subsection{Implementing a quiver framework for phenomenological
studies}\label{sec:implementation}
In order to perform the RGE evolution of the models' parameters and masses and 
compute the resulting low-energy particle spectra, we implemented the two model
variants into the publicly available 
Mathematica package \texttt{SARAH 3.3} \cite{Staub:2008uz,Staub:2010jh,Staub:2012pb}. \texttt{SARAH} is a
``spectrum generator generator'', which includes 
a library of models that may communicate with HEP tools that are widely used in
most phenomenological studies
\cite{Porod:2003um,Porod:2011nf,Hahn:2000kx,Hahn:2001rv,Hahn:1998yk,
Pukhov:1999gg,Bechtle:2008jh,Bechtle:2011sb,Bechtle:2013gu,
Kilian:2007gr,Belanger:2006is,Camargo-Molina:2013qva}. 
In particular, \texttt{SARAH} performs the task of generating Fortran routines compatible
with the \texttt{SPheno} spectrum generator \cite{Staub:2010jh}.

In order to implement our model, we have used the possiblity offered by the
package to implement and link two different 
``regimes''.  These regimes correspond to those introduced in
section~\ref{sec:gaugeSB}, each being characterised by a set of gauge groups, a
particle content and a superpotential that need to be specified. 
Regime 1 includes, for both models MI and MII, the full $G_A \times G_B \times
SU(3)_c$ gauge group along with the full quiver particle content, while
the superpotential is given by Eqns.\eqref{eq:MSSM} and \eqref{eq:Quiver} for MI
and by Eqns.\eqref{eq:MSS2} and \eqref{eq:Quiver} for MII.
In both cases in regime 2 we have the MSSM, which we supplement with an
effective action to account for 
and study the effects of the non-decoupled $D$-terms, which are of crucial
importance, as discussed in section~\ref{sec:linkingtoMSSM}. These terms are properly included in all loop calculations, self energies, branching ratios and vertices of regime two. 
We would have preferred to implement our model using a single regime such that these terms would be automatically
generated, and as there are other terms for other fields, however we found it was not
practical to integrate out so many fields in full. Moreover it was preferable to include an additional regime with
the MSSM gauge and matter configuration to ease communication of \texttt{SPheno} with
packages such as \texttt{HiggsBounds}, used to check the compatibility of the model with
experimental constraints as discussed below.

In addition to these ingredients, we need to specify on the one hand boundary
conditions for all soft parameters at the messenger scale $M$ and on the other
hand matching conditions for the parameters of the two regimes at the $G_A
\times G_B$ breaking scale which we typically take to coincide with $v$, the
linking field vev. The boundary conditions are applied according to the
discussion and relations given in section \ref{sec:SUSYbreak}, while the
matching conditions follow the lines described in section
\ref{sec:linkingtoMSSM}.

The renormalization group equations for both regimes are then calculated by
\texttt{SARAH} at two loops and appropriate Fortran routines 
are generated that can then be taken over by \texttt{SPheno} to perform the
numerical analysis. Concretely, the implementation 
includes full one- and two- loop RGEs for five gauge groups,  mixing matrices
for all fields in the quiver and the MSSM 
including associating goldstones with massive gauge bosons and gauge fixings,
full two-loop RGEs for the vev of the 
linking fields themselves, two-loop RGEs for $B\mu$, loop-level solutions for
the tadpole equations, one- and two- loop 
anomalous dimensions for all fields and two-loop RGEs for all soft breaking
parameters, linear, bilinear and trilinear. 
The MSSM particle masses are computed at one loop, however the full
two-loop corrections to the Higgs mass are implemented in \texttt{SPheno} following the
calculation in 
Refs.~\cite{Degrassi:2001yf,Brignole:2001jy,Brignole:2002bz,Dedes:2002dy}.
\\
\\
All in all, the RGE evolution is described by three energy scales 
\be
M_{\text{messenger}}\longrightarrow  T_{\text{scale}}	\longrightarrow  M_{ew} \nonumber.
\ee
Following standard practice, the highest energy scale of the theory is taken to
be the messenger scale $M$, where all boundary 
conditions resulting from the quiver structure, including exact formulae for
GMSB soft masses have been implemented and 
imposed. Again as usual, the running ends at the 
electroweak scale, which is used as an input scale for the MSSM parameters.
The intermediate 
mass scale $T_{\text{scale}}$ is associated with the quiver breaking scale as it
separates the two regimes, and at this scale appropriate matching boundary
conditions are applied including finite shifts that result from integrating out
the heavy fields of the theory (see appendix \ref{app:thresholdeffects}). We
choose it to be equal to the vev of the linking fields, $T_{\text{scale}}=v$.
All soft terms of regime 2 are matched to regime 1 as described in section
\ref{sec:linkingtoMSSM} and for the soft masses of the winos and binos in
appendix~\ref{app:fermions}. Note that the gluino finite shifts are also
accounted for as given in appendix~\ref{app:thresholdeffects}.

It is important to stress that the high-scale boundary conditions
themselves may be seen as being separable from the model (the matter 
content, gauge groups and superpotential) and may be changed with ease, if one
wished to explore, for example, 
different supersymmetry  breaking scenarios.
\\
\\
The implementation detailed above, i.e.~the construction of a tailor-made
spectrum generator for our model, allows us to create a model file for \texttt{SPheno},
which further permits us to study a quiver model in a complete manner, as we can
study the influence of RG effects of all the gauge groups and matter content in
the highest regime to the low energy spectrum, at the two-loop order.
On the practical level, this model can also serve as a first step for the
implementation of more complete or complicated setups, such as a model including
an additional $SU(3)$. It would further be trivial to change
the representation assignments of the Higgs fields in order to study chiral non-decoupled
D-terms \cite{Craig:2012bs} or flavour models, with the same precision.
However, we point out that the implementation of the electroweak only quiver
studied here leads to an interesting phenomenology in its own right,  allowing
for naturally (although moderately) heavier squarks relative to sleptons.


\subsection{Parameter space and constraints}\label{sec:PspaceConstraints}
It is now useful to describe the process through which we choose our parameter
space and the regions we will study in the next section. We moreover describe
some preliminary findings that could be of interest for model-building purposes.
\subsubsection{Choosing the parameter space}
The electroweak quiver we consider can be described by a basic set of six
parameters
\be
M, \Lambda, v, \tan \beta, \theta_1,\theta_2, 
\label{eq:TheParameters}
\ee
where $M$ is the messenger scale, the SUSY breaking scale $\Lambda = F/M$,
$F$ being the SUSY breaking F-term, $\tan \beta$ is the Higgs vev ratio and
$\theta_1$, $\theta_2$ are the mixing angles between sites A and B for 
$U(1)$ and $SU(2)$ respectively.

As an initial step, we performed extended scans over large regions of parameter
space for MI, imposing the full set of GMSB boundary conditions described
in section \ref{sec:SUSYbreak}, including the exact relations
\refe{GMSBsoftMasses} for the linking field soft masses. We focused in
particular on regions where $v \lesssim 40$ TeV, where according to the
discussion of section \ref{sec:linkingtoMSSM} the non-decoupled D-terms
should be most efficient in lifting the tree-level Higgs boson mass. A
first finding of these searches is that we could not find viable points
when the linking field vev was much below $10$ TeV, as often here either the
A-site couplings become non-perturbative before the messenger scale, the
electroweak vacuum becomes unstable or the RGE code simply wouldn't converge for
the numerical precision requirements imposed (a relative error of 0.5\%). Where
none of these issues occur, the values of $\Lambda$ are typically low and close to $v$,
such that the linking fields soft masses 
$m_L$ are too low for the D-terms to have an important impact on the Higgs
mass.

Motivated by the perturbativity issues, we implemented MII where we expect that
by removing some of the matter fields from site A the RGE running will be
reduced~\cite{Auzzi:2012dv}. We found that although the situation does
improve, it still seems to be quite difficult to achieve substantial
contributions to the Higgs tree-level mass from the non-decoupled D-terms due to
the fact that $m_L^2$ is again driven too low.

These results lead us to slightly enlarge our parameter space by promoting
the linking field squared soft mass $m_L^2$ to be a free parameter instead of
being given by \refe{GMSBsoftMasses}. This is interesting from a
theoretical perspective, as it motivates pursuing models that might provide the
additional contribution to $m_L^2$ needed in order to achieve a substantial
D-term contribution to the Higgs mass. For example, it would be interesting to
study whether this can be realized in extensions to the model including 
an additional $SU(3)$ or $U(1)$ kinetic mixing.
Within the scope of our work, the choice to make
$m_L^2$ free can be seen as a phenomenological parametrization of the linking
field soft masses along the lines of similar choices made in many supersymmetry
breaking mediation schemes.

With this small modification, we find that it is indeed perfectly possible to
achieve the required D-term size in order to reproduce the observed Higgs mass
while keeping the stop masses well below $2$ TeV. Furthermore as expected, the
mediation of SUSY breaking along the quiver acts as a suppression mechanism for
the uncoloured sparticle masses, yielding electroweakinos and sleptons lying
roughly in the range $[400,1000]$ GeV, which is on the boundary of being
within the LHC reach~\cite{ATLAS:2013hta,CMS:2013xfa}. At this point,
due to the differing bounds on coloured and non-coloured sparticles at the
LHC, we introduce a second modification to the original setup that consists of
dissociating the scale $\Lambda_3$ from $\Lambda_{1,2}$. Note however that this is a
minor modification as the two scales will not differ by orders of magnitude but
only by ${\cal{O}}(1)$ multiplicative factors.
We will see that this setup allows for a rich phenomenology with interesting
features.
\subsubsection{Constraints}
We carried out extensive scans of the parameter space described in the
previous paragraph within generous intervals. We are interested in areas of parameter
space which are characterised by low values of $v$ and moderate
splittings between $v$ and $m_L$, such that the additional D-terms do not decouple from the
low-energy theory and the uncoloured scalars are light. In
what follows, we will therefore present results that concern a subregion of the
parameter space that meets a series of requirements.

First, we wish to obtain a Higgs mass lying in the range $[122.5, 128.5]$ GeV.
This interval envelops on the one hand the experimental uncertainty in the Higgs
mass measurement~\cite{Chatrchyan:2013lba,Aad:2012tfa,Chatrchyan:2012ufa}, while
being sufficiently generous to account for uncertainties in the theoretical mass
spectrum determination \cite{Degrassi:2002fi,Arbey:2012dq}\footnote{Throughout
our calculations we assume a constant moderate top quark mass of $m_t = 173$ GeV
\citep{Beringer:1900zz}.}.
For naturalness reasons, we require this value for the Higgs mass to be
achieved for stop masses as low as possible. The stop mass is governed by
$\Lambda_3$, which also controls the masses of the lower generation squarks and
the gluino. Strong exclusion limits on these masses
arise from ATLAS and CMS null searches for jets plus missing energy, 
e.g.~$m_{\tilde{g}} > 1600$ GeV for $m_{\tilde{q}_{1,2}} > 2000$
GeV~\cite{ATLAS-CONF-2013-047,Chatrchyan:2012lia}. We are guided by these bounds
in choosing a lower limit for $\Lambda_3$.
Note that in this work, we choose not to quantify the amount of fine-tuning for
each point of the models we study, which constitutes a work in its own right
involving numerous subtleties (see for example the recent discussion in
\cite{Baer:2013gva}). It is however at least clear that qualitively, having
stops lighter than benchmark minimal GMSB improves the relative naturalness of
the model, and this motivates our choice of upper limit on $\Lambda_3$.

At the same time, according to the comments made in section \ref{sec:linkingtoMSSM}, 
we should avoid reintroducing excessive fine-tuning via the non-decoupled $D$-terms. 
Moreover, in order for the setup to be realistic, we must satisfy the condition $m_{L}^2 <
m_{v}^2$, but not approach the limit $m_{L}^2 \ll m_{v}^2$ where the quiver-induced
$D$-terms decouple. These requirements lead us to choose $m_{L}^2$ within the range
$[10^7,10^{8}]  \phantom{0} \text{GeV}^2$.
Also note that as mentioned above, for very low values of $v$
\texttt{SPheno} faces convergence issues.
The parameter $\Lambda_{1,2}$ is mainly subject to constraints from searches for
charged sleptons and charginos at LEP,
i.e. 92 GeV for charginos degenerate with the lightest neutralino, and 103.5 GeV
otherwise~\cite{LEP2}. Lower limits on sleptons staus and sneutrinos of 68 and 51 GeV 
respectively were also obtained at LEP~\cite{Barate:1999gm,Barate:2000tu,Decamp:1991uy}. 
Finally, given that the non-decoupled $D$-terms
contribute a shift to the Higgs mass as $m_Z$ does, i.e.~with a factor
$\cos 2\beta$ (\refe{eq:mh0QEW}), as opposed to the factor $\sin 2\beta$ in the
NMSSM (\refe{eq:mh0NMSSM}), we explore a rather standard MSSM-like range for
$\tan\beta$.

From our numerical analysis we find that this set of requirements is satisfied
by adopting the following parameter value ranges
\be
2.1\times 10^5 \phantom{0} \text{GeV} \le M\le  3.0\times 10^5  \phantom{0}
\text{GeV}
\ee
\be
4.0\times 10^4 \phantom{0} \text{GeV} \le \Lambda_{1,2} \le  1.9\times 10^5 
\phantom{0} \text{GeV} \nonumber
\ee
\be
1.9\times 10^5 \phantom{0} \text{GeV} \le \Lambda_{3} \le  2.1\times 10^5 
\phantom{0} \text{GeV} \nonumber
\ee
\be
1\times 10^7 \phantom{0} \text{GeV}^2 \le m_{L}^2 \le 1 \times 10^{8} 
\phantom{0} \text{GeV}^2 \nonumber
\ee
\be
1.5\times 10^4\phantom{0} \text{GeV} \le v \le 4 \times 10^{4}  \phantom{0}
\text{GeV} \nonumber
\ee
\be
5 \le \tan \beta \le  30   \nonumber
\ee
\be
0.8 \le\theta_1,\theta_2 \le  1.4   \nonumber
\ee
We have moreover chosen $\rm{sign}\mu = +1$, a low value for the messenger
index, $N = 1$, and a fixed common value for the $A$ and $K$ field Yukawa
couplings $Y_A=Y_K=0.8$. 

Apart from the theoretical and experimental constraints so far mentioned, the
low-energy spectrum is subject to further bounds. In the Higgs sector, in
addition to obtaining the lightest Higgs boson mass within the observed region,
we must ensure that its properties and decay modes comply to current LHC
observations. As an example, it is known that enhancing the Higgs mass
through non-decoupled $D$-terms enhances simultaneously the Higgs boson
couplings to down-type quarks \cite{Blum:2012ii,Craig:2012bs,Huo:2012tw}. 
In order to test whether the Higgs sector is compatible with the
constraints coming from the LHC and the TeVatron, we have linked \texttt{SPheno} to 
\texttt{HiggsBounds-4.0.0}~\citep{Bechtle:2013gu}. Taking our analysis a step further, 
we have also linked \texttt{SPheno} to \texttt{HiggsSignals-1.0.0}~\cite{Bechtle:2013xfa}, which
allows us to test in particular whether the lightest Higgs boson properties are in agreement with 
all relevant existing mass and signal strength measurements from the LHC and TeVatron.

\begin{table}
\begin{center} 
\begin{tabular}{|c|c|} 
\hline \hline 
 Observable & Accepted range \\
\hline 
$B_s \rightarrow X_s \gamma$ &  $[2.78, 4.32]\times10^{-4}$ \cite{Amhis:2012bh}
\\
$\delta a_\mu$ &  $ < 20 \times 10^{-10}$ \citep{Beringer:1900zz} \\
$\Delta\rho$ &  $< 1.2 \times 10^{-3}$ \citep{Beringer:1900zz}  \\
$BR(B_s \rightarrow \mu^+ \mu^-)$ &  $< 7.7 \times 10^{-9}$ \cite{Aaij:2012nna}
\\
\hline \hline
\end{tabular} \caption{Low-energy observable constraints imposed in our
analysis.  \label{tab:LowEnergyObs}}
\end{center} 
\end{table}

Finally, we use the in-built functionalities of \texttt{SPheno} in order to apply a set
of necessary low-energy constraints, all of which are taken at $3\sigma$: the
SUSY contributions to the muon anomalous magnetic moment $\delta a_\mu$ and the
branching ratios $BR(B_s \rightarrow X_s \gamma)$ and $BR(B_s \rightarrow \mu^+
\mu^-)$ and, due to the presence of relatively light sfermions in our
spectra, the $\rho$ parameter. The allowed ranges used for these observables
are shown in table \ref{tab:LowEnergyObs}, where theoretical uncertainties
and experimental errors are added in quadrature.
\section{Results}
\label{sec:results}
\renewcommand{\arraystretch}{1.1}
\begin{table}
\begin{center} 
\begin{tabular}{|c|c|c|c|} 
\hline \hline 
 & MI & MIIa & MIIb \\
\hline 
\hline 
  \multicolumn{4}{|c|}{Input values}\\ 
\hline 
M &  233 TeV & 288 TeV  & 260 TeV\\
$\Lambda_{1,2}$ &  44.9 TeV  & 85.6 TeV & 111 TeV\\
$\Lambda_3$ &  190 TeV & 206 TeV& 208 TeV  \\
$m_{L}^2$ &  47.3 TeV$^2$ & 83.3 TeV$^2$ & 86.2 TeV$^2$\\
$v$ &  26.2 TeV  &   26.5 TeV &   25.4 TeV  \\
$\theta_1,\theta_2 $ &  $1.18, 1.13 $  & 1.09,1.33 & 1.05,1.04  \\
$\tan \beta$ & 16 &12 & 28\\
\hline 
  \multicolumn{4}{|c|}{ Squark sector }\\ 
\hline 
$m_{\tilde{t}_1}$ & 1.84 TeV  & 1.99 TeV & 409 GeV\\
$m_{\tilde{t}_2}$ & 1.98 TeV  & 2.06 TeV & 3.49 TeV\\
$A_t$  & -442 GeV & -146 GeV & -141 GeV \\
$m_{\tilde{b}_R}$ & 1.95 TeV  & 2.05 TeV & 2.56 TeV\\
$m_{\tilde{q}_{12,L}}$ & 2.05 TeV  &  2.12 TeV &  2.19 TeV\\
$m_{\tilde{q}_{12,R}}$ & 1.97 TeV  &  2.10 TeV &  2.14 TeV\\
\hline
  \multicolumn{4}{|c|}{ Slepton sector}\\ 
\hline 
$m_{\tilde{l}_{12,L}}$ & 738 GeV  &  314 GeV &  515 GeV\\
$m_{\tilde{l}_{3,L}}$  & 736 GeV  &  315 GeV &  440 GeV\\
$m_{\tilde{l}_{12,R}}$ & 901 GeV  &  183 GeV &  262 GeV\\
$m_{\tilde{l}_{3,R}}$  & 899 GeV  &  110 GeV &  4.31 TeV\\
\hline
  \multicolumn{4}{|c|}{Gaugino sector }\\
\hline 
$m_{\tilde{\chi}_1^0}$ & 53.2 GeV  & 116  GeV & 154 GeV  \\
$m_{\tilde{\chi}_2^0}$ & 99.3 GeV  & 242  GeV & 306 GeV  \\

$m_{\tilde{\chi}_3^0}$ & 187 GeV  & 750  GeV & 818 GeV  \\

$m_{\tilde{\chi}_4^0}$ & 222 GeV  & 755  GeV & 823 GeV  \\

$m_{\tilde{\chi}_1^{\pm}}$ & 96.8 GeV  &   242   GeV  & 306 GeV         \\
$m_{\tilde{\chi}_2^{\pm}}$ & 225 GeV  &   756   GeV  & 823 GeV         \\
$m_{\tilde{g}}$ & $1.62$ TeV  &  $1.66$ TeV & $1.75$ TeV  \\
\hline 
  \multicolumn{4}{|c|}{Higgs sector }\\ 
\hline 
$m_{h_0}$   & $125$ GeV  &    127 GeV	& 125 GeV		 \\
$m_{H_0}$   & $720$ GeV  &    792 GeV	& 885 GeV     \\
$m_{A_0}$   & $721$ GeV  &    796 GeV	& 894 GeV		\\
$m_{H_\pm}$ & $726$ GeV  &    799 GeV	& 893 GeV
\\
\hline \hline
\end{tabular} \caption{Mass spectra of three example points for MI
and MII,
along with the associated input parameters as defined in section
\ref{sec:PspaceConstraints}.
Note that $m_{\tilde{f}_{12,L/R}}$ and $m_{\tilde{f}_{12,L/R}}$ are the masses
of the lower and third generation 
left/right-handed squarks and sleptons, and that the sneutrino and sbottom
masses can be inferred 
via $m_{\tilde{\nu}_{i}}\sim m_{\tilde{f}_{i,L}}$, $m_{\tilde{b}_{L}}\sim
m_{\tilde{t}_{L}}$.
\label{tab:ExampleSpectra}}
\end{center} 
\end{table}
\renewcommand{\arraystretch}{}
Having described our model, and how it is implemented in \texttt{SARAH}, we turn to study
the low scale spectrum,
which we find has several interesting features. 
Examples of complementary representative points, one for MI and two for MII, are
given in table~\ref{tab:ExampleSpectra}.
\clearpage
\noindent
In table~\ref{tab:ExampleSpectra} we observe that particles of the electroweak 
sector can be substantially
lighter than those of
the coloured sector.
This arises due to the quiver structure of the model, as explained in
section~\ref{sec:SUSYbreak},
 which provides a suppression factor $s(x,y)$ for the 
non-coloured scalar masses, for details see appendix~\ref{app:SQUIVER}.
The suppression is further enhanced by the fact that we have chosen to study
the range of parameter space where $\Lambda_{1,2}<\Lambda_3$.
Therefore it is possible for the masses of electroweakinos, sleptons and the
heavy Higgs bosons 
to lie well below 1 TeV.
One observes in table~\ref{tab:ExampleSpectra} that this results in 
the next-to-lightest supersymmetric particles (NLSPs), being either 
a neutralino, as in MI and MIIb, or stau, as in MIIa.
A sneutrino NLSP is also possible as will be discussed in detail 
later in this section.
As we only consider a single $SU(3)$ gauge group at the high scale, the masses
of the coloured sparticles do not experience this suppression.
This means that the coloured sector lies in general between 1.5 and 2.5 TeV,
the stops being the lightest squarks. However, in MII, a splitting is generated
between the left and right stop soft masses, for reasons discussed below, 
as shown in point MIIb of table~\ref{tab:ExampleSpectra}.
In the following we will study the spectra of these models in terms of their
compatibility with the current experimental constraints described in the previous section and the 
prospects for detecting signs of TeV-scale sparticles in the near future. 

\subsection{The Higgs mass and couplings}\label{sec:HiggsMass}
As the non-decoupled D-terms lift the tree-level Higgs mass, 
as in Eqn.~\eqref{eq:mh0QEW}, here we investigate the range of stop masses
in these models for which $m_h$ lies in the desired range,
and how the stop contribution compares to that of the non-decoupled D-terms.
The same non-decoupled D-terms can affect the couplings of the Higgs,
so we further investigate these couplings in light of current and future experimental
measurements.

\subsubsection{The Higgs mass}
\begin{figure}[t!]
\begin{center}
\includegraphics[width=0.49\textwidth]{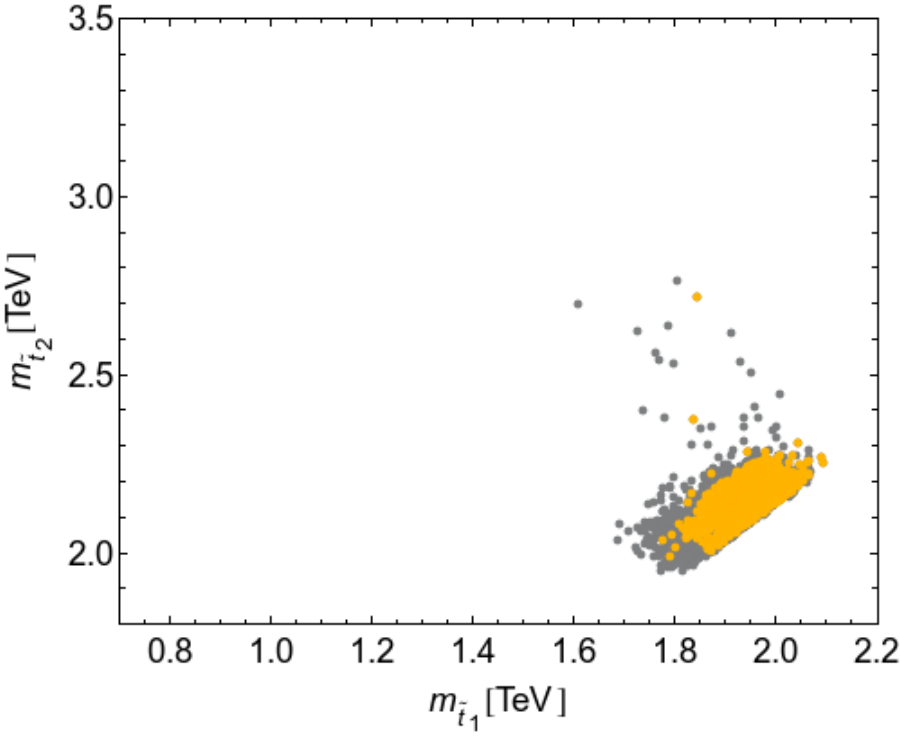}
\includegraphics[width=0.49\textwidth]{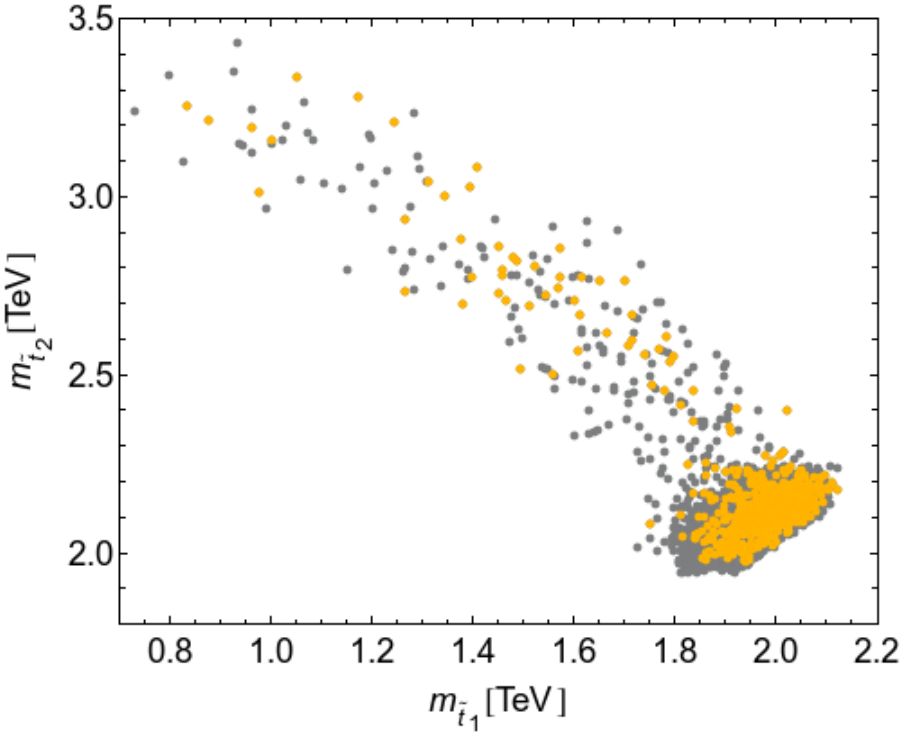}
\caption{The mass of the heavy stop $m_{\tilde{t}_2}$ as a function of
$m_{\tilde{t}_1}$ for MI (left) and MII (right). 
Points satisfying low-energy and the Higgs mass constraints are shown in yellow,
whereas the remaining excluded points are
shown in grey.} 
\label{fig:Mst1vsMst2}
\end{center}
\end{figure}

As mentioned in section \ref{sec:PspaceConstraints}, we have chosen $\Lambda_3$
such that the masses of
gluino and the first and second generation squarks lie above the LHC exclusion
limits.
In MI, this translates into the stop masses being close to 2 TeV, which means
that
the shift in the tree-level Higgs mass required in order to obtain $m_h\sim
125.5$ GeV is small, 
and the required soft mass of the linking field remains below $10$ TeV.
The situation is fairly similar in MII, however we find that there is a slight
tendency for a
splitting to arise between the left and right-handed stop soft masses,
due to the RGEs driving the left-handed mass downwards, and the right-handed
mass upwards.
This can be understood in terms of the differences between the RGE equations for
the two models, where in MII above the 
quiver-breaking scale 
the Higgs soft masses are only affected by the third generation squarks, whereas
for MI the Higgs soft mass
RGE equations contain all generations. This results in a larger splitting
between the up and down type Higgs soft 
masses which further generates a larger splitting between the left and right
handed stop.
The distribution of the masses of the light and heavy stops
i.e.~$m_{\tilde{t}_1}$ and $m_{\tilde{t}_2}$ as defined in appendix~\ref{app:fermions}
for the two models are displayed in figure~\ref{fig:Mst1vsMst2}. 
Here the allowed points are shown in yellow and those points excluded 
by the various constraints described in section~\ref{sec:PspaceConstraints} 
in grey. We clearly observe that for MII a larger splitting between the stops
is possible, and the lighter stop may be as light as 400 GeV, as seen 
in the benchmark point MIIb in table~\ref{tab:ExampleSpectra}.

\begin{figure}[t!]
\begin{center}
\includegraphics[width=0.49\textwidth]{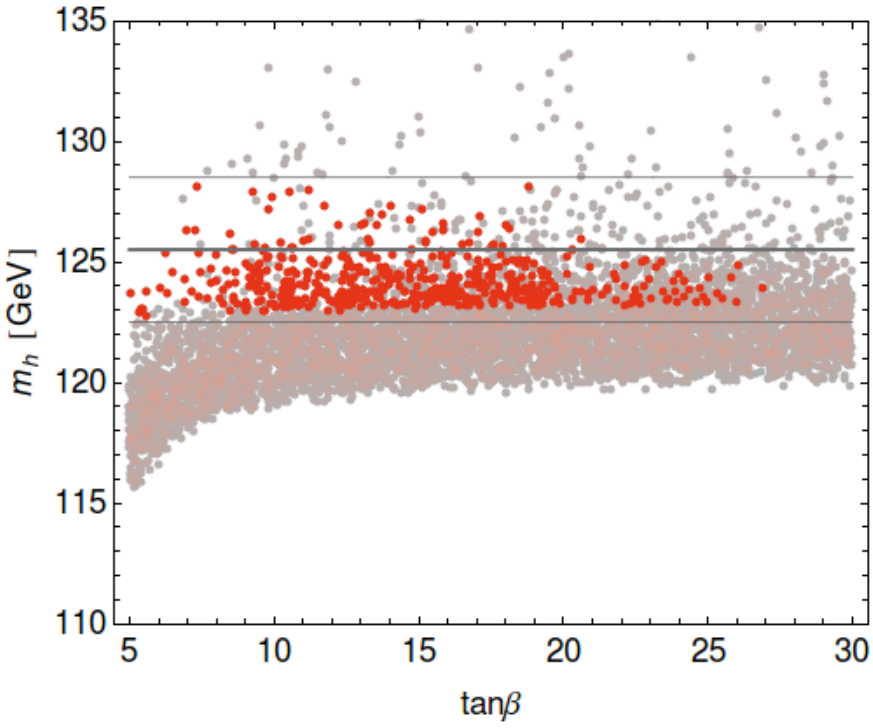}
\includegraphics[width=0.49\textwidth]{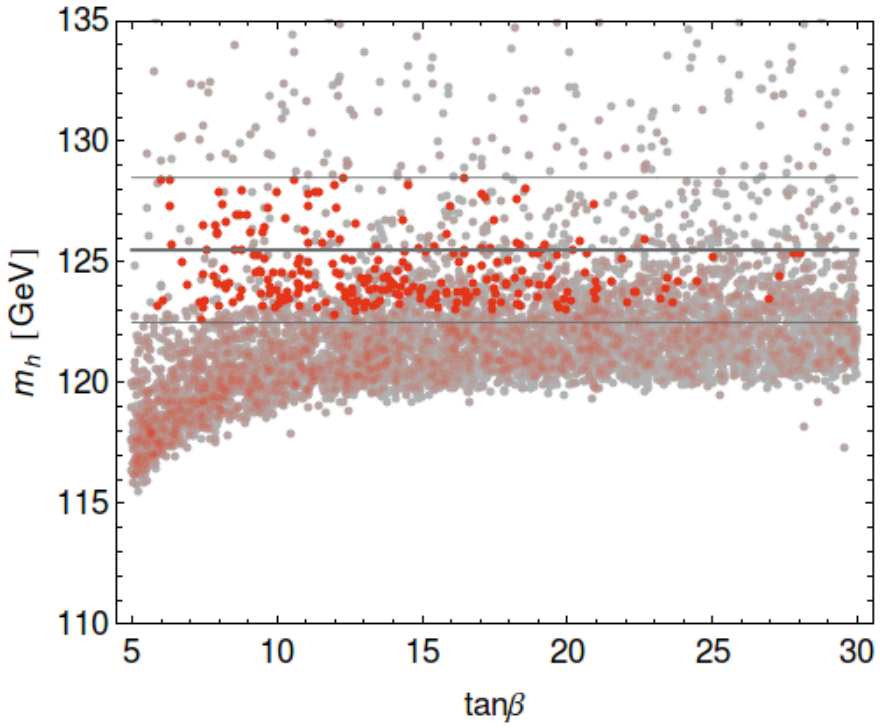}
\caption{The Higgs mass as a function of $\tan\beta$ for MI (left) and MII
(right). Points satisfying all constraints 
and $m_{\tilde{t}_1}<2$ TeV are bright red, 
whereas paler points indicate that stop masses are in the range
$2\,\mathrm{TeV}<m_{\tilde{t}_1}<2.3$ TeV and only low-energy constraints are satisfied.
Grey points are excluded.
The thick (thin) grey lines denote the central value (uncertainty) on $m_h$.} 
\label{fig:mHvsTb}
\end{center}
\end{figure}

In figure~\ref{fig:mHvsTb} we have plotted the Higgs mass as a function of $\tan\beta$
for the two variants of our model. 
Here the bright red points respect $m_{\tilde{t}_1}<2$ TeV and all constraints 
imposed, the pale red points only comply with the low-energy constraints 
and the grey points are excluded.
The full two-loop corrections to the Higgs mass are implemented in \texttt{SPheno}
following the calculation in 
Refs.~\cite{Degrassi:2001yf,Brignole:2001jy,Brignole:2002bz,Dedes:2002dy}.
We conclude that a Higgs mass within the limits $\sim125.5\pm 3.0$ GeV is
achievable in both MI and MII. Note that the larger range in $m_h$ for MII can be 
explained by the larger range in stop masses.
Indeed, when the left and right handed stop soft masses are not equal, as
shown in figure~\ref{fig:Mst1vsMst2} 
for MII, the simplified expression
for the one-loop Higgs mass given in Eqn.~\eqref{eq:higgsmass} is no longer
valid.
An additional correction must be added to Eqn.~\eqref{eq:higgsmass} of the form
\cite{Haber:1996fp} 
\be
\label{eq:HiggsShift}
\Delta m_{h,1}^2 = \frac{3 m_Z^2}{16\pi^2v_{ew}^2}(1-\frac{8}{3}\sin\theta_W^2)
\cos 2\beta\,
m_t^2\,\ln\left(\frac{m_{\tilde{q}^3_L}^2}{m_{\tilde{u}^3_R}}\right) ,\\
\ee
which for the case $m_{\tilde{q}^3_L}^2<m_{\tilde{u}^3_R}$ in MII induces an
enhancement to the Higgs mass of around $1$-$2$ GeV.
Note that the sfermion mixing matrix is defined in
appendix~\ref{app:HiggsSfermions}.
As the bright red points correspond to $m_{\tilde{t}_1}<2$ TeV, this further
demonstrates that the effect of
the non-decoupled D-term seems to reduce the fine tuning by allowing lighter
stops than in standard GMSB.

\begin{figure}[t!]
\begin{center}
\includegraphics[width=0.49\textwidth]{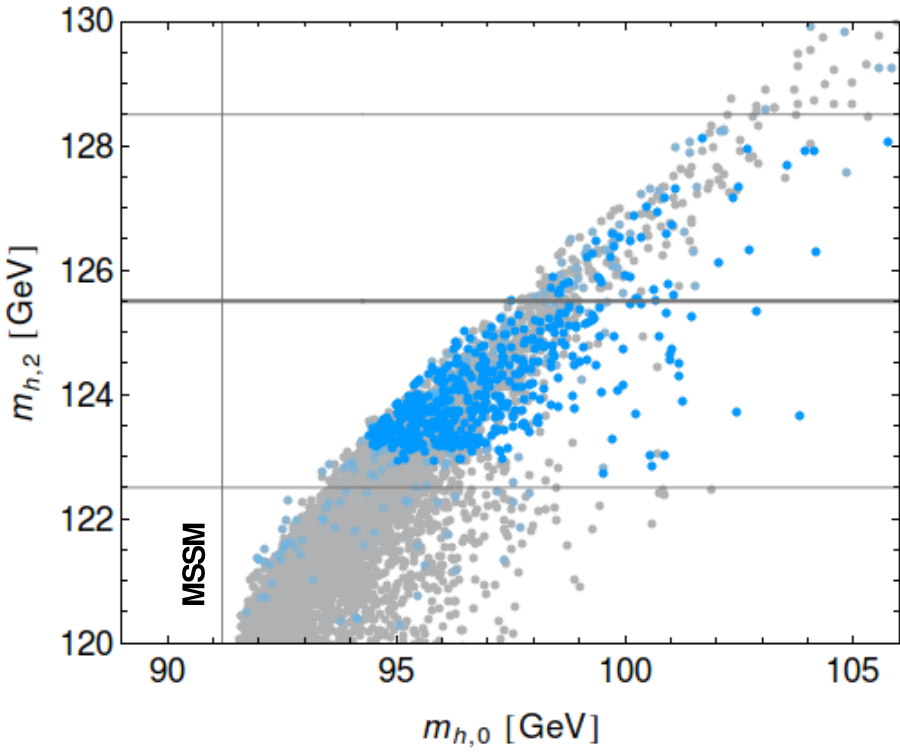}
\includegraphics[width=0.49\textwidth]{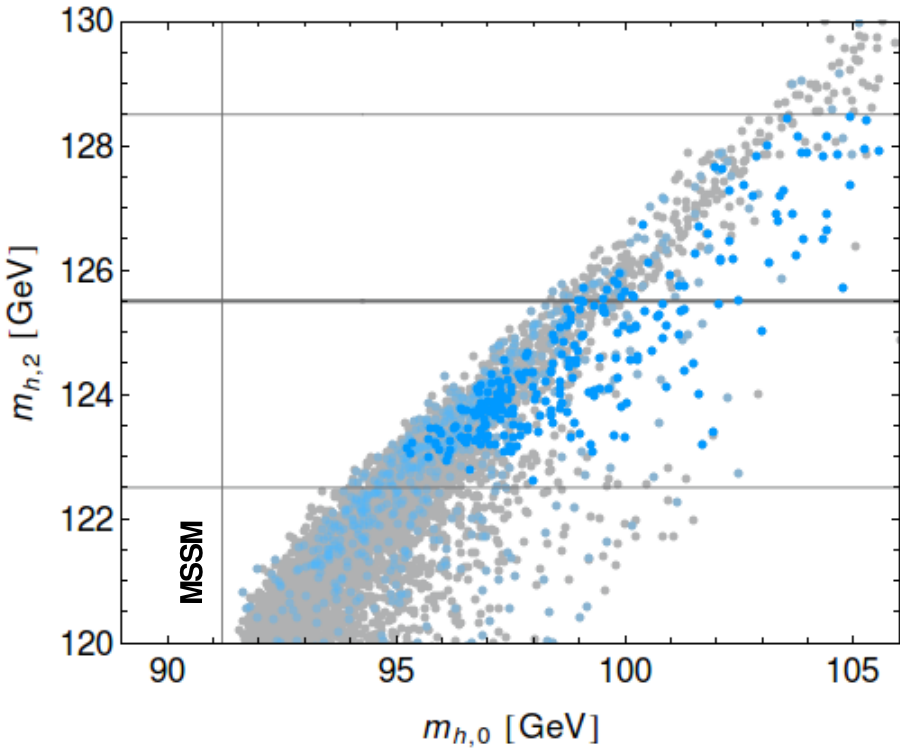}
\caption{The two-loop Higgs mass ($m_{h,2}$) as a function of the tree-level
Higgs mass ($m_{h,0}$) for MI (left) 
and MII (right). Points satisfying all constraints for which
$m_{\tilde{t}_1}<2$ TeV are bright blue, 
paler points indicate low-energy constraints are satisfied and stop masses are in the range 2 TeV$<m_{\tilde{t}_1}<2.3$ TeV,
and points excluded by
low energy constraints are shown in grey. The vertical line indicates the MSSM
bound on $m_{h,0}$. 
The thick (thin) grey lines denote the central value (uncertainty) on $m_h$.} 
\label{fig:mH-2Lvs0L}
\end{center}
\end{figure}

To make the distinction between the non-decoupled D-term and radiative,
i.e. stop sector, contributions 
to the Higgs mass clearer, 
we compare the tree-level result $m_{h,0}$ to the full two-loop result
$m_{h,2}$ in figure~\ref{fig:mH-2Lvs0L}.
Here the bright blue points respect $m_{\tilde{t}_1}<2$ TeV and all constraints 
imposed, the pale blue points only satisfy low-energy restrictions 
and the grey points are excluded.
As opposed to the mGMSB result, where $m_{h,0}$ is bounded by $m_Z$,
here we observe that a shift of up to 10 GeV is possible for both MI and
MII, while keeping $m_L<10$ TeV.
This in turn means that the contribution of the radiative corrections required
to achieve $m_{h}\sim125.5$ GeV is diminished,
rendering the model more natural.
Interestingly, the splitting of the stops observed in MII results in a distinct
difference between the two plots in 
figure~\ref{fig:mH-2Lvs0L}, which can be understood from
Eqn.~\eqref{eq:HiggsShift}. Despite the fact that the range 
in $\Lambda_3$ is the same for both MI and MII,
the stop splitting enhances the size of the radiative corrections, resulting in
a smaller shift in the tree 
level Higgs mass
required to obtain a value of $m_h$ in agreement with experiment.


\subsubsection{The Higgs couplings}

Since the discovery of the Higgs boson, not only has its mass been used to 
discriminate between supersymmetric models but also its couplings, see
e.g.~\cite{Benbrik:2012rm,Bechtle:2012jw,Blum:2012ii}.
The deviation of these couplings from the SM can be parameterised via the set of
ratios $r_i$, for $i=b,\gamma,g$ etc, where
\be
r_i=\frac{\Gamma_{\rm MSSM}(h\to ii)}{\Gamma_{\rm SM}(h\to ii)}. 
\ee
The $r_i$ are further related to the signal strengths $\mu_{i}$ normally quoted
by ATLAS and CMS, 
see e.g.~Refs.~\cite{Aad:2013wqa,CMS:yva}.
Note that the errors on the measured signal strengths are still too large to
make detailed interpretations
about the potential underlying SUSY model, and at present the $\mu_{i}$ are all
SM compatible, 
and therefore we will not tackle a precise calculation of the various 
signal strengths in this work.
As mentioned, we however make sure that the lightest Higgs boson signal 
strengths are in agreement with the existing LHC and TeVatron measurements 
within 3$\sigma$, employing the \texttt{HiggsSignals} code.
Recent studies by both ATLAS and CMS \cite{ATLAS:2013hta,CMS:2013xfa} have found
that with a luminosity of 
300~$\rm fb^{-1}$ at the 14 TeV LHC, an uncertainty on the measurement of $r_b$ 
should only be 10--13\%.
A more sensitive determination however should be possible at the international
linear collider (ILC), where 
for a centre of mass energy ($\sqrt{s}$) of 500 GeV, 500~$\rm fb^{-1}$ and
polarized beams ${\it P}(e^+,e^-)=(-0.8,+0.3)$,
a precision of 1.8\% is quoted in Ref.~\cite{Baer:2013cma}.

\begin{figure}[t!]
\begin{center}
\includegraphics[width=0.49\textwidth]{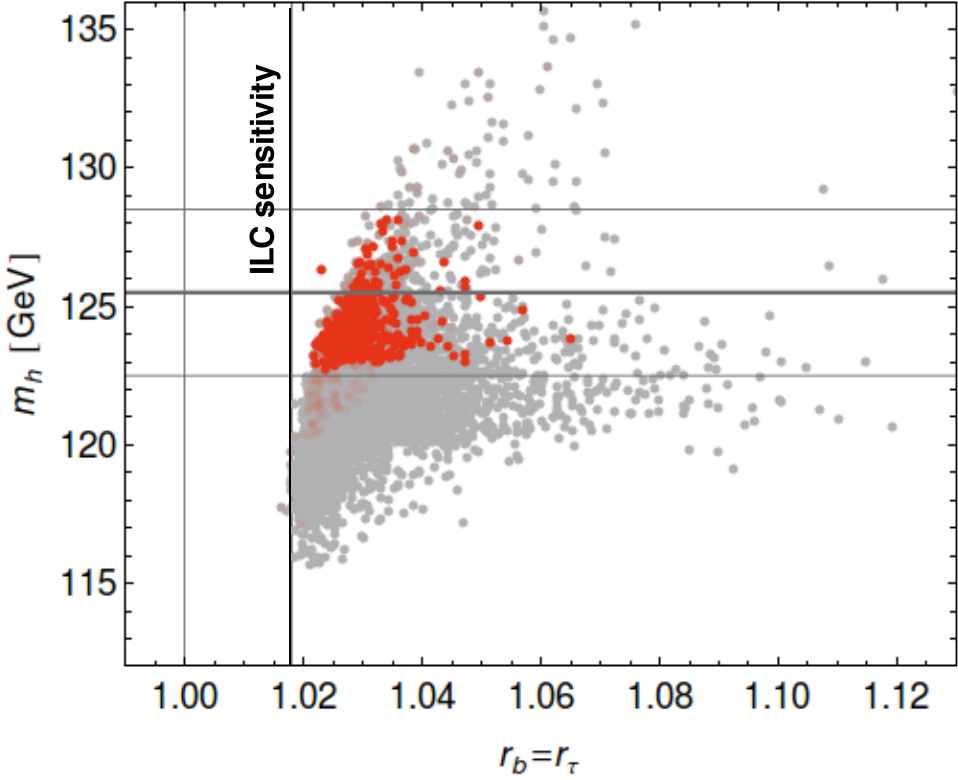}
\includegraphics[width=0.49\textwidth]{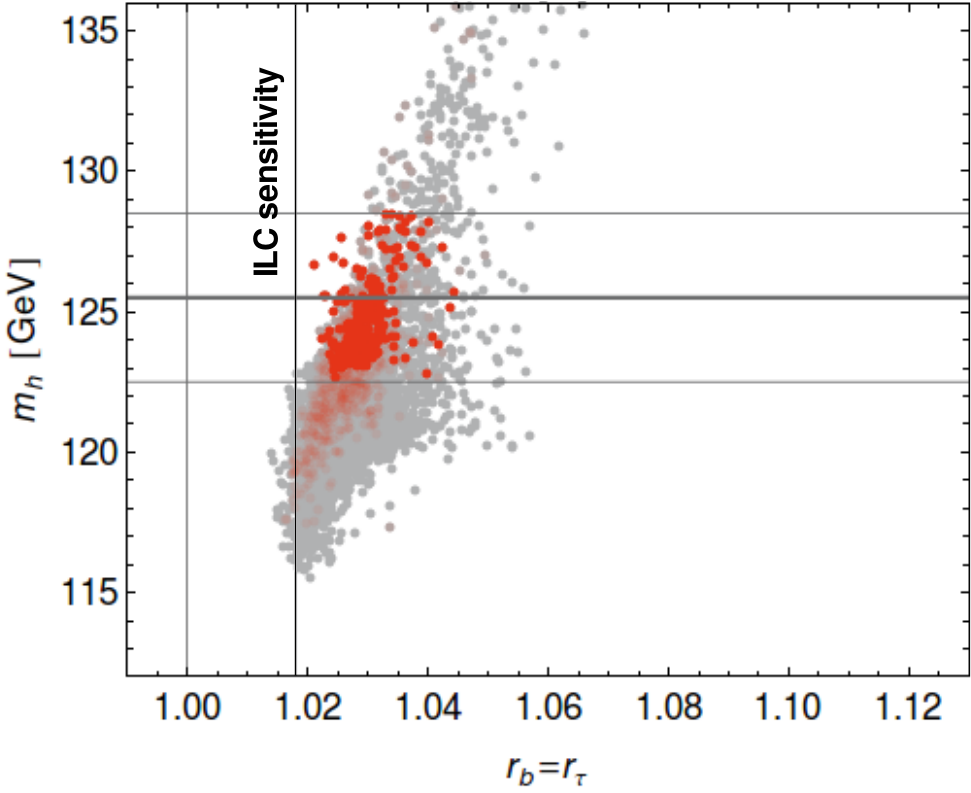}
\caption{The Higgs mass as a function of $r_b=r_\tau$  for MI (left) and MII
(right). 
 Points satisfying all constraints 
and $m_{\tilde{t}_1}<2$ TeV are bright red, 
whereas paler points indicate that stop masses are in the range
$2\,\mathrm{TeV}<m_{\tilde{t}_1}<2.3$ TeV and only low-energy constraints are satisfied.
Grey points are excluded.
The thick (thin) grey horizontal lines denote the central value (uncertainty) on $m_h$ and
the vertical lines show the SM value $r_b=1$ and the projected ILC uncertainty
of $2\%$ (see text).} 
\label{fig:mHvsRb}
\end{center}
\end{figure}

In our model, the non-decoupled D-terms result in a tree level contribution 
to the Higgs coupling to down-type fermions.
The ratio $r_b$ ($=r_\tau$ at tree level) takes the form
\be
r_b=-\frac{\sin\alpha}{\cos\beta},
\ee
where $\alpha$ is the angle between the two Higgs doublets in the MSSM, and is
defined by~\cite{Auzzi:2012dv}
\be
\tan 2\alpha=\frac{m_{A_0}^2\cos 2\beta+m_{h,0}^2}{m_{A_0}^2\cos
2\beta-m_{h,0}^2}\tan2\beta.
\ee
Here $m_{A_0}$ is the pseudoscalar Higgs mass of the MSSM and $m_{h,0}$ is the
tree-level Higgs mass given 
in Eqn.~\eqref{eq:mh0QEW}.
We plot the Higgs mass as a function of $r_b$ in figure~\ref{fig:mHvsRb},
where again the bright red points respect $m_{\tilde{t}_1}<2$ TeV and all constraints 
imposed, the pale red points only satisfy low energy constraints 
(i.e.~they do not comply with our requirements in the Higgs sector)
and the grey points are excluded.
We find that only a $\sim 2\%$ change in $r_{b/\tau}$ is required for
MI, 
and a $\sim 4\%$ change for MII in order to obtain a Higgs mass of 125.5 GeV,
with $m_{\tilde{t}_1}<2$ TeV.
Note that in our model, the enhanced coupling to down-type fermions results in a
suppression of 
the signal strength $\mu_\gamma$~\cite{Blum:2012ii,Craig:2012bs}, which was not
favoured by initial measurements at the LHC~\cite{ATLAS:2012wma}.
However, as data has collected, the results for $\mu_\gamma$ appear more and
more SM-like~\cite{Aad:2013wqa,CMS:yva}.
As in this work the tree-level shift in the Higgs mass only needs to be under 10
GeV, 
we consider small values of $\Delta_{1,2}$, for which the deviation in 
the coupling of the Higgs to down-type sfermions are
well within the current LHC bounds (see e.g. Ref.~\cite{Aad:2013wqa}) as shown
in figure~\ref{fig:mHvsRb}.
Such deviations should start to become detectable at the $\sqrt{s}=500$ GeV
ILC. 

\subsection{Sparticle searches at the LHC}\label{sec:Sparticles}

As mentioned in section~\ref{sec:PspaceConstraints}, the choice of $\Lambda_3$ ensures 
that the masses of the gluino and lower generation squarks approximately 
respect the limits from direct searches at the LHC.
On the other hand, as the scale of the electroweak sector is set by
$\Lambda_{1,2}$,
by allowing $\Lambda_{1,2}<\Lambda_3$ we explore a range of parameter space
for which 
the electroweak sector has a greater chance of being observed at the LHC.
Further, as discussed previously, the quiver structure means that at the high
scale the masses of uncoloured scalar particles lying on
site A are suppressed, which can in particular result in light higgsinos or
sleptons compared to minimal GMSB.

The phenomenology of the model depends decisively on the nature of the NLSP, 
as this decides which SM particle is present in the final state along with 
the gravitino $\tilde{G}$.
ATLAS and CMS have recently made much progress on constraining gauge mediated
models, where they 
study final states containing missing transverse energy ($E_T^{\rm miss}$) due
to the 
gravitino ($\tilde{G}$) escaping the detector. 
Bino-like NLSPs decay via $\tilde{\chi_1}\to\tilde{G}\gamma$, such that the
signature is $\gamma\gamma+E_T^{\rm miss}$, 
along with additional jets depending on whether the production process is
$\tilde{g}\tilde{g}$
or $\tilde{\chi}_1^0\tilde{\chi}_1^0$~\cite{Aad:2012zza}.
When higgsino-like, the NLSP instead decays to a Higgs which can be detected via
$b$ jets, 
and a mixed higgsinos-bino NLSP can be searched for
via a $\gamma b\bar{b}+E_T^{\rm miss}$ signature~\cite{Aad:2012jva}.
For stau or sneutrino NLSPs the $\tau$ or $\nu$ must be searched for in the final state.

\begin{figure}[t!]
\begin{center}
\includegraphics[width=0.49\textwidth]{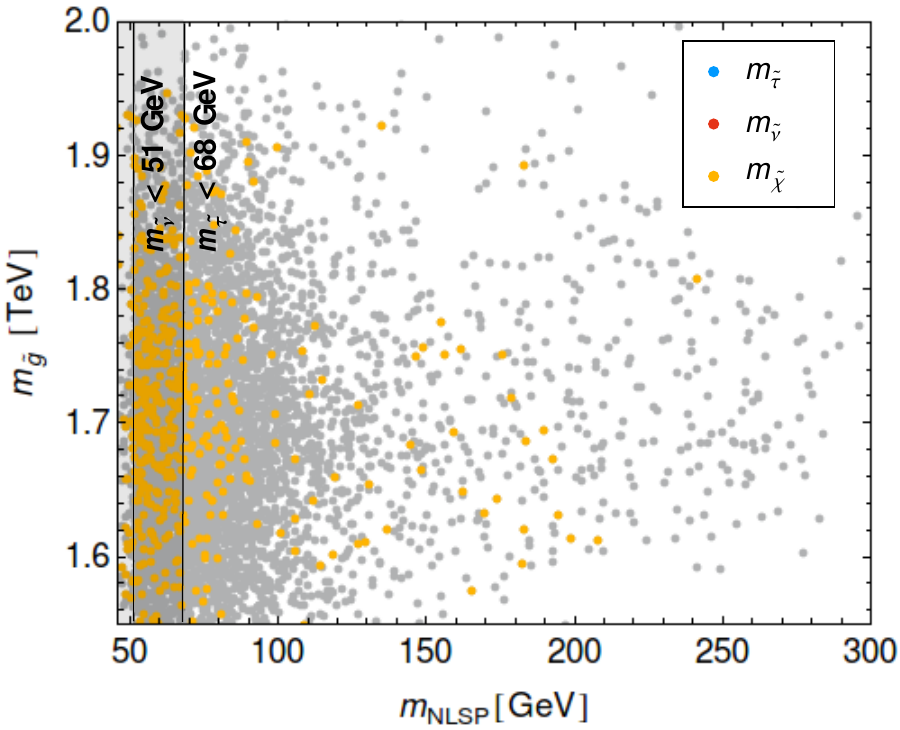}
\includegraphics[width=0.49\textwidth]{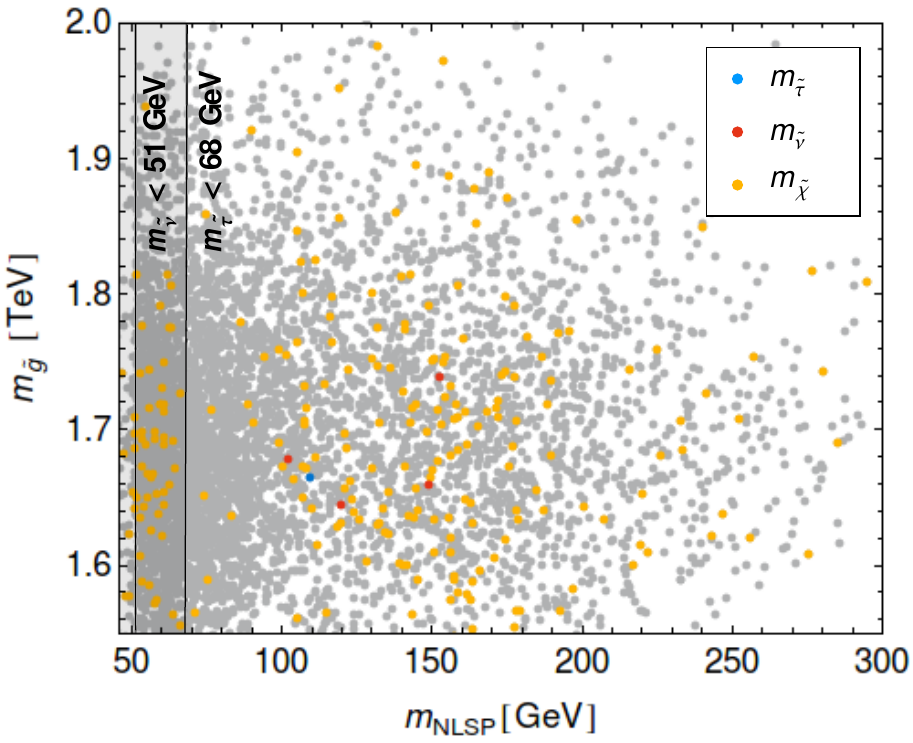}
\caption{The gluino mass ($m_{\tilde{g}}$) as a function of NLSP mass ($m_{\rm
NLSP}$) for MI (left) and MII (right), where the colours indicate the type of NLSP as shown in the legend.
The LEP exclusion limit for the case of the $\tilde{\tau}$ and $\tilde{\nu}$ NLSP is clearly 
marked, whereas the limit for $m_{\tilde{\chi}^0_1}$ is given by the y-axis. The grey points are excluded by experimental constraints
as described in the text} 
\label{fig:mgVmnlsp}
\end{center}
\end{figure}

\begin{figure}[t!]
\begin{center}
\includegraphics[width=0.49\textwidth]{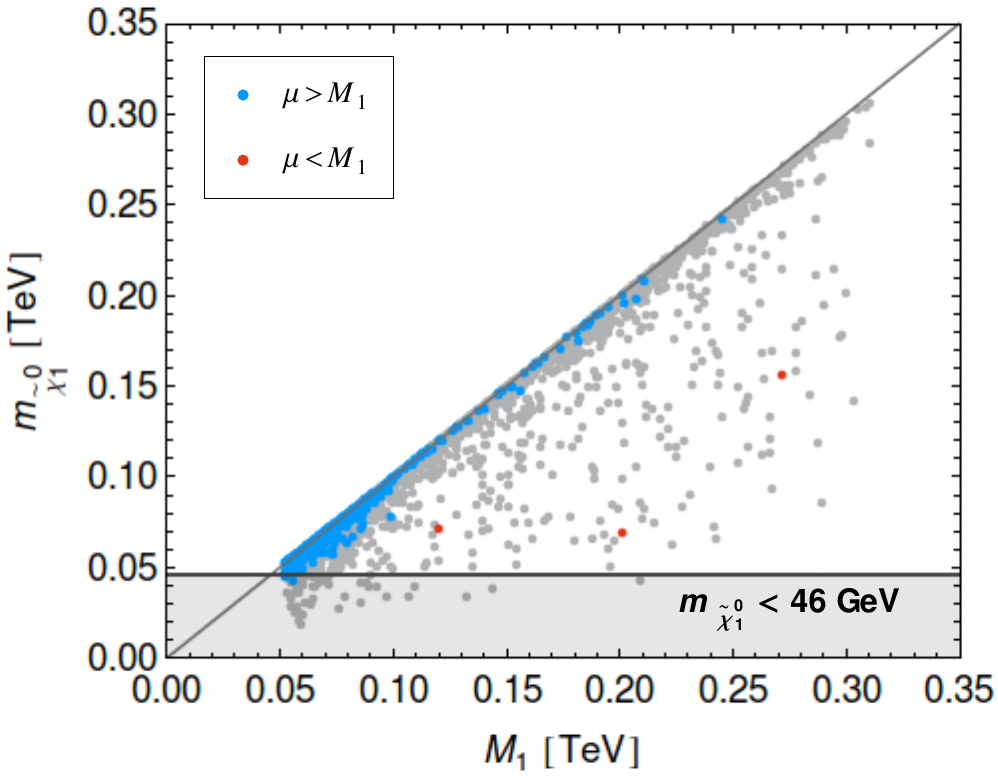}
\includegraphics[width=0.49\textwidth]{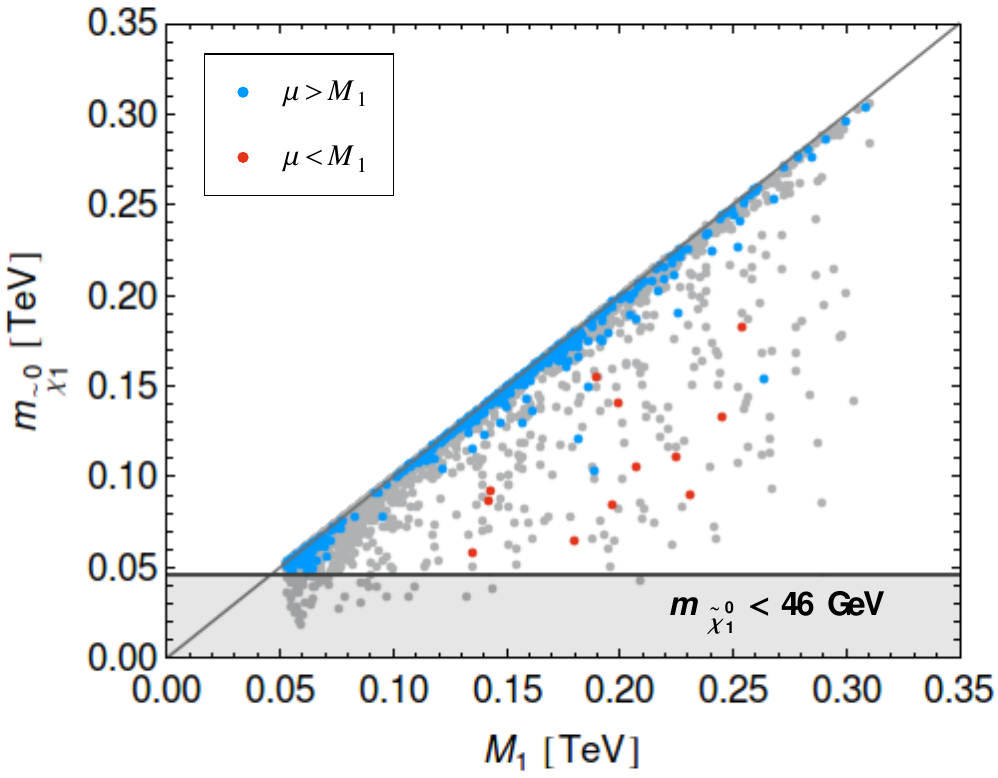}
\caption{The lightest neutralino mass ($m_{\tilde{\chi}_1^0}$) as a function of
$M_1$ for MI (left) and MII (right). The grey points are excluded by experimental constraints.
The grey diagonal line indicates $m_{\tilde{\chi}_1^0}=M_1$, and the horizontal line indicates 
the LEP exclusion limit as described in the text.} 
\label{fig:mneu1Vm1}
\end{center}
\end{figure}

In order to determine which experimental searches are relevant for these models,
in figure~\ref{fig:mgVmnlsp} we examine the region of the 
$m_{\tilde{g}}$--$m_{\rm NLSP}$ plane accessed by our scans, indicating the
type of NLSP for each point, which we find may be the neutralino, stau or sneutrino. 
The LEP exclusion limits (see section~\ref{sec:PspaceConstraints}) for both the $\tilde{\tau}$ 
and $\tilde{\nu}$ NLSP are clearly 
marked, whereas the limit for $m_{\tilde{\chi}^0_1}$ is given by the y-axis. 
We find that for MII there are allowed points for which either the sneutrino or
stau are the NLSP, however for MI no such points were found. 
In MI, the generations of sfermions are 
treated equally, such that the staus lie close to the other sleptons.
On the other hand in the case of MII, as mentioned earlier, above symmetry
breaking the third generation sfermions are on site A, whereas the lower generation ones 
on site B. This has the result that, as in the stop sector, the left-handed stau soft
mass may be lower than the right-handed one, such that a sneutrino NLSP is possible.
Therefore, although points in both models were found where the NLSP is the stau or sneutrino,
only in MII do points survive the demanding constraints imposed on the Higgs mass and couplings 
due to measurements at 
the LHC, as illustrated in figure~\ref{fig:mneu1Vm1}.

As the lightest neutralino $\tilde{\chi}_1^0$ appears to be the favoured candidate for the NLSP, it is interesting to explore 
its composition as this will enlighten us as to which decay modes are preferred.
Therefore in figure~\ref{fig:mneu1Vm1}, we show the lightest neutralino mass, $m_{\tilde{\chi}_1^0}$, as a function
of $M_1$. From this plot one can deduce whether $\tilde{\chi}_1^0$ is higgsino or bino-like respectively, depending on the
higgsino and bino masses, approximately given by $\mu$ and $M_1$ respectively. 
The ubiquitous blue points indicate $\mu>M_1$ whereas the more rare red points, 
which are even absent for MI, show $\mu<M_1$. The grey points are excluded by Higgs and low-energy
constraints, and the horizontal line demarcates the LEP-excluded region for $m_{\tilde{\chi}_1^0}$.
In both MI and MII, the higgsino is rarely lighter than
the bino, such that the NLSP is mostly bino-like or mixed bino-higgsino, 
while both can easily be below 300 GeV.
This is interesting in light of the fact that experiments are sensitive to 
the nature of the neutralino NLSP, and the searches would therefore involve 
photons and/or Higgs bosons
and missing transverse energy.
Note that the feature of $\mu$ being light results in the model being less
fine-tuned.

\begin{figure}[t!]
\begin{center}
\includegraphics[width=0.49\textwidth]{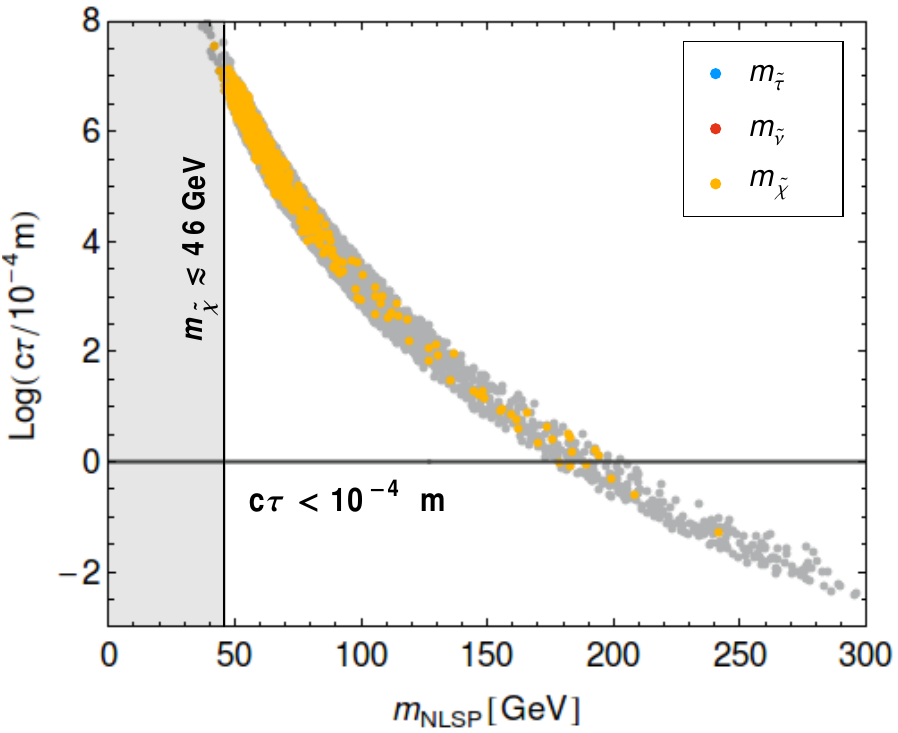}
\includegraphics[width=0.49\textwidth]{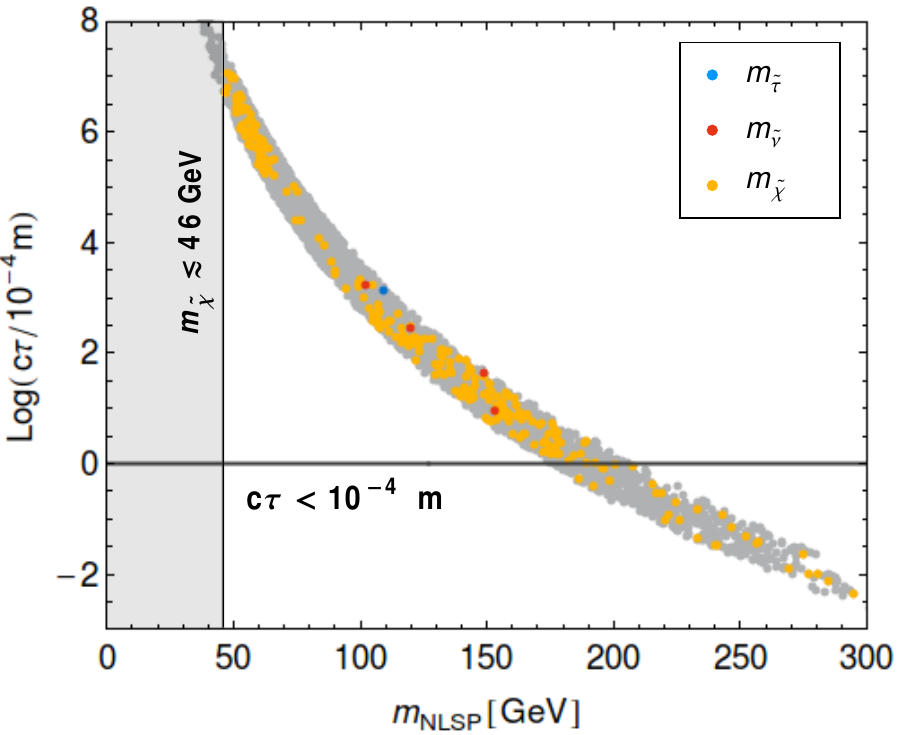}
\caption{The decay length ($c\tau$) of the NLSP as a function of its mass ($m_{\rm
NLSP}$) for MI (left) and MII (right). 
Points for which $m_{\tilde{t}_1}<2$ TeV and which satisfy all constraints imposed are shown in yellow, whereas 
the remaining excluded points are shown in grey. 
The vertical lines indicate the exclusion limits for staus and sneutrinos from LEP.} 
\label{fig:lVmnlsp}
\end{center}
\end{figure}

The experimental search strategy is not only dependent on the SM particle in the final state,
but also on the decay length of the NLSP, $c\tau$.
This can be approximated by~\cite{Giudice:1998bp},
\be
c\tau\sim\frac{16 \pi F^2}{m^5_{NLSP}},
\ee
where we take $F=\Lambda M$.
In the region of parameter space considered in this paper, the NLSP decays within the detector.
For the case of the neutralino NLSP decaying to a photon and gravitino, which is the prevalent case in both MI and MII, 
the excellent time measurement 
of the electromagnetic calorimeter in both ATLAS and CMS means that the time of arrival of the photon can be measured.
If the NLSP decays immediately, i.e. if $c\tau<10^{-4}$ m, then the decay 
is characterised as prompt, but otherwise it is
non-prompt and it may be possible to deduce its decay length~\cite{Aad:2013oua,Chatrchyan:2012jwg}.
We therefore show the decay length of the NLSP in figure~\ref{fig:lVmnlsp}, as a 
function of the mass $m_{\rm NLSP}$.
From this figure we can confirm that the lightest neutralino NLSP may undergo both promp or non-prompt decays
to the photon and the gravitino, although in MI fewer points survive for which the neutralino decays promptly.

The most important channels for these models at the LHC are therefore 
searches for photons and missing transverse energy, where the photons may be 
prompt or non-prompt. 
Here the dominant production would be electroweak, as our choice of $\Lambda_3$
is such that the gluon and squark pair production is suppressed.
Studies so far by CMS have concentrated on strong production of the 
bino-like NLSP~\cite{Chatrchyan:2012bba,Chatrchyan:2012jwg}, whereas
ATLAS has considered the diphoton and missing transverse energy final state from 
direct electroweakino production, for the case of
both promptly decaying~\cite{Aad:2012zza} and long-lived neutralinos~\cite{Aad:2013oua}.
However, the bounds obtained by ATLAS are not directly applicable here, as they are 
presented for a specific point SPS8~\cite{Allanach:2002nj}, where the neutralino NLSP 
decays predominantly to the photon and gravitino which is not necessarily the case in our model, 
especially due to the fact that the higgsino is often light.
Therefore the bounds on final states including Higgs bosons, 
studied in Refs.~\cite{ATLAS:2012crz,CMS:2013zzz}, must be taken into account.
In order to constrain MII, one must further consider the stau and sneutrino NLSP, such that
final states involving $\tau$s and missing transverse energy are of interest~\cite{ATLAS:2013ama,CMS:2013zzz}.
It would be of great interest to combine all these excluded cross-sections to extract precise 
exclusion bounds (along the lines of e.g. Ref.~\cite{Bharucha:2013epa}), but this is beyond the scope of this paper.
It nonetheless seems that for gauge mediated models 
an interesting region of parameter space is starting to be probed, and we 
eagerly await further results.

\section{Conclusions and discussion}\label{sec:conclusions}
In this paper we have examined phenomenological aspects of a minimal gauge extension of the 
MSSM containing two copies of the electroweak gauge group. 
Using state-of-the art publicly available HEP tools, we have computed 
the two-loop RG equations for \textit{all} parameters of two variants MI and MII 
of this basic setup, characterised by different assignments for the representation of 
the MSSM chiral superfields. 
Although the model may be amenable to any set of soft term boundary conditions, we have chosen 
to work within the framework of gauge mediatied supersymmetry breaking and we 
performed the RG evolution from the messenger scale down to the electroweak scale 
in order to compute the sparticle spectrum. We calculated the corresponding sparticle masses 
at one loop, the predicted Higgs boson mass at two loops and 
further investigated the predictions of the models MI and MII for the most relevant 
experimental observables.

As the extended gauge structure results in non-decoupled D-terms
which increase the tree-level Higgs mass, the resulting spectrum can be 
more natural than in minimal GMSB. We further found that in order to be in agreement 
with Higgs mass constraints, while keeping the stop masses below $2$ TeV, 
one must generate sufficiently large $\Delta$'s
(\refe{eq:nondecoupled}). This requires the linking soft mass $m_L$ to be 
${\cal{O}}(3-10)$ times higher than expected from the exact GMSB boundary conditions, which
indicates a useful direction in which to extend this work on a theoretical
level. 

We also found that both variants MI and MII of the model would have interesting 
phenomenological consequences at colliders, since they could be probed either 
indirectly through Higgs couplings measurements or via direct sparticle production.
The Higgs couplings to down-type fermions deviate from the SM 
due to the $\Delta$'s by $r_b\lesssim 6\%$ for MI and $r_b\lesssim4\%$ for MII, and 
although at present this is well within 
experimental limits, such deviations should be measurable at a linear collider. 
As we have focused on the region of 
parameter space where the coloured sector is $\sim2$ TeV in order to evade bounds 
on squark and gluino production from the LHC,
while the electroweak sector is kept below 1 TeV, the most promising 
production channel at the LHC is the direct production of electroweakinos. As the 
predominant NLSP is the bino-like neutralino, 
diphoton and missing transverse energy searches offer the most 
promising search perspectives, though for MII, the NLSP is not limited to the bino 
such that finals states containing $\tau$ or $h$ and missing transverse energy 
are also relevant. As the LHC exclusions are presented in terms 
of specific models, we are therefore keen to reinterpret these 
in order to understand how these bounds translate in the case of our model.

There are a number of ways in which this work may be extended.
A first step, as mentioned earlier, would be to determine 
whether larger $\Delta$'s can be realized without making $m_L$ a free parameter 
by including $U(1)$ kinetic mixing or, more ambitiously, an additional $SU(3)$.
This could be achieved by means of the tools we have developed with the help 
of the publicly available package \texttt{SARAH}.
It would further be of interest to study related models of flavour, or models with
chiral non-decoupled D-terms at the same level of precision. By moving Higgs 
fields or generations onto different quiver sites, such models are relatively 
straightforward to implement in our setup. 
It would also be ideal to construct single regime models, in
particular for cases where the phenomenology of additional light states may become
relevant. Furthermore, not only are there $SO(N)$ and $Sp(N)$ gauge extensions, but even more 
general quiver constructions, such as those with 3 or more sites
\cite{McGarrie:2010qr,Auzzi:2011wt}, may also be implemented in full. The study
of these models and their GUT completion is also a noble task from the
perspective of string phenomenology which has so far been rather neglected. 
\appendix

\acknowledgments  We would all like to thank Florian Staub for numerous
communications on \texttt{SARAH} and Werner Porod on \texttt{SPheno}.  M.M. would like to thank
Martin Winkler, Jan Hajer, Kazuki Sakurai,  Lisa Zeune, Florian Domingo, Kai
Schmidt-Hoberg, Felix Br\"ummer and J\"org J\"ackel for very useful discussions.
A.G. would like to acknowledge enlightening discussions with Fawzi Boudjema,
Genevi\`eve Belanger, Bj\"orn Herrmann, Antonio Mariano, Kirtimaan Mohan, Pierre
Salati and Paul Sorba. Our model package is available on request.  M.M. is
funded by the Alexander von Humboldt Foundation.
\appendix

\section{Some further comments on the
implementation}\label{sec:moreimplementation}
In this section we include some useful comments on the implementation of the
model.  
\subsection{Fermion mixing and soft term matching}\label{app:fermions}
In regime 1 there are many fermions that may mix together.  The uncharged
fermions are the diagonal linking fermions, the light and massive bino-type
fermions, the uncharged light and massive wino type fermions and finally the
diagonal fermion of the $A$ adjoint superfield as well as the K singlet fermion:
\be
\left(\chi_{L}^{1},\chi_{\tilde{L}}^{1}
,\chi_{L}^{4},\chi_{\tilde{L}}^{4},\tilde{B}_L ,\tilde{B}_M,\tilde{W}^3_L
,\tilde{W}^3_M,\psi_A^1,\psi_K \right)\label{eq:mixing1}.
\ee
The lightest two of these states become the MSSM bino and uncharged wino.  The
mass matrices may be found in the .pdf of the model file in \texttt{SARAH}.

The charged fermions that mix together are the off-diagonal linking fermions,
the charged wino-type light and massive gauginos and the off-diagonal $A$
superfield fermions:
\be
\left(\chi_{L}^{2},\chi_{\tilde{L}}^{2}
,\chi_{L}^{3},\chi_{\tilde{L}}^{3},\tilde{W}^1_L
,\tilde{W}^1_M,\tilde{W}^2_L
,\tilde{W}^2_M,\psi_A^2,\psi_A^3 \right)\label{eq:mixing2}
\ee
the lightest two of these become the MSSM charged winos.  The rest of the
states, both scalar and fermion, of the linking fields $K,A, L,\tilde{L}$ are
integrated out at the threshold scale between the first and second regime.
\subsection{MSSM Higgs and sfermion mixing matrices}
\label{app:HiggsSfermions}
The non-decoupled D-terms of section \ref{sec:linkingtoMSSM} appear in the
tadpole equations 
as well as the Higgs mixing matrices. For the real components \( \left(\phi_{d},
\phi_{u}\right), \left(\phi_{d}, \phi_{u}\right) \) we get
\begin{equation} 
m^2_{h} = \!\!\left( 
\begin{array}{cc}
m_{11}&\!\!\!\! -\frac{1}{4} g_{12}^2 v_d v_u \! - \!\text{Re}[B_{\mu}]\\ 
\!\!\!-\frac{1}{4} g_{12}^2 v_d v_u \!\! - \text{Re}[B_{\mu}]  &m_{22}\end{array} 
\!\! \right) 
 \end{equation} 
where $g_1=g'$ and for convenience we use $g_{12}^2=\Big(g_{1}^{2} + g_{2}^{2} + g_1^2\Delta_1^{2}
+g_2^2 \Delta_2^{2}\Big)$.
\begin{align} 
m_{11} &= \frac{1}{8} \Big(8 m_{H_d}^2  + 8 |\mu|^2  + g_{12}^2\Big(3 v_{d}^{2}  -
v_{u}^{2} \Big)\Big)\\ 
m_{22} &= \frac{1}{8} \Big(8 m_{H_u}^2  + 8 |\mu|^2  - g_{12}^2\Big(-3 v_{u}^{2}  +
v_{d}^{2}\Big)\Big)
\end{align} 
while for for pseudo-scalar Higgses \( \left(\sigma_{d}, \sigma_{u}\right),
\left(\sigma_{d}, \sigma_{u}\right) \) the relevant expressions are
 
\begin{equation} 
m^2_{A^0} = \left( 
\begin{array}{cc}
m_{11} &\text{Re}[B_{\mu}]\\ 
\text{Re}[B_{\mu}]&m_{22}\end{array} 
\right) +  \xi_{Z}m_Z^2 
 \end{equation} 
\begin{align} 
m_{11} &= \frac{1}{8} \Big(8 m_{H_d}^2  + 8 |\mu|^2  + g_{12}^2\Big(- v_{u}^{2}  +
v_{d}^{2}\Big)\Big)\\ 
m_{22} &= \frac{1}{8} \Big(8 m_{H_u}^2  + 8 |\mu|^2  - g_{12}^2\Big(- v_{u}^{2}  +
v_{d}^{2}\Big)\Big)
\end{align} 
The mass matrix for the charged Higgses \( \left(H_d^-, H_u^{+,*}\right),
\left(H_d^{-,*}, H_u^+\right) \) reads
 
\begin{equation} 
m^2_{H^-} = \left( 
\begin{array}{cc}
m_{11} &\frac{1}{4} \Big(4 B_{\mu}^*  + \Big(g_{2}^{2} +
g_2^2\Delta_2^{2}\Big)v_d v_u \Big)\\ 
\frac{1}{4} \Big(4 B_{\mu}  + \Big(g_{2}^{2} + g_2^2\Delta_2^{2}\Big)v_d v_u
\Big) &m_{22}\end{array} 
\right) +  \xi_{W^-}m_{W^-}^2 
 \end{equation} 
\begin{align} 
m_{11} &= \frac{1}{8} \Big(8 m_{H_d}^2  + 8 |\mu|^2  + g_{12}^2v_{d}^{2}  + \hat{g}^2_{12} v_{u}^{2}
\Big)\\ 
m_{22} &= \frac{1}{8} \Big(8 m_{H_u}^2  + 8 |\mu|^2  + g_{12}^2v_{u}^{2}  + \hat{g}^2_{12} v_{d}^{2}
\Big)
\end{align} 
where we have used the abbreviation $\hat{g}_{12}^2=\Big(-
g_{1}^{2}  - g_1^2\Delta_1^{2}  + g_{2}^{2} + g_2^2\Delta_2^{2}\Big)$, and
in all the above expressions, the $\xi$-terms are gauge-dependent
contributions (and we work in Feynman gauge throughout this paper).

For completeness, we also include the mixing matrix $M_{\tilde{f}}$ of a generic
sfermion $\tilde{f}$ which
may be a squark or charged slepton.
This matrix takes the form:
\begin{equation}
M_{\tilde{f}}=
\left( \begin{array}{cc}
m_{\tilde{f}_L}^2+m_{f}^2+\hat{M}_{Z}^2 (I^f_3-Q_f s_W^2) & m_f X^{\ast}_f  \\[.5em]
m_f X_f  & m_{\tilde{f}_R}^2+m_f^2+\hat{M}_{Z}^2\,Q_f s_W^2
\end{array} \right),
\label{eq:sfermion}
\end{equation}
for $s_w=\sin\theta_W$ where $\theta_W$ is the Weinberg weak mixing angle, 
and we make use of the abbreviation $\hat{M}_{Z}^2\equiv m_Z^2\cos{2\beta}$.
The off-diagonal element $X_f$ is defined in terms of the trilinear
coupling $A_f$ via
\begin{equation}
 X_f=A_f-\mu^{\ast} \left\{\cot\beta,\tan\beta\right\},
\end{equation}
where $\cot\beta$ applies for the up-type quarks, $f=u,c,t$, and
$\tan\beta$ applies for the down-type fermions, $f=d,s,b,e,\mu,\tau$. 
Note that $m_f$, $Q_f$ and $I_3^f$ are the mass,
charge and isospin projection 
of the fermion $f$, respectively.
On diagonalization of this matrix one obtains the light and heavy sfermion
masses
$m_{\tilde{f}_1}$ and $m_{\tilde{f}_2}$.

\subsection{Renormalisation group equations}
We evolved the model down from the messenger scale $M$, to a threshold scale
$T_{\text{scale}}$, which is associated with the masses of the linking field 
states $O(m^2_v)$.   The two-loop renormalisation group equations were used in
both regimes 1 and 2, along with one-loop finite energy corrections and two-loop
anomalous dimensions.
\\
\\
The beta functions of the gauge couplings of the first regime at one loop are
\be
\beta_{g_{a}}\equiv\frac{d}{dt}g_{a}=\frac{b_{a}}{16\pi^2}g_{a}^3  \ \  \
\text{with}   \ \ \  b_{a}=(\frac{39}{5},\frac{6}{5},-2,3,-3)
\ee
where $a=U(1)_A,U(1)_B,SU(2)_A,SU(2)_B,SU(3)$, 
which may be compared with the MSSM regime where
\be
b_a=(33/5,1,-3).
\ee
Let us also track the top Yukawas using the ``only third family approximation'',
\be
\beta^1_{y_t}\equiv\frac{d}{dt}y_{t}\simeq \frac{y_t}{16\pi^2}\left[ 4y_t^* y_t 
+y^*_by_b-\frac{16}{3}g_3^2 -3g^2_{A2}-\frac{13}{15}g^2_{A1}\right].
\ee
In the second regime these become
\be
\beta_{y_t}\equiv\frac{d}{dt}y_{t}\simeq \frac{y_t}{16\pi^2}\left[ 6y_t^* y_t 
+y^*_b y_b-\frac{80}{15}g_3^2 -3g^2_2-\frac{13}{15}g^2_1\right].
\ee
In the first regime  we find the trilinear $A_t$ coupling to be 
\begin{align}
\nonumber 16\pi^2 \frac{d}{dt}A_t\simeq& A_t\left[9y^*_ty_t
+y^*_by_b-\frac{16}{3}g_3^2-3g^2_{A2} -\frac{13}{15}g^2_{A1}\right]\\
 &+y_t\left[\frac{32}{3}g^2_3m_{\tilde{g}}+6g^2_{A2}m_{\tilde{W}_A}+\frac{26}{15}
g^2_{A1}m_{\tilde{B}_A}\right]+2a_b y^*_b y_t
\end{align}
whereas in the MSSM
\begin{align}
\nonumber 16\pi^2 \frac{d}{dt}A_t\simeq & A_t\left[18y^*_ty_t
+y^*_by_b-\frac{16}{3}g_3^2-3g^2_2 -\frac{13}{15}g^2_1\right]\\
 &+y_t\left[\frac{32}{3}g^2_3m_{\tilde{g}}+6g^2_{2}m_{\tilde{W}}+\frac{26}{15}
g^2_{1}m_{\tilde{B}}\right]+2a_b y^*_b y_t.
\end{align}
Let us also look at how the gauginos obtain soft mases.  The one-loop beta
functions for the B-site gaugino soft masses are given by \be
\beta^1_{m_{\tilde{B}_{B}}}=\frac{12}{5}g^2_{B1}m_{\tilde{B}_{B}}  \ \ \ \ 
\beta^1_{m_{\tilde{W}_{B}}}=-4 g^2_{B2}m_{\tilde{W}_{B}} \ \  \ 
\beta^1_{m_{\tilde{g}}}=-6g^2_{3}m_{\tilde{g}}.
\ee
For the A-site gauginos they are given by
\be
\beta^1_{m_{\tilde{B}_{A}}}=\frac{78}{5}g^2_{A1}m_{\tilde{B}_{A}}  \ \ \ \ 
\beta^1_{m_{\tilde{W}_{A}}}=6 g^2_{A2}m_{\tilde{W}_{A}}.
\ee
Even though the A-site gaugino masses are vanishing at the messenger scale $M$
the two loop-contributions  which typically depend on all the other gaugino soft
mases, feedback into the one-loop contributions. Finally the supersymmetric
Dirac masses associated with the quiver structure will lift their mass
eigenstates.  The two-loop equations are given by 
\begin{align}
\beta^{(2)}_{m_{\tilde{B}_{B}}}=&\frac{6}{25}g^2_{B1}\Big(12g_{B1}^2m_{\tilde{B}
_{B}} +
30g^2_{2A}(m_{\tilde{W}_{A}}+m_{\tilde{W}_{B}})+30g^2_{B2}(m_{\tilde{B}_{B}}+m_{
\tilde{W}_{B}})
\nonumber\\
 &+ 6g_{A1}^{2} (m_{\tilde{B}_A} + m_{\tilde{B}_B})-30 Y_{A}^* (m_{\tilde{B}_B}
Y_{A}  - T_{A} )
 +5Y_{K}^* (T_{K}- m_{\tilde{B}_B} Y_{K}   ) \Big) \\
\beta_{m_{\tilde{B}_A}}^{(2)}  =&  
\frac{4}{75} g_{A1}^{2} \Big(-\frac{270}{2} Y_{A}^* (m_{\tilde{B}_A} Y_{A}  -
T_{A} )-\frac{45}{2} Y_{K}^* (m_{\tilde{B}_A} Y_{K}  - T_{K} )\nonumber
 + 620 g_{3}^{2} m_{\tilde{B}_A}  \\&\nonumber+650 g_{A1}^{2} m_{\tilde{B}_A}+315
g_{A2}^{2} m_{\tilde{B}_A} +27 g_{B1}^{2} m_{\tilde{B}_A} +135 g_{B2}^{2}
m_{\tilde{B}_A} +315 g_{A2}^{2} m_{\tilde{W}_A}  \nonumber \\ &+27 g_{B1}^{2}
m_{\tilde{B}_B}
 +135 g_{B2}^{2} m_{\tilde{W}_B} +620 g_{3}^{2} m_{\tilde{g}} \\ &-35
m_{\tilde{B}_A} \mbox{Tr}({Y_d  Y_{d}^{\dagger}}) -135 m_{\tilde{B}_A}
\mbox{Tr}({Y_e  Y_{e}^{\dagger}}) -65 m_{\tilde{B}_A} \mbox{Tr}({Y_u 
Y_{u}^{\dagger}}) \nonumber \\ 
 &+35 \mbox{Tr}({Y_{d}^{\dagger}  T_d}) +135 \mbox{Tr}({Y_{e}^{\dagger}  T_e})
+65 \mbox{Tr}({Y_{u}^{\dagger}  T_u}) \Big)
\end{align}

\begin{align}
 \beta_{m_{\tilde{W}_A}}^{(2)}  =&  
\frac{2}{5} g_{A2}^{2} \Big(15 g_{A1}^{2} m_{\tilde{B}_A} +120 g_{3}^{2}
m_{\tilde{W}_A} +15 g_{A1}^{2} m_{\tilde{W}_A} +390 g_{A2}^{2} m_{\tilde{W}_A}
+6 g_{B1}^{2} m_{\tilde{W}_A}\nonumber \\ 
 & +30 g_{B2}^{2} m_{\tilde{W}_A} +6 g_{B1}^{2} m_{\tilde{B}_B} +30 g_{B2}^{2}
m_{\tilde{W}_B} +120 g_{3}^{2} m_{\tilde{g}}\nonumber \\
&-30 Y_{A}^* (m_{\tilde{W}_A} Y_{A} 
- T_{A} )+Y_{K}^* (-5 m_{\tilde{W}_A} Y_{K}  + 5 T_{K} ) \nonumber \\
 &-10 m_{\tilde{W}_A}\mbox{Tr}({Y_d  Y_{d}^{\dagger}}) 
 -10 m_{\tilde{W}_A} \mbox{Tr}({Y_e  Y_{e}^{\dagger}}) 
 -10 m_{\tilde{W}_A} \mbox{Tr}({Y_u  Y_{u}^{\dagger}}) +10
\mbox{Tr}({Y_{d}^{\dagger}  T_d})  \nonumber \\
&+10 \mbox{Tr}({Y_{e}^{\dagger}  T_e}) +10
\mbox{Tr}({Y_{u}^{\dagger}  T_u}) \Big) \\
\beta_{m_{\tilde{W}_B}}^{(2)}  =&  
\frac{2}{5} g_{B2}^{2} \Big(6 g_{A1}^{2} m_{\tilde{B}_A} +30 g_{A2}^{2}
m_{\tilde{W}_A} +6 g_{B1}^{2} m_{\tilde{B}_B} +6 g_{A1}^{2} m_{\tilde{W}_B} +30
g_{A2}^{2} m_{\tilde{W}_B} \nonumber \\ 
 &+6 g_{B1}^{2} m_{\tilde{W}_B} +140 g_{B2}^{2} m_{\tilde{W}_B} -70 Y_{A}^*
(m_{\tilde{W}_B} Y_{A}  - T_{A} ) \nonumber \\
&+Y_{K}^* (-5 m_{\tilde{W}_B} Y_{K}  + 5 T_{K}
)\Big)\\ 
\beta_{m_{\tilde{g}}}^{(2)}  =&  
\frac{2}{15} g_{3}^{2} \Big(33 g_{A1}^{2} m_{\tilde{B}_A} +135 g_{A2}^{2}
m_{\tilde{W}_A} +420 g_{3}^{2} m_{\tilde{g}} +33 g_{A1}^{2} m_{\tilde{g}} +135
g_{A2}^{2} m_{\tilde{g}}\nonumber \\ 
 & -20 m_{\tilde{g}} \mbox{Tr}({Y_d  Y_{d}^{\dagger}}) -20 m_{\tilde{g}}
\mbox{Tr}({Y_u  Y_{u}^{\dagger}}) +20 \mbox{Tr}({Y_{d}^{\dagger}  T_d}) +20
\mbox{Tr}({Y_{u}^{\dagger}  T_u}) \Big).
\end{align}
The soft masses for the quiver module run too:
\begin{align}
\beta_{m_{A}^2}^{(1)}  &=  
-8 g_{B2}^{2} |m_{\tilde{W}_B}|^2  +4 (m_{A}^2 + m_{L}^2 +
m_{\tilde{L}}^2)|Y_{A}|^2  +4 |T_{A}|^2
\\
\beta_{m_{K}^2}^{(1)} & =   + 2(m_{K}^2 + m_{L}^2 + m_{\tilde{L}}^2)|Y_{K}|^2  +
2|T_{K}|^2 \label{Mk}\\ 
\beta_{m_{\tilde{L}}^2}^{(1)} & =  
\frac{1}{10} \Big(-12 g_{A1}^{2} |m_{\tilde{B}_A}|^2 -60 g_{A2}^{2}
|m_{\tilde{W}_A}|^2 -12 g_{B1}^{2} |m_{\tilde{B}_B}|^2 -60 g_{B2}^{2}
|m_{\tilde{W}_B}|^2 \nonumber \\ & \ \ \ \ +30( m_{A}^2 + m_{L}^2  
 +m_{\tilde{L}}^2 )|Y_{A}|^2 +5 (m_{K}^2 +m_{L}^2+m_{\tilde{L}}^2 )|Y_{K}|^2\nonumber\\ &
\ \ \ \ +30 |T_{A}|^2 +5 |T_{K}|^2 +6 g_{A1}^2 \sigma_{1,3} \Big),
\end{align}
where by $\sigma_{1,3}$ we denote the soft mass combination
\begin{align}
\sigma_{1,3} & =  -2 \mbox{Tr}({m_u^2})  -2 m_{L}^2  + 2 m_{\tilde{L}}^2  -
m_{H_d}^2  + m_{H_u}^2 + \mbox{Tr}({m_d^2})\nonumber\\& + \mbox{Tr}({m_e^2}) -
\mbox{Tr}(m_l^2)  + \frac{1}{3}\mbox{Tr}(m_q^2).
\end{align}
The vev of the linking fields runs as well:
\be
\beta^{(1)}_v=\frac{v}{20}(-30|Y_A|^2 - 5|Y_K|^2 + 3g_{A1}^2 + 15g_{A2}^2 +
3g_{B1}^2 + 15g_{B2}^2 )(1 + \xi)  .
\ee
Further equations may be found in the pdf for this model, including all
anomalous dimensions and beta functions.  At the electroweak scale one finds 
\begin{align}
\nonumber|\mu|^2 =&\frac{1}{8(v_d^2 - v_u^2)} (-8m_{H_d}^2 v_d^2 + 8 m_{H_u}^2v_u^2-
g_1^2\Delta_1^2v_d^4 - g_2^2\Delta_2^2v_d^4  \\+&g_1^2 \Delta_1^2 v_u^4 +
g_2^2\Delta_2^2v_u^4   +           
  g_2^2v_u^4 - g_2^2v_d^4  - g_1^2 v_d^4 +g_1^2  v_u^4)
\end{align}
and
\be
B_{\mu} = -\frac{v_d v_u }{4(v_d^2 - v_u^2)}\left( 4m_{H_d}^2 - 4m_{H_u}^2 +
(g_1^2 + g_2^2 +g_1^2 \Delta_1^2 +g_2^2 \Delta^2_2)(v_d^2 -v_u^2)\right).
\ee
which are after all used to minimise the electroweak tadpole equations.
\subsection{Threshold effects}\label{app:thresholdeffects}
We integrate out various states at the threshold between the two regimes.   
These include the fermions discussed above as well as all linking scalars and
scalars of $K$ and $A$.  To implement this correctly, with two-loop RGEs, we
edited by hand the \texttt{SPheno} code to properly account for the finite shifts and the
mass orderings of the particles integrated out, given by 
\be
g_{i}\rightarrow g_{i}\Big[1\pm\frac{g_{i}^2 b^i_{state}(R)
}{8\pi^2}\ln\left(\frac{M_{state}}{M_{T}}\right)\Big],
\ee
to account for our particular matter content and
\be
m_{\tilde{g}}\rightarrow m_{\tilde{g}}\Big[1\pm\frac{g_{i}^2 b^i_{state}(R)
}{8\pi^2}\ln\left(\frac{M_{state}}{M_{T}}\right)\Big]
\ee
for the gluino shift between regimes. The other soft mass parameters for the
bino and winos are matched as the lightest states as explained in
appendix~\ref{app:fermions}. The shifts for each field component are given by
\be 
b_{state}(R) = \{11/3,-2/3,-1/3,-1/6   \}\times\frac{T(R)}{D(R)}.
\ee
The numbers are associated to a gauge boson, weyl fermion, complex  and real
scalar respectively. $T(R)$ is the index (half the Dynkin index $I(R)$), such
that $T(\Box)=1/2, T(Adj)=N_c$).  It is divided by the dimension of the
representation D(R) as each shift is for the component of the field and not the
full multiplet, in \texttt{SARAH}.   The massive gauge fields are integrated out either
on their own or by including them with the finite shifts of the real eaten
goldstone modes.
\subsection{Soft mass function}\label{app:SQUIVER}
\begin{figure}[t!]
\begin{center}
\includegraphics[width=0.49\textwidth]{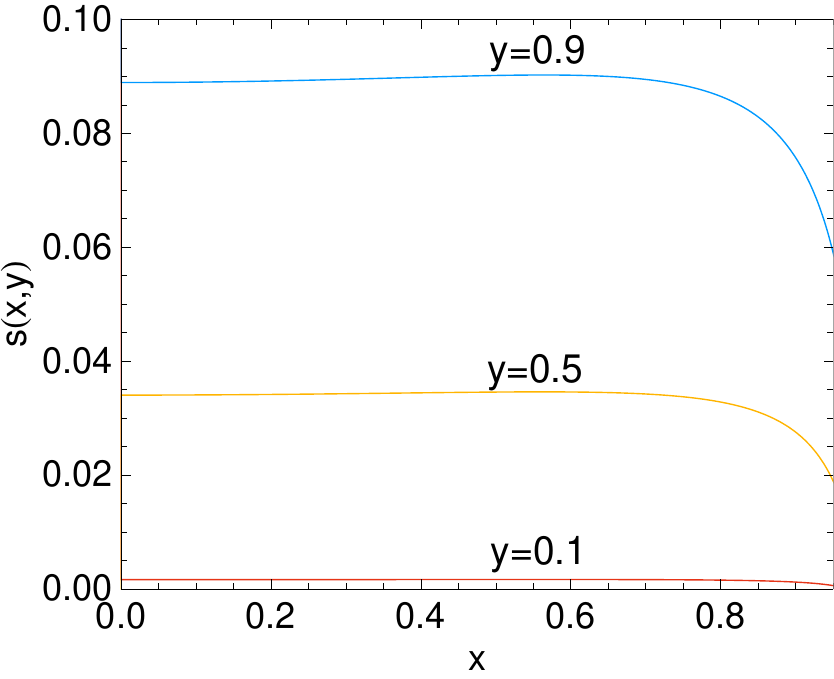}
\caption{We show $s(x,y)$ as a function of the parameter $x$ for fixed values of
$y$ as indicated.} 
\label{fig:SQUIVER}
\end{center}
\end{figure}

In the \texttt{SARAH} package we included, in the form of a Fortran function, the
generalisation of the usual mGMSB formula $f(x)$ 
to the case of a two site quiver model: $s(x,y)$, as pictured in figure \ref{fig:SQUIVER}.
The analytic expression is given by 
\be
s(x,y)=\frac{1}{2x^2}\left(s_0 +\frac{s_1+ s_2}{y^2} + s_3 +s_4 +s_5 \right)
 + \, (x \rightarrow -x)\, , 
\ee
where
\be\nonumber
s_0=2(1+x) \left( \log (1+x) -2 {\rm Li}_2  \left( \frac{ x}{1+x}\right)
+\frac{1}{2} {\rm Li}_2 \left( \frac{2 x}{1+x}\right) \right)   \, , 
\ee
\be
s_1=- 4 x^2   - 2 x(1+x) \log^2(1+x) - x^2 \, {\rm Li}_2(x^2) \, ,
\nonumber\ee
\be
s_2=8 \left(1+x\right)^2 h\left(\frac{y^2}{1+x},1\right)-4 x \left(1+x\right)
h\left(\frac{y^2}{1+x},\frac{1}{1+x}\right)
\nonumber\ee
  \be
   -4 x    h\left(y^2,1+x \right)-8 h\left(y^2,1\right) \, ,
\nonumber\ee
\be
s_3= -2 h\left(\frac{1}{y^2},\frac{1}{y^2}\right)
-2 x \,   h\left(\frac{1+x}{y^2},\frac{1}{y^2}\right) +
2(1+ x) h\left(\frac{1+x}{y^2},\frac{1+x}{y^2}\right) \, ,
\nonumber \ee
\be
s_4=(1+x) \left(  2  h\left(\frac{y^2}{1+x},\frac{1}{1+x}\right)
 - h\left(\frac{y^2}{1+x},1\right)-
h\left(\frac{y^2}{1+x},\frac{1-x}{1+x}\right)  \right) \, ,
\nonumber \ee
\be
s_5= 2 h\left(y^2,1+x\right)-2 h\left(y^2,1\right) \, . 
\ee
The function $h$ is given by the integral
\be
h(a,b)= \int_0^1 dx  \left( 1+ {\rm Li}_2 (1-\mu^2) -\frac{\mu^2}{1-\mu^2}  \log
\mu^2 \right) \, .
\ee
The dilogarithm is defined as ${\rm Li}_2(x)=-\int_0^1 \frac{dt}{t}\log(1-xt)$ 
with
\be
\mu^2=\frac{a x + b(1-x)}{x(1-x)} \, \ \ , \ \   a=m_1^2/m_0^2   \ \ , \ \
b=m_2^2/m_0^2.
\ee
So as not to introduce IR divergent pieces it is best to first  evaluate terms
with massless propagators.  In that case the  the function $h$  simplifies to 
$h(0,b)=1+\rm{Li}_2 (1-b)$ and has a symmetry $h(b,0)$=$h(0,b)$.   For four
massive poles, the analytic expression for $h$ is used in \texttt{SARAH}
\begin{align}
\nonumber h(a,b)=&1-\frac{\log a \log b}{2} -\frac{a+b-1}{\sqrt{\Delta}} \left( {\rm Li}_2
\left (-\frac{u_2}{v_1} \right) + {\rm Li}_2 \left (-\frac{v_2}{u_1} \right)
\right.\\
&\left. + \frac{1}{4} \log^2 \frac{u_2}{v_1}  +  \frac{1}{4} \log^2
\frac{v_2}{u_1} +    \frac{1}{4} \log^2 \frac{u_1}{v_1} -    \frac{1}{4} \log^2
\frac{u_2}{v_2} + \frac{\pi^2}{6} \right) \, ,
\end{align}
where
\be
\Delta= 1-2 (a+b) +(a-b)^2 \, ,  \qquad u_{1,2}= \frac{1+b-a \pm
\sqrt{\Delta}}{2} \, ,
\ee
\be v_{1,2}=\frac{1-b+ a \pm \sqrt{\Delta} }{2} \, .
\ee
For illustration, in figure \ref{fig:SQUIVER} we depict $s(x,y)$ as a function of the parameter $x$ for some indicative, fixed values of $y$.

\subsection{Generalising non decoupled D-terms}
Previously non decoupled D-terms have been used to explore both vector-like and Chiral D-terms for the MSSM Higgses \cite{Batra:2003nj,Craig:2012bs}. It is actually the case that this effective action effects all fields charged under the relevant symmetries and as a result, there will be effective terms for squarks and sleptons too.  This point has so far not been mentioned in the literature. We therefore supply a more general derivation for the two site quiver, whose main result is \refe{superoperator}, although we only include the Higgs contributions in our study. It may also be extend to the three site case.  

For two abelian or non-abelian gauge groups $G_A\times G_B$ that break to the diagonal, one may write canonical kinetic terms for Chiral superfields charged under only site A or site B:
\be
\mathcal{L}\supset \int d^4\theta \Big( \sum_i A^{\dagger}_i e^{g_aV_a}A_i+ \sum_j B^{\dagger}_j e^{g_bV_b}B_j \Big).
\ee
After breaking to the diagonal, there is a massless and massive vector multiplet
\be
V_G=\frac{g_aV_b+g_bV_a}{\sqrt{g^2_a+g^2_b}}   \ \ \  \  \   V_H=\frac{-g_aV_a+g_bV_b}{\sqrt{g^2_a+g^2_b}}  
\ee
and usefully for computing the equation of motion, there is a mass term in the K\"ahler potential
\be
\mathcal{L}\supset \int d^4 \theta \  m_V^2V^2_H +...
\ee
One may in fact add a number of soft mass terms
\be
\mathcal{L}\supset \int d^4 \theta \ (m_{\chi}m_V^2 \theta^2 + \bar{m}_{\chi}m_V^2 \bar{\theta}^2 -\frac{1}{2}m_V^2m_s^2 \theta^4  )V^2+\int d^2 \theta m_{\lambda}W^2_{\alpha} +\int d^2 \bar{\theta} \bar{m}_{\lambda}\bar{W}^2_{\dot{\alpha}}
\ee
to parameterise the soft breaking fermion $\chi$, the real uneaten scalar, and the Majorana soft mass for  $\lambda$, respectively. In terms of standard current multiplets satisfying $D^2 \mathcal{J}=0$,  the K\"ahler potential may be written to leading order in $V_H$ as
\be
 K_{H}\supset g_d\left(\frac{g_a}{g_b}\right)\mathcal{J}_a V_H +  g_d\left(\frac{g_b}{g_a}\right)\mathcal{J}_b V_H + ...
\ee
where $\mathcal{J}_{a/b}$ are the currents that contain all the fields charged under site a or site b:
\be {\cal J}^A = J^A + i \theta j^A - i \bar \theta  \bar j^A -
\theta\sigma^\mu\bar \theta j_\mu^A +
\frac{1}{2}\theta\theta\bar\theta\bar \sigma^\mu\partial_\mu j^A -
\frac{1}{2}\bar\theta\bar\theta\theta \sigma^\mu\partial_\mu \bar
j^A - \frac{1}{4}\theta\theta\bar\theta\bar\theta \Box J^A,\ee
with the leading term being the scalar current $J^A = \sum_i \phi_i^\dag T^A \phi_i$. The diagonal gauge group coupling is $g_d=g_{\text{SM}}$. The effective lagrangian after integrating out $V_H$ will then be of the form
\be
\mathcal{L}_{\text{eff}}=\int d^4\theta \left( \sum_i A^{\dagger}_i e^{g_dV_d}A_i+ \sum_j B^{\dagger}_j e^{g_dV_d}B_j \right)+\mathcal{O}
\ee
where $\mathcal{O}$ is the effective super operator
\be
\mathcal{O}=g_d^2\int d^4 \theta \left(\frac{1}{m_V^2}-\frac{m_s^2\theta^4}{m_V^2 + m_s^2}\right)\sum_A  \Big[\left(\frac{g_a}{g_b}\right)\mathcal{J}^A_a  - \left(\frac{g_b}{g_a}\right)\mathcal{J}^A_b\Big]^2 \label{superoperator}
\ee
with a sum over A generators. It is this effective super operator \refe{superoperator} that is the most general expression for the non decoupled D-term of the two site quiver, and produces both the Chiral and Vector-like non decoupled D-terms as limiting cases.  Explicitly for model MI the scalar currents are given by
\be
J_{U(1)_A}=\frac{1}{2}H^{\dagger}_uH_u-\frac{1}{2} H^{\dagger}_dH_d-\frac{1}{2}\tilde{l}^{\dagger}\tilde{l}+\frac{1}{6}\tilde{q}^{\dagger}\tilde{q}+\frac{1}{3}\tilde{d}^{\dagger}\tilde{d}-\frac{2}{3}\tilde{u}^{\dagger}\tilde{u} +\tilde{e}^{\dagger}\tilde{e} \ \ , \ \ J_{U(1)_B}=0
\ee
\be
J^A_{SU(2)_A}=\frac{1}{2}\left( H^{\dagger}_u\sigma^A H_u+ H^{\dagger}_d\sigma^A H_d+\tilde{q}^{\dagger}\sigma^A \tilde{q} +\tilde{l}^{\dagger}\sigma^A \tilde{l}\right) \ \ , \  \ J^A_{SU(2)_B}=0
\ee
with all flavour, and colour indices implicitly traced.
For MII one finds
\be
J_{U(1)_A}=\frac{1}{2}H^{\dagger}_uH_u-\frac{1}{2} H^{\dagger}_dH_d + \left[-\frac{1}{2}\tilde{l}^{\dagger}\tilde{l}+\frac{1}{6}\tilde{q}^{\dagger}\tilde{q}+\frac{1}{3}\tilde{d}^{\dagger}\tilde{d}-\frac{2}{3}\tilde{u}^{\dagger}\tilde{u} +\tilde{e}^{\dagger}\tilde{e}\right]_{3} , \nonumber \ee
\be
 J_{U(1)_B}=\left[-\frac{1}{2}\tilde{l}^{\dagger}\tilde{l}+\frac{1}{6}\tilde{q}^{\dagger}\tilde{q}+\frac{1}{3}\tilde{d}^{\dagger}\tilde{d}-\frac{2}{3}\tilde{u}^{\dagger}\tilde{u} +\tilde{e}^{\dagger}\tilde{e} \right]_{1,2}
\ee
\be
J^A_{SU(2)_A}=\frac{1}{2}\left( H^{\dagger}_u\sigma^A H_u+ H^{\dagger}_d\sigma^A H_d\right)+\frac{1}{2}\left[\tilde{q}^{\dagger}\sigma^A \tilde{q} +\tilde{l}^{\dagger}\sigma^A \tilde{l}\right]_{3} \ ,  \nonumber \ee
\be J^A_{SU(2)_B}=+\frac{1}{2}\left[\tilde{q}^{\dagger}\sigma^A \tilde{q} +\tilde{l}^{\dagger}\sigma^A \tilde{l}\right]_{1,2} .
\ee
The effective action containing all fields charged under the gauge groups may  be included in regime two of the SARAH model file and due to the square in \refe{superoperator} these terms generate both mass shifts for all charged squarks and sleptons as well as additional quartic vertices.  These additional contributions to branching ratios would need to be included in a precision study involving Higgs and sfermion decays, as might be accessible to an $e^+,e^{-}$ collider such as the ILC.  This effect, albeit subtle, if measured precisely enough would determine which gauge groups each and every matter field is charged under and therefore uncover the  full structure of the model.
\bibliographystyle{JHEP}
\bibliography{susyscan}

\providecommand{\href}[2]{#2}\begingroup\raggedright\begin{thebibliography}{10}

\bibitem{Aad:2012tfa}
{\bf ATLAS} Collaboration, G.~Aad et~al., {\it {Observation of a new particle
  in the search for the Standard Model Higgs boson with the ATLAS detector at
  the LHC}},  {\em Phys.Lett.} {\bf B716} (2012) 1--29,
  [\href{http://xxx.lanl.gov/abs/1207.7214}{{\tt arXiv:1207.7214}}].

\bibitem{Chatrchyan:2012ufa}
{\bf CMS} Collaboration, S.~Chatrchyan et~al., {\it {Observation of a new boson
  at a mass of 125 GeV with the CMS experiment at the LHC}},  {\em Phys.Lett.}
  {\bf B716} (2012) 30--61, [\href{http://xxx.lanl.gov/abs/1207.7235}{{\tt
  arXiv:1207.7235}}].

\bibitem{Batra:2003nj}
P.~Batra, A.~Delgado, D.~E. Kaplan, and T.~M. Tait, {\it {The Higgs mass bound
  in gauge extensions of the minimal supersymmetric standard model}},  {\em
  JHEP} {\bf 0402} (2004) 043,
  [\href{http://xxx.lanl.gov/abs/hep-ph/0309149}{{\tt hep-ph/0309149}}].

\bibitem{Dine:2007xi}
M.~Dine, N.~Seiberg, and S.~Thomas, {\it {Higgs physics as a window beyond the
  MSSM (BMSSM)}},  {\em Phys.Rev.} {\bf D76} (2007) 095004,
  [\href{http://xxx.lanl.gov/abs/0707.0005}{{\tt arXiv:0707.0005}}].

\bibitem{Blum:2012ii}
K.~Blum, R.~T. D'Agnolo, and J.~Fan, {\it {Natural SUSY Predicts: Higgs
  Couplings}},  {\em JHEP} {\bf 1301} (2013) 057,
  [\href{http://xxx.lanl.gov/abs/1206.5303}{{\tt arXiv:1206.5303}}].

\bibitem{Csaki:2001em}
C.~Csaki, J.~Erlich, C.~Grojean, and G.~D. Kribs, {\it {4-D constructions of
  supersymmetric extra dimensions and gaugino mediation}},  {\em Phys.Rev.}
  {\bf D65} (2002) 015003, [\href{http://xxx.lanl.gov/abs/hep-ph/0106044}{{\tt
  hep-ph/0106044}}].

\bibitem{Cheng:2001an}
H.~Cheng, D.~Kaplan, M.~Schmaltz, and W.~Skiba, {\it {Deconstructing gaugino
  mediation}},  {\em Phys.Lett.} {\bf B515} (2001) 395--399,
  [\href{http://xxx.lanl.gov/abs/hep-ph/0106098}{{\tt hep-ph/0106098}}].

\bibitem{Batra:2004vc}
P.~Batra, A.~Delgado, D.~E. Kaplan, and T.~M. Tait, {\it {Running into new
  territory in SUSY parameter space}},  {\em JHEP} {\bf 0406} (2004) 032,
  [\href{http://xxx.lanl.gov/abs/hep-ph/0404251}{{\tt hep-ph/0404251}}].

\bibitem{Delgado:2004pr}
A.~Delgado, {\it {Raising the Higgs mass in supersymmetric models}},
  \href{http://xxx.lanl.gov/abs/hep-ph/0409073}{{\tt hep-ph/0409073}}.

\bibitem{DeSimone:2008gm}
A.~De~Simone, J.~Fan, M.~Schmaltz, and W.~Skiba, {\it {Low-scale gaugino
  mediation, lots of leptons at the LHC}},  {\em Phys.Rev.} {\bf D78} (2008)
  095010, [\href{http://xxx.lanl.gov/abs/0808.2052}{{\tt arXiv:0808.2052}}].

\bibitem{Medina:2009ey}
A.~D. Medina, N.~R. Shah, and C.~E. Wagner, {\it {A Heavy Higgs and a Light
  Sneutrino NLSP in the MSSM with Enhanced SU(2) D-terms}},  {\em Phys.Rev.}
  {\bf D80} (2009) 015001, [\href{http://xxx.lanl.gov/abs/0904.1625}{{\tt
  arXiv:0904.1625}}].

\bibitem{McGarrie:2010qr}
M.~McGarrie, {\it {General Gauge Mediation and Deconstruction}},  {\em JHEP}
  {\bf 1011} (2010) 152, [\href{http://xxx.lanl.gov/abs/1009.0012}{{\tt
  arXiv:1009.0012}}].

\bibitem{Craig:2011yk}
N.~Craig, D.~Green, and A.~Katz, {\it {(De)Constructing a Natural and Flavorful
  Supersymmetric Standard Model}},  {\em JHEP} {\bf 1107} (2011) 045,
  [\href{http://xxx.lanl.gov/abs/1103.3708}{{\tt arXiv:1103.3708}}].

\bibitem{Auzzi:2012dv}
R.~Auzzi, A.~Giveon, S.~B. Gudnason, and T.~Shacham, {\it {A Light Stop with
  Flavor in Natural SUSY}},  {\em JHEP} {\bf 1301} (2013) 169,
  [\href{http://xxx.lanl.gov/abs/1208.6263}{{\tt arXiv:1208.6263}}].

\bibitem{Huo:2012tw}
R.~Huo, G.~Lee, A.~M. Thalapillil, and C.~E. Wagner, {\it {$SU(2)\otimes SU(2)$
  Gauge Extensions of the MSSM Revisited}},  {\em Phys.Rev.} {\bf D87} (2013)
  055011, [\href{http://xxx.lanl.gov/abs/1212.0560}{{\tt arXiv:1212.0560}}].

\bibitem{d2013fitting}
R.~T. D'Agnolo, E.~Kuflik, and M.~Zanetti, {\it {Fitting the Higgs to Natural
  SUSY}},  {\em JHEP} {\bf 1303} (2013) 043,
  [\href{http://xxx.lanl.gov/abs/1212.1165}{{\tt arXiv:1212.1165}}].

\bibitem{Staub:2008uz}
F.~Staub, {\it {SARAH}},  \href{http://xxx.lanl.gov/abs/0806.0538}{{\tt
  arXiv:0806.0538}}.

\bibitem{Staub:2010jh}
F.~Staub, {\it {Automatic Calculation of supersymmetric Renormalization Group
  Equations and Self Energies}},  {\em Comput.Phys.Commun.} {\bf 182} (2011)
  808--833, [\href{http://xxx.lanl.gov/abs/1002.0840}{{\tt arXiv:1002.0840}}].

\bibitem{Staub:2012pb}
F.~Staub, {\it {SARAH 3.2: Dirac Gauginos, UFO output, and more}},  {\em
  Computer Physics Communications} {\bf 184} (2013) pp. 1792--1809,
  [\href{http://xxx.lanl.gov/abs/1207.0906}{{\tt arXiv:1207.0906}}].

\bibitem{Muller:1996dj}
D.~J. Muller and S.~Nandi, {\it {Top flavor: A Separate SU(2) for the third
  family}},  {\em Phys.Lett.} {\bf B383} (1996) 345--350,
  [\href{http://xxx.lanl.gov/abs/hep-ph/9602390}{{\tt hep-ph/9602390}}].

\bibitem{Malkawi:1996fs}
E.~Malkawi, T.~M. Tait, and C.~Yuan, {\it {A Model of strong flavor dynamics
  for the top quark}},  {\em Phys.Lett.} {\bf B385} (1996) 304--310,
  [\href{http://xxx.lanl.gov/abs/hep-ph/9603349}{{\tt hep-ph/9603349}}].

\bibitem{Malkawi:1999sa}
E.~Malkawi and C.~Yuan, {\it {New physics in the third family and its effect on
  low-energy data}},  {\em Phys.Rev.} {\bf D61} (2000) 015007,
  [\href{http://xxx.lanl.gov/abs/hep-ph/9906215}{{\tt hep-ph/9906215}}].

\bibitem{Auzzi:2010mb}
R.~Auzzi and A.~Giveon, {\it {The Sparticle spectrum in Minimal gaugino-Gauge
  Mediation}},  {\em JHEP} {\bf 1010} (2010) 088,
  [\href{http://xxx.lanl.gov/abs/1009.1714}{{\tt arXiv:1009.1714}}].

\bibitem{McGarrie:2011dc}
M.~McGarrie, {\it {Hybrid Gauge Mediation}},  {\em JHEP} {\bf 1109} (2011) 138,
  [\href{http://xxx.lanl.gov/abs/1101.5158}{{\tt arXiv:1101.5158}}].

\bibitem{Craig:2012hc}
N.~Craig, S.~Dimopoulos, and T.~Gherghetta, {\it {Split families unified}},
  {\em JHEP} {\bf 1204} (2012) 116,
  [\href{http://xxx.lanl.gov/abs/1203.0572}{{\tt arXiv:1203.0572}}].

\bibitem{Allanach:2001kg}
B.~Allanach, {\it {SOFTSUSY: a program for calculating supersymmetric
  spectra}},  {\em Comput.Phys.Commun.} {\bf 143} (2002) 305--331,
  [\href{http://xxx.lanl.gov/abs/hep-ph/0104145}{{\tt hep-ph/0104145}}].

\bibitem{modelpdf}
A.~Bharucha, A.~Goudelis, and M.~McGarrie, ``Electroweak quiver gauge model.''
  See supplementary material, October, 2013.

\bibitem{Baer:1993ae}
H.~Baer, F.~E. Paige, S.~D. Protopopescu, and X.~Tata, {\it {Simulating
  Supersymmetry with ISAJET 7.0 / ISASUSY 1.0}},
  \href{http://xxx.lanl.gov/abs/hep-ph/9305342}{{\tt hep-ph/9305342}}.

\bibitem{Djouadi:2002ze}
A.~Djouadi, J.-L. Kneur, and G.~Moultaka, {\it {SuSpect: A Fortran code for the
  supersymmetric and Higgs particle spectrum in the MSSM}},  {\em
  Comput.Phys.Commun.} {\bf 176} (2007) 426--455,
  [\href{http://xxx.lanl.gov/abs/hep-ph/0211331}{{\tt hep-ph/0211331}}].

\bibitem{Porod:2003um}
W.~Porod, {\it {SPheno, a program for calculating supersymmetric spectra, SUSY
  particle decays and SUSY particle production at e+ e- colliders}},  {\em
  Comput.Phys.Commun.} {\bf 153} (2003) 275--315,
  [\href{http://xxx.lanl.gov/abs/hep-ph/0301101}{{\tt hep-ph/0301101}}].

\bibitem{Skands:2003cj}
P.~Z. Skands, B.~Allanach, H.~Baer, C.~Balazs, G.~Belanger, et~al., {\it {SUSY
  Les Houches accord: Interfacing SUSY spectrum calculators, decay packages,
  and event generators}},  {\em JHEP} {\bf 0407} (2004) 036,
  [\href{http://xxx.lanl.gov/abs/hep-ph/0311123}{{\tt hep-ph/0311123}}].

\bibitem{Allanach:2008qq}
B.~Allanach, C.~Balazs, G.~Belanger, M.~Bernhardt, F.~Boudjema, et~al., {\it
  {SUSY Les Houches Accord 2}},  {\em Comput.Phys.Commun.} {\bf 180} (2009)
  8--25, [\href{http://xxx.lanl.gov/abs/0801.0045}{{\tt arXiv:0801.0045}}].

\bibitem{Porod:2011nf}
W.~Porod and F.~Staub, {\it {SPheno 3.1: Extensions including flavour,
  CP-phases and models beyond the MSSM}},  {\em Comput.Phys.Commun.} {\bf 183}
  (2012) 2458--2469, [\href{http://xxx.lanl.gov/abs/1104.1573}{{\tt
  arXiv:1104.1573}}].

\bibitem{Maloney:2004rc}
A.~Maloney, A.~Pierce, and J.~G. Wacker, {\it {D-terms, unification, and the
  Higgs mass}},  {\em JHEP} {\bf 0606} (2006) 034,
  [\href{http://xxx.lanl.gov/abs/hep-ph/0409127}{{\tt hep-ph/0409127}}].

\bibitem{Drees:2004jm}
M.~Drees, R.~Godbole, and P.~Roy, {\it {Theory and phenomenology of sparticles:
  An account of four-dimensional N=1 supersymmetry in high energy physics}}, .

\bibitem{Bando:1987br}
M.~Bando, T.~Kugo, and K.~Yamawaki, {\it {Nonlinear Realization and Hidden
  Local Symmetries}},  {\em Phys.Rept.} {\bf 164} (1988) 217--314.

\bibitem{ArkaniHamed:2001ca}
N.~Arkani-Hamed, A.~G. Cohen, and H.~Georgi, {\it {(De)constructing
  dimensions}},  {\em Phys.Rev.Lett.} {\bf 86} (2001) 4757--4761,
  [\href{http://xxx.lanl.gov/abs/hep-th/0104005}{{\tt hep-th/0104005}}].

\bibitem{Hill:2000mu}
C.~T. Hill, S.~Pokorski, and J.~Wang, {\it {Gauge invariant effective
  Lagrangian for Kaluza-Klein modes}},  {\em Phys.Rev.} {\bf D64} (2001)
  105005, [\href{http://xxx.lanl.gov/abs/hep-th/0104035}{{\tt
  hep-th/0104035}}].

\bibitem{ArkaniHamed:2001nc}
N.~Arkani-Hamed, A.~G. Cohen, and H.~Georgi, {\it {Electroweak symmetry
  breaking from dimensional deconstruction}},  {\em Phys.Lett.} {\bf B513}
  (2001) 232--240, [\href{http://xxx.lanl.gov/abs/hep-ph/0105239}{{\tt
  hep-ph/0105239}}].

\bibitem{McGarrie:2012ks}
M.~McGarrie, {\it {General Resonance Mediation}},  {\em JHEP} {\bf 1303} (2013)
  093, [\href{http://xxx.lanl.gov/abs/1207.4484}{{\tt arXiv:1207.4484}}].

\bibitem{McGarrie:2013hca}
M.~McGarrie, {\it {5D Maximally Supersymmetric Yang-Mills in 4D Superspace:
  Applications}},  {\em JHEP} {\bf 1304} (2013) 161,
  [\href{http://xxx.lanl.gov/abs/1303.4534}{{\tt arXiv:1303.4534}}].

\bibitem{McGarrie:2010yk}
M.~McGarrie and D.~C. Thompson, {\it {Warped General Gauge Mediation}},  {\em
  Phys.Rev.} {\bf D82} (2010) 125034,
  [\href{http://xxx.lanl.gov/abs/1009.4696}{{\tt arXiv:1009.4696}}].

\bibitem{McGarrie:2012fi}
M.~McGarrie, {\it {Holography for General Gauge Mediation}},  {\em JHEP} {\bf
  1302} (2013) 132, [\href{http://xxx.lanl.gov/abs/1210.4935}{{\tt
  arXiv:1210.4935}}].

\bibitem{McGarrie:2010kh}
M.~McGarrie and R.~Russo, {\it {General Gauge Mediation in 5D}},  {\em
  Phys.Rev.} {\bf D82} (2010) 035001,
  [\href{http://xxx.lanl.gov/abs/1004.3305}{{\tt arXiv:1004.3305}}].

\bibitem{Hahn:2000kx}
T.~Hahn, {\it {Generating Feynman diagrams and amplitudes with FeynArts 3}},
  {\em Comput.Phys.Commun.} {\bf 140} (2001) 418--431,
  [\href{http://xxx.lanl.gov/abs/hep-ph/0012260}{{\tt hep-ph/0012260}}].

\bibitem{Hahn:2001rv}
T.~Hahn and C.~Schappacher, {\it {The Implementation of the minimal
  supersymmetric standard model in FeynArts and FormCalc}},  {\em
  Comput.Phys.Commun.} {\bf 143} (2002) 54--68,
  [\href{http://xxx.lanl.gov/abs/hep-ph/0105349}{{\tt hep-ph/0105349}}].

\bibitem{Hahn:1998yk}
T.~Hahn and M.~Perez-Victoria, {\it {Automatized one loop calculations in
  four-dimensions and D-dimensions}},  {\em Comput.Phys.Commun.} {\bf 118}
  (1999) 153--165, [\href{http://xxx.lanl.gov/abs/hep-ph/9807565}{{\tt
  hep-ph/9807565}}].

\bibitem{Pukhov:1999gg}
A.~Pukhov, E.~Boos, M.~Dubinin, V.~Edneral, V.~Ilyin, et~al., {\it {CompHEP: A
  Package for evaluation of Feynman diagrams and integration over multiparticle
  phase space}},  \href{http://xxx.lanl.gov/abs/hep-ph/9908288}{{\tt
  hep-ph/9908288}}.

\bibitem{Bechtle:2008jh}
P.~Bechtle, O.~Brein, S.~Heinemeyer, G.~Weiglein, and K.~E. Williams, {\it
  {HiggsBounds: Confronting Arbitrary Higgs Sectors with Exclusion Bounds from
  LEP and the Tevatron}},  {\em Comput.Phys.Commun.} {\bf 181} (2010) 138--167,
  [\href{http://xxx.lanl.gov/abs/0811.4169}{{\tt arXiv:0811.4169}}].

\bibitem{Bechtle:2011sb}
P.~Bechtle, O.~Brein, S.~Heinemeyer, G.~Weiglein, and K.~E. Williams, {\it
  {HiggsBounds 2.0.0: Confronting Neutral and Charged Higgs Sector Predictions
  with Exclusion Bounds from LEP and the Tevatron}},  {\em Comput.Phys.Commun.}
  {\bf 182} (2011) 2605--2631, [\href{http://xxx.lanl.gov/abs/1102.1898}{{\tt
  arXiv:1102.1898}}].

\bibitem{Bechtle:2013gu}
P.~Bechtle, O.~Brein, S.~Heinemeyer, O.~Stal, T.~Stefaniak, et~al., {\it
  {Recent Developments in HiggsBounds and a Preview of HiggsSignals}},  {\em
  PoS} {\bf CHARGED2012} (2012) 024,
  [\href{http://xxx.lanl.gov/abs/1301.2345}{{\tt arXiv:1301.2345}}].

\bibitem{Kilian:2007gr}
W.~Kilian, T.~Ohl, and J.~Reuter, {\it {WHIZARD: Simulating Multi-Particle
  Processes at LHC and ILC}},  {\em Eur.Phys.J.} {\bf C71} (2011) 1742,
  [\href{http://xxx.lanl.gov/abs/0708.4233}{{\tt arXiv:0708.4233}}].

\bibitem{Belanger:2006is}
G.~Belanger, F.~Boudjema, A.~Pukhov, and A.~Semenov, {\it {MicrOMEGAs 2.0: A
  Program to calculate the relic density of dark matter in a generic model}},
  {\em Comput.Phys.Commun.} {\bf 176} (2007) 367--382,
  [\href{http://xxx.lanl.gov/abs/hep-ph/0607059}{{\tt hep-ph/0607059}}].

\bibitem{Camargo-Molina:2013qva}
J.~Camargo-Molina, B.~O'Leary, W.~Porod, and F.~Staub, {\it {Vevacious: A Tool
  For Finding The Global Minima Of One-Loop Effective Potentials With Many
  Scalars}},  \href{http://xxx.lanl.gov/abs/1307.1477}{{\tt arXiv:1307.1477}}.

\bibitem{Degrassi:2001yf}
G.~Degrassi, P.~Slavich, and F.~Zwirner, {\it {On the neutral Higgs boson
  masses in the MSSM for arbitrary stop mixing}},  {\em Nucl.Phys.} {\bf B611}
  (2001) 403--422, [\href{http://xxx.lanl.gov/abs/hep-ph/0105096}{{\tt
  hep-ph/0105096}}].

\bibitem{Brignole:2001jy}
A.~Brignole, G.~Degrassi, P.~Slavich, and F.~Zwirner, {\it {On the
  O(alpha(t)**2) two loop corrections to the neutral Higgs boson masses in the
  MSSM}},  {\em Nucl.Phys.} {\bf B631} (2002) 195--218,
  [\href{http://xxx.lanl.gov/abs/hep-ph/0112177}{{\tt hep-ph/0112177}}].

\bibitem{Brignole:2002bz}
A.~Brignole, G.~Degrassi, P.~Slavich, and F.~Zwirner, {\it {On the two loop
  sbottom corrections to the neutral Higgs boson masses in the MSSM}},  {\em
  Nucl.Phys.} {\bf B643} (2002) 79--92,
  [\href{http://xxx.lanl.gov/abs/hep-ph/0206101}{{\tt hep-ph/0206101}}].

\bibitem{Dedes:2002dy}
A.~Dedes and P.~Slavich, {\it {Two loop corrections to radiative electroweak
  symmetry breaking in the MSSM}},  {\em Nucl.Phys.} {\bf B657} (2003)
  333--354, [\href{http://xxx.lanl.gov/abs/hep-ph/0212132}{{\tt
  hep-ph/0212132}}].

\bibitem{Craig:2012bs}
N.~Craig and A.~Katz, {\it {A Supersymmetric Higgs Sector with Chiral
  D-terms}},  {\em JHEP} {\bf 1305} (2013) 015,
  [\href{http://xxx.lanl.gov/abs/1212.2635}{{\tt arXiv:1212.2635}}].

\bibitem{ATLAS:2013hta}
{\bf ATLAS} Collaboration, {\it {Physics at a High-Luminosity LHC with ATLAS}},
   \href{http://xxx.lanl.gov/abs/1307.7292}{{\tt arXiv:1307.7292}}.

\bibitem{CMS:2013xfa}
{\bf CMS} Collaboration, {\it {Projected Performance of an Upgraded CMS
  Detector at the LHC and HL-LHC: Contribution to the Snowmass Process}},
  \href{http://xxx.lanl.gov/abs/1307.7135}{{\tt arXiv:1307.7135}}.

\bibitem{Chatrchyan:2013lba}
{\bf CMS} Collaboration, S.~Chatrchyan et~al., {\it {Observation of a new boson
  with mass near 125 GeV in pp collisions at sqrt(s) = 7 and 8 TeV}},  {\em
  JHEP} {\bf 06} (2013) 081, [\href{http://xxx.lanl.gov/abs/1303.4571}{{\tt
  arXiv:1303.4571}}].

\bibitem{Degrassi:2002fi}
G.~Degrassi, S.~Heinemeyer, W.~Hollik, P.~Slavich, and G.~Weiglein, {\it
  {Towards high precision predictions for the MSSM Higgs sector}},  {\em
  Eur.Phys.J.} {\bf C28} (2003) 133--143,
  [\href{http://xxx.lanl.gov/abs/hep-ph/0212020}{{\tt hep-ph/0212020}}].

\bibitem{Arbey:2012dq}
A.~Arbey, M.~Battaglia, A.~Djouadi, and F.~Mahmoudi, {\it {The Higgs sector of
  the phenomenological MSSM in the light of the Higgs boson discovery}},  {\em
  JHEP} {\bf 1209} (2012) 107, [\href{http://xxx.lanl.gov/abs/1207.1348}{{\tt
  arXiv:1207.1348}}].

\bibitem{Beringer:1900zz}
{\bf Particle Data Group} Collaboration, J.~Beringer et~al., {\it {Review of
  Particle Physics (RPP)}},  {\em Phys.Rev.} {\bf D86} (2012) 010001.

\bibitem{ATLAS-CONF-2013-047}
{\bf ATLAS} Collaboration, {\it Search for squarks and gluinos with the atlas
  detector in final states with jets and missing transverse momentum and 20.3
  fb$^{-1}$ of $\sqrt{s}=8$ tev proton-proton collision data},  Tech. Rep.
  ATLAS-CONF-2013-047, CERN, Geneva, May, 2013.

\bibitem{Chatrchyan:2012lia}
{\bf CMS} Collaboration, S.~Chatrchyan et~al., {\it {Search for new physics in
  the multijet and missing transverse momentum final state in proton-proton
  collisions at $\sqrt{s} = 7$ TeV}},  {\em Phys.Rev.Lett.} {\bf 109} (2012)
  171803, [\href{http://xxx.lanl.gov/abs/1207.1898}{{\tt arXiv:1207.1898}}].

\bibitem{Baer:2013gva}
H.~Baer, V.~Barger, and D.~Mickelson, {\it {How conventional measures
  overestimate electroweak fine-tuning in supersymmetric theory}},
  \href{http://xxx.lanl.gov/abs/1309.2984}{{\tt arXiv:1309.2984}}.

\bibitem{LEP2}
{\bf ALEPH, DELPHI, L3 and OPAL} Collaboration, LEPSUSYWG, {\it {notes
  LEPSUSYWG/02-04.1 and LEPSUSYWG/01-03.1}},  2001.

\bibitem{Barate:1999gm}
{\bf ALEPH} Collaboration, R.~Barate et~al., {\it {Search for gauge mediated
  SUSY breaking topologies at $S^{(1/2)}$ similar to 189-GeV}},  {\em
  Eur.Phys.J.} {\bf C16} (2000) 71--85.

\bibitem{Barate:2000tu}
{\bf ALEPH} Collaboration, R.~Barate et~al., {\it {Search for supersymmetric
  particles in $e^{+} e^{-}$ collisions at $\sqrt{s}$ up to 202-GeV and mass
  limit for the lightest neutralino}},  {\em Phys.Lett.} {\bf B499} (2001)
  67--84, [\href{http://xxx.lanl.gov/abs/hep-ex/0011047}{{\tt
  hep-ex/0011047}}].

\bibitem{Decamp:1991uy}
{\bf ALEPH} Collaboration, D.~Decamp et~al., {\it {Searches for new particles
  in $Z$ decays using the ALEPH detector}},  {\em Phys.Rept.} {\bf 216} (1992)
  253--340.

\bibitem{Bechtle:2013xfa}
P.~Bechtle, S.~Heinemeyer, O.~St{\aa}l, T.~Stefaniak, and G.~Weiglein, {\it
  {HiggsSignals: Confronting arbitrary Higgs sectors with measurements at the
  Tevatron and the LHC}},  \href{http://xxx.lanl.gov/abs/1305.1933}{{\tt
  arXiv:1305.1933}}.

\bibitem{Amhis:2012bh}
{\bf Heavy Flavor Averaging Group} Collaboration, Y.~Amhis et~al., {\it
  {Averages of B-Hadron, C-Hadron, and tau-lepton properties as of early
  2012}},  \href{http://xxx.lanl.gov/abs/1207.1158}{{\tt arXiv:1207.1158}}.

\bibitem{Aaij:2012nna}
{\bf LHCb} Collaboration, R.~Aaij et~al., {\it {First Evidence for the Decay
  $B^0_s \to \mu^+\mu^-$}},  {\em Phys.Rev.Lett.} {\bf 110} (2013) 021801,
  [\href{http://xxx.lanl.gov/abs/1211.2674}{{\tt arXiv:1211.2674}}].

\bibitem{Haber:1996fp}
H.~E. Haber, R.~Hempfling, and A.~H. Hoang, {\it {Approximating the radiatively
  corrected Higgs mass in the minimal supersymmetric model}},  {\em Z.Phys.}
  {\bf C75} (1997) 539--554,
  [\href{http://xxx.lanl.gov/abs/hep-ph/9609331}{{\tt hep-ph/9609331}}].

\bibitem{Benbrik:2012rm}
R.~Benbrik, M.~Gomez~Bock, S.~Heinemeyer, O.~Stal, G.~Weiglein, et~al., {\it
  {Confronting the MSSM and the NMSSM with the Discovery of a Signal in the two
  Photon Channel at the LHC}},  {\em Eur.Phys.J.} {\bf C72} (2012) 2171,
  [\href{http://xxx.lanl.gov/abs/1207.1096}{{\tt arXiv:1207.1096}}].

\bibitem{Bechtle:2012jw}
P.~Bechtle, S.~Heinemeyer, O.~Stal, T.~Stefaniak, G.~Weiglein, et~al., {\it
  {MSSM Interpretations of the LHC Discovery: Light or Heavy Higgs?}},  {\em
  Eur.Phys.J.} {\bf C73} (2013) 2354,
  [\href{http://xxx.lanl.gov/abs/1211.1955}{{\tt arXiv:1211.1955}}].

\bibitem{Aad:2013wqa}
{\bf ATLAS} Collaboration, G.~Aad et~al., {\it {Measurements of Higgs boson
  production and couplings in diboson final states with the ATLAS detector at
  the LHC}},  {\em Phys.Lett.} {\bf B} (2013)
  [\href{http://xxx.lanl.gov/abs/1307.1427}{{\tt arXiv:1307.1427}}].

\bibitem{CMS:yva}
{\bf CMS} Collaboration, {\it {Combination of standard model Higgs boson
  searches and measurements of the properties of the new boson with a mass near
  125 GeV}},  Tech. Rep. CMS-PAS-HIG-12-045, CERN, Geneva, 2012.

\bibitem{Baer:2013cma}
H.~Baer, T.~Barklow, K.~Fujii, Y.~Gao, A.~Hoang, et~al., {\it {The
  International Linear Collider Technical Design Report - Volume 2: Physics}},
  \href{http://xxx.lanl.gov/abs/1306.6352}{{\tt arXiv:1306.6352}}.

\bibitem{ATLAS:2012wma}
{\bf ATLAS} Collaboration, {\it {Coupling properties of the new Higgs-like
  boson observed with the ATLAS detector at the LHC}},  Tech. Rep.
  ATLAS-CONF-2012-127, ATLAS-COM-CONF-2012-161, CERN, Geneva, 2012.

\bibitem{Aad:2012zza}
{\bf ATLAS} Collaboration, G.~Aad et~al., {\it {Search for diphoton events with
  large missing transverse momentum in 7 TeV proton-proton collision data with
  the ATLAS detector}},  {\em Phys.Lett.} {\bf B718} (2012) 411--430,
  [\href{http://xxx.lanl.gov/abs/1209.0753}{{\tt arXiv:1209.0753}}].

\bibitem{Aad:2012jva}
{\bf ATLAS} Collaboration, G.~Aad et~al., {\it {Search for supersymmetry in
  events with photons, bottom quarks, and missing transverse momentum in
  proton-proton collisions at a centre-of-mass energy of 7 TeV with the ATLAS
  detector}},  {\em Phys.Lett.} {\bf B719} (2013) 261--279,
  [\href{http://xxx.lanl.gov/abs/1211.1167}{{\tt arXiv:1211.1167}}].

\bibitem{Giudice:1998bp}
G.~Giudice and R.~Rattazzi, {\it {Theories with gauge mediated supersymmetry
  breaking}},  {\em Phys.Rept.} {\bf 322} (1999) 419--499,
  [\href{http://xxx.lanl.gov/abs/hep-ph/9801271}{{\tt hep-ph/9801271}}].

\bibitem{Aad:2013oua}
{\bf ATLAS} Collaboration, G.~Aad et~al., {\it {Search for non-pointing photons
  in the diphoton and $E_T^miss$ final state in sqrt(s) = 7 TeV proton-proton
  collisions using the ATLAS detector}},  {\em Phys.Rev.} {\bf D88} (2013)
  012001, [\href{http://xxx.lanl.gov/abs/1304.6310}{{\tt arXiv:1304.6310}}].

\bibitem{Chatrchyan:2012jwg}
{\bf CMS} Collaboration, S.~Chatrchyan et~al., {\it {Search for long-lived
  particles decaying to photons and missing energy in proton-proton collisions
  at $\sqrt{s}=7$ TeV}},  {\em Phys.Lett.} {\bf B722} (2013) 273--294,
  [\href{http://xxx.lanl.gov/abs/1212.1838}{{\tt arXiv:1212.1838}}].

\bibitem{Chatrchyan:2012bba}
{\bf CMS} Collaboration, S.~Chatrchyan et~al., {\it {Search for new physics in
  events with photons, jets, and missing transverse energy in $pp$ collisions
  at $\sqrt{s}=7$ TeV}},  {\em JHEP} {\bf 1303} (2013) 111,
  [\href{http://xxx.lanl.gov/abs/1211.4784}{{\tt arXiv:1211.4784}}].

\bibitem{Allanach:2002nj}
B.~Allanach, M.~Battaglia, G.~Blair, M.~S. Carena, A.~De~Roeck, et~al., {\it
  {The Snowmass points and slopes: Benchmarks for SUSY searches}},  {\em
  Eur.Phys.J.} {\bf C25} (2002) 113--123,
  [\href{http://xxx.lanl.gov/abs/hep-ph/0202233}{{\tt hep-ph/0202233}}].

\bibitem{ATLAS:2012crz}
{\bf ATLAS} Collaboration, {\it {Search for supersymmetry in final states with
  jets, missing transverse momentum and a $Z$ boson at $\sqrt{s}=8$ TeV with
  the ATLAS detector}},  Tech. Rep. ATLAS-CONF-2012-152, CERN, Geneva, Nov,
  2012.

\bibitem{CMS:2013zzz}
{\bf CMS} Collaboration, {\it {A search for anomalous production of events with
  three or more leptons using 19.5/fb of sqrt(s)=8 TeV LHC data}},  Tech. Rep.
  CMS-PAS-SUS-13-002, CERN, Geneva, 2013.

\bibitem{ATLAS:2013ama}
{\bf ATLAS} Collaboration, {\it {Search for Supersymmetry in Events with Large
  Missing Transverse Momentum, Jets, and at Least One Tau Lepton in 21
  $fb^{-1}$ of $\sqrt{s}$ = 8 TeV Proton-Proton Collision Data with the ATLAS
  Detector}},  Tech. Rep. ATLAS-CONF-2013-026, CERN, Geneva, Mar, 2013.

\bibitem{Bharucha:2013epa}
A.~Bharucha, S.~Heinemeyer, and F.~von~der Pahlen, {\it {Direct
  Chargino-Neutralino Production at the LHC: Interpreting the Exclusion Limits
  in the Complex MSSM}},  \href{http://xxx.lanl.gov/abs/1307.4237}{{\tt
  arXiv:1307.4237}}.

\bibitem{Auzzi:2011wt}
R.~Auzzi, A.~Giveon, and S.~B. Gudnason, {\it {Mediation of Supersymmetry
  Breaking in Quivers}},  {\em JHEP} {\bf 1112} (2011) 016,
  [\href{http://xxx.lanl.gov/abs/1110.1453}{{\tt arXiv:1110.1453}}].

\end{thebibliography}\endgroup

\end{document}